\documentclass[aps,prb,twocolumn,showpacs,floatfix,twoside,eqsecnum]{revtex4-1}

\usepackage{amssymb}
\usepackage{graphicx}
\usepackage{amsmath}
\usepackage{color}
\usepackage{xcolor}
\usepackage{float}
\usepackage{subcaption}
\usepackage[justification=centerlast,font={small}]{caption}
\usepackage{natbib}

\begin{document}

\title{Mean-field thermodynamic quantum time-space crystal: spontaneous\\
breaking of time-translation symmetry in a macroscopic fermion\\
system.}
\author{Konstantin B. Efetov}
\affiliation{Ruhr University Bochum, Faculty of Physics and Astronomy, Bochum, 44780,
Germany }
\affiliation{National University of Science and Technology \textquotedblleft
MISiS\textquotedblright , Moscow, 119049, Russia}
\affiliation{International Institute of Physics, UFRN, 59078-400 Natal, Brazil}
\date{\today }

\begin{abstract}
A model demonstrating existence of a thermodynamically stable quantum
time-space crystal has been proposed and studied. This state is
characterized by an order parameter periodic in both real and imaginary
times. The average of the order parameter over phases of the oscillations
vanishes but correlation functions of two or more order parameters show
non-decaying oscillations. An alternative interpretation of the results is
based on a concept of an operator order parameter introduced for this
purpose. The model studied here has been suggested previously, in
particular, for describing the pseudogap state in superconducting cuprates.
Although many properties of the time-space crystal considered here are close
to those of a well known DDW state, static magnetic moments oscillating at $%
\left( \pi ,\pi \right) $ do not exist. Instead, $\delta$-peaks at finite
energies are predicted in the cross-section of inelastic spin-polarized
neutron scattering.
\end{abstract}

\maketitle

\section{Introduction}

\subsection{What are thermodynamic quantum time-crystals?}

Crystalline structures play a very important role in modern condensed matter
physics and material science. These can be periodic arrays of atoms in
metals and insulators but very often they arise as a result of a sharp phase
transition. Antiferromagnets, charge density waves and many other states of
matter can serve as well known examples of space crystals. If a space
crystal appears at a certain critical temperature or a critical parameter
characterizing the system one can expect sharp changes of physical
quantities at the critical point. The standard way of describing a phase
transition is based on the concept of an order parameter introduced by
Landau \cite{landau}. This quantity equals zero in the disordered phase but
is finite in the ordered one.

At the same time, one should be a little bit more careful with this
definition because the formal average of the order parameter over
thermodynamic states can strictly speaking be equal to zero due to the
degeneracy of the free energy functional at the minimum. Indeed, the average
magnetic moment of an antiferromagnet or the charge of a charge density wave
(CDW) equals zero because the energies of the states of the structures
shifted by, e.g., half a period are equal to each other and the average
vanishes. We write this property in the form
\begin{equation}
\int \rho \left( \mathbf{r-r}_{0}\right) d\mathbf{r}_{0}=0,  \label{i0}
\end{equation}%
where, for example, $\rho \left( \mathbf{r}\right) =\rho _{0}\cos \left(
\mathbf{Qr}\right) $ can be the order parameter of a CDW. The integration is
performed over the period along a primitive vector and Eq. (\ref{i0}) should
be valid for any direction. Of course, depending on the system under
consideration there can be additional types of averaging over the
degeneracies of the ground state. The energy of a superconductor does not
depend on phase and averaging over the phase gives zero, the direction of
the antiferromagnetic moment can be arbitrary and the average over the
directions equals zero. One can easily continue the list of examples but, to
simplify the discussion, we will have in mind the averaging as it is written
in Eq. (\ref{i0}).

One may ask whether it makes a sense or not to average over the degeneracies
of the ground state. Actually, this depends on the type of the experiment
designed to probe the material. In many situations, one considers systems
with contacts selecting one of the states. However, probing a system with
the help of e.g. X-ray scattering one detects a contribution coming from
many parts of a sample, and the order parameter can have different phases in
different parts of it. Then, one should average over the position of CDW.

In order to avoid the ambiguity of the definition of the order parameter one
speaks usually of a long-range order. For example, CDW is characterized by a
correlation function
\begin{equation}
K\left( \mathbf{r}\right) =\int \rho \left( \mathbf{r}_{0}\right) \rho
\left( \mathbf{r-r}_{0}\right) dV_{0},  \label{i1}
\end{equation}%
where the integration is performed over the elementary cell. The long-range
order is determined by a non-decaying asymptotic behavior of the correlation
function $K\left( \mathbf{r}\right) $ at infinity%
\begin{equation}
\lim_{\left\vert \mathbf{r}\right\vert \rightarrow \infty }K\left( \mathbf{r}%
\right) \propto \cos \left( \mathbf{Qr}\right) ,  \label{i2}
\end{equation}%
where $\mathbf{Q}$ is vector of the CDW oscillations. It is the long-range
order that characterizes any crystalline structure unambiguously.

Space and time play in many respects a similar role, and it looks quite
natural to extend the notion of the space crystals to (thermodynamic)
time-space crystals just adding an additional time coordinate to the above
definitions and using coordinates $R=(t,\mathbf{r})$ on equal footing.
Suppose, one comes to an order parameter $B\left( R\right) $ oscillating
both in time and space. Again, one should assume in analogy with Eq. (\ref%
{i0}) that
\begin{equation}
\int B\left( R-R_{0}\right) dR_{0}=0,  \label{i3}
\end{equation}%
where $R_{0}=\left( t_{0},\mathbf{r}_{0}\right) .$ The integration in Eq. (%
\ref{i3}) is performed over the period in time-space along a primitive
vector. Now the primitive vectors include the one directed along the time
axis. The integral (\ref{i3}) is assumed to be zero for the integration
along any primitive vector including integration over time.

It is clear that already averaging over $t_{0}$ must give zero (or a
time-independent constant) if one speaks of a state in the thermodynamic
equilibrium. Indeed, there cannot be any selected $t_{0}$ in the
equilibrium, although a certain time $t_{0}$ appears naturally in
non-equilibrium situations marking the beginning of a process. This fact can
be emphasized explicitly by the following integral
\begin{equation}
\int_{\mathrm{period}}B\left( t-t_{0},\mathbf{r-r}_{0}\right) dt_{0}=0.
\label{i3b}
\end{equation}%
In Eq. (\ref{i3b}) the integration is performed over the period of the
function $B\left( t\right) .$

The long-range order both in space and time is introduced using the
correlation function%
\begin{equation}
K\left( R\right) =\int B\left( -R_{0}\right) B\left( R-R_{0}\right) d\Omega
_{0},  \label{i3a}
\end{equation}%
where the integration is performed over the elementary cell in time-space.
The long-range order in time-space is determined by the following asymptotic
behavior
\begin{equation}
\lim_{\left\vert \mathbf{r}\right\vert \rightarrow \infty ,\left\vert
t\right\vert \rightarrow \infty }K\left( R\right) \propto \cos \left(
\mathbf{Qr}\right) \cos \left( \Gamma t\right) ,  \label{i4}
\end{equation}%
and $\Gamma $ is a characteristic energy.

Here, only macroscopic systems with the volume of the system $V\rightarrow
\infty $ in the thermodynamic equilibrium are considered, and Eqs. (\ref{i4}%
) should be valid for arbitrarily large volumes and times. Both $\mathbf{Q}$
and $\Gamma $ in Eq. (\ref{i4}) are supposed to be independent on the volume
$V\rightarrow \infty .$ Oscillations in time of two-times correlation
functions are very well known in, e.g., two-or more level systems. However,
in the limit $V\rightarrow \infty $ the level spacing $\Delta $ in such
systems goes to zero and the frequency of the oscillations vanishes, which
contrasts Eq. (\ref{i4}) written for finite $\Gamma $ dependent on internal
parameters of the model but not on the volume. The systems possessing these
properties are classified here as `thermodynamic quantum time-space
crystals'. To the best of my knowledge, this type of behavior has not been
known so far.

Although Eqs. (\ref{i3}-\ref{i4}) written in analogy with Eqs. (\ref{i1}-\ref%
{i2}) are just a guess, they, if accepted, can exclude certain types of
proposals on how to realize the time crystals. Quantum-mechanical averaging
of an operator can lead to classical oscillations of, e.g., currents, and
this would result in a radiation and a loss of energy, which is not possible
in a thermodynamically stable state. The energy can also be lost due
radiation of phonons in solid, etc.. However, as follows from the present
discussion, any time-dependent order parameter of a thermodynamically stable
state can appear only in the form of $B\left( t-t_{0},\mathbf{r}\right) $
with arbitrary $t_{0}$. Then, integrating over $t_{0}$ like it is done in
Eq. (\ref{i3b}) gives zero (or constant), and one cannot have anything like
currents oscillating in time. (If the integral in Eq. (\ref{i3b}) is a
time-independent constant one can subtract the corresponding constant from
the definition of $B\left( t,\mathbf{r}\right) $). All this means that Eq. (%
\ref{i3b}) is a necessary condition for any model proposed for the
thermodynamically stable time crystal, and the latter is an essentially
quantum phenomenon. At the same time, Fourier transform of two-time
correlation functions determines scattering amplitudes, and therefore there
should be possibility to observe the thermodynamically stable time crystals
experimentally.

It is important to emphasize that, in this introductory section, we merely
want to discuss some possibilities of introducing the thermodynamic quantum
time crystals (TQTC) without violating laws of the nature, and, at the same
time, make this notion very similar to the space crystals. The discussion of
this section is not necessary for performing explicit calculations within a
model introduced later in the paper, and serves only for visualizing results
in simple terms.

Several years ago Wilczek \cite{wilczek} has proposed a concept of quantum
time crystals using a rather simple model that possessed a state with a
current oscillating in time. However, a more careful consideration of the
model \cite{bruno} has led to the conclusion that this was not the
equilibrium state. These publications were followed by a hot discussion of
the possibility of realization of a thermodynamically stable quantum time
crystal \cite{wilczek1,li,bruno1,bruno2,nozieres,wilczek2}. More general
arguments against thermodynamically stable quantum time crystals in a
macroscopic system have been presented later \cite{watanabe}. As a result, a
consensus has been achieved that thermodynamical macroscopic quantum time
crystals could not exist.

Slowly decaying oscillations in systems out of equilibrium are not forbidden
by the `no-go' theorems, and their study is definitely interesting by its
own. Recent theoretical \cite{volovik,sacha,sondhi1,sondhi2,nayak,yao} and
experimental \cite{autti,zhang,choi} works have clearly demonstrated that
this research field is very interesting and is fast growing. At present, the
term `Quantum Time Crystal' is usually used for non-equilibrium systems. It
is difficult to cite here all papers already published in this direction of
research but this activity is clearly different from the investigation of
the possibility of the thermodynamically stable time crystal presented below.

It comes as a great surprise that a time-space crystal may exist as a
thermodynamically stable state \cite{efetovPRL}. We use the term
`Thermodynamic Quantum Time Crystal' (TQTC) here to distinguish between the
equilibrium and out-of-equilibrium states.

It turns out that the TQTC state can appear as a result of the breaking of
the time-translation invariance of the original model and formation of a
time-dependenent order parameter $B\left( t\right) $ with the properties
described by Eqs. (\ref{i3}-\ref{i4}). This is a new effect. Actually, the
`no-go' theorem \cite{watanabe} is proven only for models where such a
symmetry breaking does not occur and is definitely valid for conventional
models considered in the past. However, the formation of the time-dependent
order parameter invalidates the proof and this will be discussed in detail
later.

In this paper, considering a model of interacting fermions it is
demonstrated that the system can undergo a phase transition into a state
with an order parameter oscillating in both imaginary $\tau $ and real $t$
time. Studying the behavior in imaginary time $\tau $ is necessary for
calculation of the free energy of the system, which is a standard very
convenient method in quantum field theory. The period of the oscillations in
the imaginary time equals $1/mT,$ where $m$ is integer, as required by
boundary conditions for bosonic fields. The phase of the oscillations is
arbitrary and the average over the position both in real and imaginary time
of the periodic structure equals zero, in agreement with Eqs. (\ref{i3}, \ref%
{i3b}). Therefore, the system does not lose energy, which is the necessary
condition for the thermodynamic equilibrium. The correlation function of the
order parameters at real times has the form of Eq. (\ref{i4}), and its
Fourier-transform determines the quantum scattering cross section. The TQTC
obtained here can exist in arbitrarily big volume and is a completely new
type of ordered states of matter.

Although being rather general, the model considered here has been introduced
previously in a slightly different form of spin-fermion model with
overlapping hot spots (SFMOHS) for description of underdoped superconducting
cuprates \cite{volkov1,volkov2,volkov3}. The new state of TQTC obtained
within this model is characterized by a loop currents order parameter
oscillating both in space and time. The phase of the oscillations in time is
arbitrary and averaging over the latter gives zero. As a result, the time
reversal symmetry is broken but no static magnetic moments appear. These
features may correspond to the pseudogap state \cite{timusk,norman,hashimoto}
and we make explicit calculations and obtain results having in mind this
possibility.

\subsection{Pseudogap state in superconducting cuprates.}

The pseudgap state is characterized by the loss of density of states due to
the opening of a partial gap at the Fermi level below the pseudogap
temperature $T^{\ast }>T_{c},$ where $T_{c}$ is the superconducting
temperature. This gap decreases monotonously with the hole doping, which has
been first observed in NMR (Knight shift) \cite{warren,alloul} and, more
recently, ARPES \cite{hashimoto,damascelli} and Raman \cite{benhabib,loret}
scattering studies.

However, modern experiments add a lot of unconventional details to this
picture, showing that various ordering tendencies play a crucial role in the
pseudogap state. The point-group symmetry of the $CuO_{2}$ planes is broken
in the pseudogap phase, which is seen from scanning tunneling microscopy
(STM) \cite{kohsaka,lawler} and transport studies \cite{daou,cyr} of the
pseudogap phase. More recently, magnetic torque measurements \cite{sato} of
the bulk magnetic susceptibility confirmed $C_{4}$ breaking occurring at $%
T^{\ast }$. Additionally, an inversion symmetry breaking associated with
pseudogap has been discovered by means of second harmonic optical anisotropy
measurement \cite{zhao}.

Other experiments suggest that an unconventional time-reversal symmetry
breaking can also be associated with the pseudogap. Polarized neutron
diffraction studies of different cuprate families reveal a magnetic signal
commensurate with the lattice appearing below $T^{\ast }$ and interpreted as
being due to a $\mathbf{Q}=0$ intra-unit cell magnetic order \cite%
{fauque,sidis}. The signal has been observed to start developing above $%
T^{\ast }$ with a finite correlation length \cite{mangin,mangin1} and breaks
the $C_{4}$ symmetry of such a signal. Additionally, at a temperature Tk
that is below Tc but shares a similar doping dependence, polar Kerr effect
has been observed \cite{xia,he}, which indicates \cite{kapitulnik,cho} that
time-reversal symmetry is broken. Additional signatures of a temporally
fluctuating magnetism below $T^{\ast }$ are also available from the recent
$\mu$%
Sr studies \cite{pal}.

While the signatures described above indicate that the pseudogap is a
distinct phase with a lower symmetry, there also exist experiments \cite%
{timusk,shekhter} with thermodynamic evidence for a corresponding phase
transition. Transport measurements suggest the existence of quantum critical
points (QCPs) of the pseudogap phase \cite{badoux}, accompanied by strong
mass enhancement \cite{ramshaw} in line with the existence of a QCP.

As for theory, one of the initial interpretations was that the pseudogap is
a manifestation of a fluctuating superconductivity, either in a form of
preformed Bose pairs \cite{randeria,alexandrov} or strong phase fluctuations
\cite{emery}.

However, the onset temperatures of superconducting fluctuations observed in
experiments \cite{li1,alloul1} are considerably below $T^{\ast }$ and have a
distinct doping dependence. Another scenario dating back to Ref. \cite%
{anderson} attributes the pseudogap to strong short-range correlations due
to a strong on-site repulsion. However, this scenario does not explain the
broken symmetries of the pseudogap state.

A different class of proposals for explaining the pseudogap involves a
competing symmetry-breaking order. One of the possible candidates discussed
in the literature is a $\mathbf{Q}=0$ orbital loop current order \cite%
{varma1,varma2,varma3}. While successfully describing some of the
experimentally observed phenomena, in particular, those of Refs. \cite%
{fauque,sidis,mangin,mangin1}, it does not lead to a gap on the Fermi
surface at the mean-field level. Numerical studies of the three-band Hubbard
model give arguments both for \cite{weber,weber1} and against \cite%
{thomale,nishimoto,kung} this type of order. Moreover, in contrast to the
experiments \cite{fauque,sidis,mangin,mangin1} the loop currents at $\mathbf{%
Q}=0$ have not been observed in recent experiments \cite{croft} on neutron
scattering, which is in agreement with negative results of experiments on
nuclear magnetic resonance and spin rotation measurements.

More recent studies focused on the important role of the interplay between
CDW and superconducting fluctuations \cite{efetov2013}, preemptive orders
\cite{wang2014}, CDW phase fluctuations \cite{caprara}, and hypothetical
SU(2) symmetry \cite{pepin2014,kloss}. Breaking of the time reversal and $%
C_{4}$ symmetries is not easily obtained from those models, though.

An interesting possibility is the d-density wave (DDW) state \cite%
{chakravarty} (also known as orbital flux phase \cite{affleck,marston})
which is characterized by a pattern of bond currents modulated with the
wavevector $\mathbf{Q}=\left( \pi ,\pi \right) $ and is not generally
accompanied by a charge modulation. This order leads to a reconstructed
Fermi surface consistent with the transport \cite{badoux,storey} and ARPES
\cite{hashimoto} signatures of the pseudogap. Moreover, the time-reversal
symmetry is also broken and a modified version of DDW can explain the polar
Kerr effect \cite{sharma} observation. Additionally, model calculations show
\cite{atkinson,makhfudz} that the system in the DDW state can be unstable to
the formation of axial CDWs.

Experimental studies \cite{mook,mook1} aimed at direct detection of magnetic
moments created by the DDW state seemed to give results in favor of the
existence of the magnetic moments. At the same time, neutron scattering \cite%
{stock} and $\mu SR$ \cite{sonier} experiments have rather unambiguously
demonstrated absence of any static magnetic order at $\mathbf{Q}=\left( \pi
,\pi \right) $. At present, there is no clear evidence for existence in the
cuprates of static magnetic moments corresponding to the DDW.

In principle, the idea of using the DDW state for explanation of the origin
of the pseudogap state might look promising \cite{chakravarty} because many
observed effects correlate with predictions of this proposal (see, e.g.
Refs. \cite{tewari,trunin,sharma}). At the same time, as already mentioned,
a magnetic structure with the vector $\mathbf{Q}=\left( \pi ,\pi \right) $
has not really been confirmed experimentally, although a time dependent
magnetism was seen recently \cite{pal} below $T^{\ast }$.

If the pseudogap state really corresponded to a temporarily oscillating loop
currents order parameter with $\mathbf{Q}=\left( \pi ,\pi \right) $, the
magnetic moments would not be seen in neutron elastic scattering
experiments. At the same time, the time reversal symmetry would be broken
with all consequences and the gap in the spectrum would exist in the
antinodal regions. The goal of this paper is to show that the TQTC state
leading to such a picture is really possible.

\subsection{Plan of the presentation.}

In this paper we will demonstrate considering a spin-fermion model that the
TQTC can exist and the properties of the order parameter correspond to those
written in Eqs. (\ref{i3}-\ref{i4}). It has been shown previously \cite%
{volkov3} that the model under consideration can give the DDW state. Now we
will show that the model allows one to obtain a state with zero static loop
currents but non-zero dynamic correlation of the currents that can be
observed in, e.g., inelastic spin-polarized neutron scattering. At the same
time, many other properties are similar to those of the DDW state.

In Section \ref{sec:Model} a spin-fermion model with overlapping hot spots
is introduced and simplified, in Section \ref{sec:HS} general formulas for
the partition function are derived decoupling of the electron-electron
interaction by integration over auxiliary fields and minimizing a free
energy functional containing these fields, while in Section \ref{sec:Free}
the free energy of the system is calculated. Section \ref{sec:Time_Crystal}
is devoted to calculation of time-dependent correlation functions using
averaging over phases of oscillations, while in Section \ref{sec:Operator}
the same correlation functions are calculated using an operator order
parameter and a quantum mechanical averaging. In Section \ref{sec:Mean_Field}%
, the main results are re-derived using the Hamiltonian formulation of the
model and the disagreement with the `no-go' theorem is explained. In Section %
\ref{sec:Experiment} possibilities of experimental observation of the
`Thermodynamic Qauntum Time-Space Crystal' are discussed, and the final
discussion of the results is presented in Section \ref{sec:Conclusion}.

\section{\label{sec:Model}Spin-fermion model with overlapping hot spots
(SFMOHS) and its simplification.}

The spin-fermion model with overlapping hot spots (SFMOHS) has been
suggested and further studied in Refs. \cite{volkov1,volkov2,volkov3} for
description of superconducting cuprates. This model originates from
previously used spin-fermion models with $8$ hot spots \cite%
{abanov2003,metlitski2010,efetov2013,wang2014,pepin2014} by the\textbf{\ }%
assumption that the hot spots on the Fermi surface are not isolated, may
overlap and form antinodal `hot regions'. This can happen when the fermion
energies are not far away from the van Hove singularities in the spectrum of
the cuprates, which corresponds to results of ARPES study \cite%
{hashimoto1,he,kaminski,anzai}.

\begin{figure*}[tph]
\begin{subfigure}[b] {0.35\linewidth}
\includegraphics[width=\linewidth]{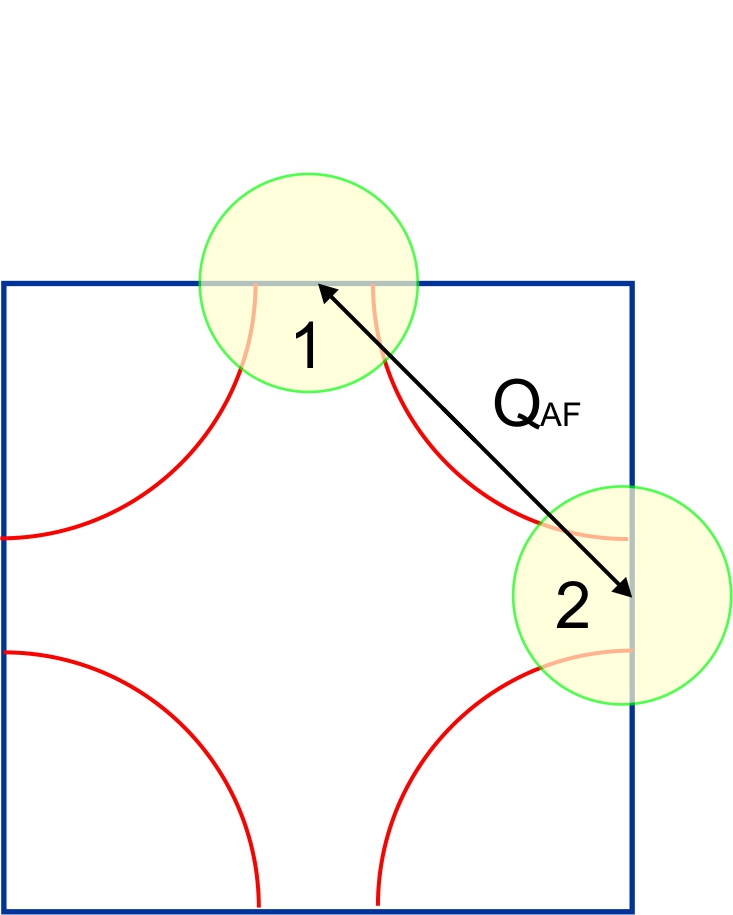}
\caption{Fermi surface and interaction.}
\end{subfigure} \hspace{0.2\linewidth} 
\begin{subfigure}[b]{0.35\linewidth}
\includegraphics[width=\linewidth]{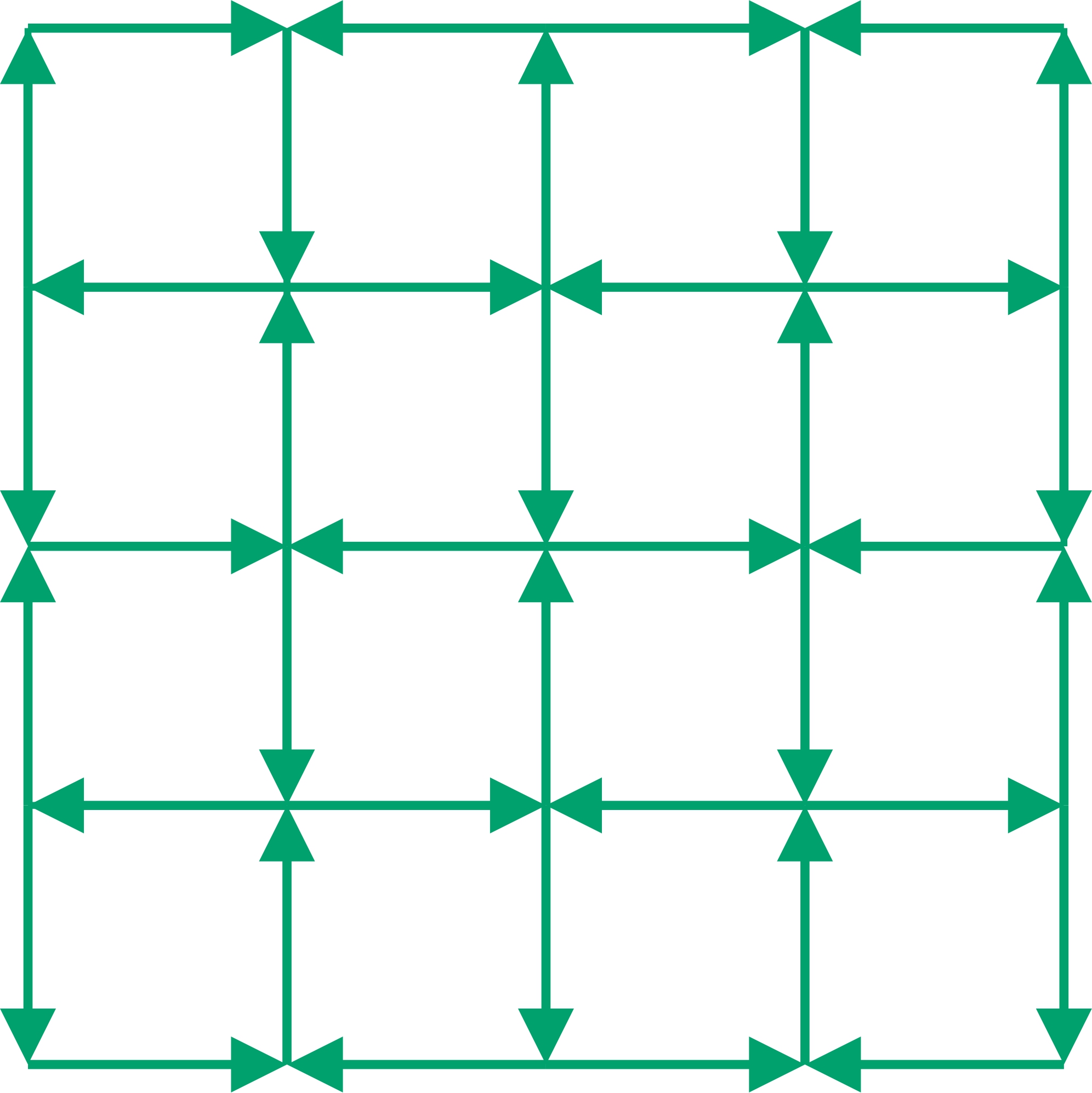}
\caption{Loop currents.}
\end{subfigure}
\caption{(Color online.) Spin-fermion model with overlapping hot spots and
loop currents.}
\label{fig:hotspots}
\end{figure*}

It is assumed that most important are momenta near the middles of the edges
of the Brillouin zone (hot regions) and the latter are numerated as $1$ and $%
2$ (see Fig. \ref{fig:hotspots}a).

The vector $\mathbf{Q}_{AF}$ connecting the middles of the edges equals the
vector of the antiferromagnetic modulation of the parent compound. This
contrasts the structure of the `conventional' spin-fermion model with $8$%
-hot spots \cite{abanov2003,metlitski2010,efetov2013,wang2014,pepin2014}.

We write the partition function $Z$ of the system in a form of a functional
integral over anticommuting vector fields $\chi _{\alpha }^{a}\left(
X\right) $ with the subscript $\alpha $ and the superscript $a$ numerating
spin and the hot regions
\begin{equation}
Z=\int \exp \left[ -S_{\mathrm{0}}\left[ \chi \right] -S_{\mathrm{int}}\left[
\chi \right] -S_{\mathrm{c}}\left[ \chi \right] \right] D\chi .  \label{k24}
\end{equation}%
As usual, the fields $\chi $ satisfy the fermionic boundary conditions
\begin{equation}
\chi \left( \tau \right) =-\chi \left( \tau +1/T\right) .  \label{k24a}
\end{equation}%
In Eq. (\ref{k24}), the action $S_{\mathrm{0}}$ of the non-interacting
particles equals

\begin{equation}
S_{\mathrm{0}}=\int \chi ^{+}\left( X\right) \left[ \partial _{\tau
}+\varepsilon ^{+}\left( -i\mathbf{\nabla }\right) +\Sigma _{3}\varepsilon
^{-}\left( -i\mathbf{\nabla }\right) \text{ }\right] \chi \left( X\right) dX.
\label{k25}
\end{equation}%
Herein, vectors $\chi $ and $\chi ^{+}$ have components%
\begin{equation}
\chi =\left(
\begin{array}{c}
\chi _{1}^{1} \\
\chi _{2}^{1} \\
\chi _{1}^{2} \\
\chi _{1}^{2}%
\end{array}%
\right) ,\quad \chi ^{+}=\left(
\begin{array}{cccc}
\chi _{1}^{1\ast } & \chi _{2}^{1\ast } & \chi _{1}^{2\ast } & \chi
_{2}^{\ast 2}%
\end{array}%
\right)  \label{e1}
\end{equation}%
and $X=\left( \tau ,\mathbf{r}\right) $ is a $4$-dimensional coordinate in
space and imaginary time $\tau $, varying in the interval $0<\tau <1/T,$ $T$
is temperature. In Eq. (\ref{k25})
\begin{equation}
\varepsilon ^{\pm }\left( \mathbf{p}\right) =\frac{1}{2}\left( \varepsilon
_{1}\left( \mathbf{p}\right) \pm \varepsilon _{2}\left( \mathbf{p}\right)
\right) ,  \label{k25c}
\end{equation}%
where $\varepsilon _{1}\left( \mathbf{p}\right) $ and $\varepsilon
_{2}\left( \mathbf{p}\right) $ are two-dimensional spectra of the fermions
in the regions $1$ and $2$ counted from the chemical potential $\mu $
(momenta are counted from the middles of edges). Matrices $\Sigma
_{1},\Sigma _{2},\Sigma _{3}$ are Pauli matrices in the space of numbers $1$
and $2$ numerating the hot regions.

The interaction via antiferromagnetic paramagnons reads

\begin{eqnarray}
&&S_{\mathrm{int}}\left[ \chi ,\chi ^{+}\right] =-\frac{\lambda ^{2}}{2}\int
D_{\mathrm{0}}\left( X-X^{\prime }\right)  \notag \\
&&\times \left( \chi ^{+}\left( X\right) \vec{\sigma}\Sigma _{1}\chi \left(
X\right) \right) \left( \chi ^{+}\left( X^{\prime }\right) \vec{\sigma}%
\Sigma _{1}\chi \left( X^{\prime }\right) \right) dXdX^{\prime },  \notag \\
&&  \label{k25a}
\end{eqnarray}%
where $\lambda $ is a coupling constant, and $D_{\mathrm{0}}$ is propagator
of critical paramagnons. Its Fourier transform can be written as
\begin{equation}
D_{\mathrm{0}}\left( \omega ,\mathbf{q}\right) =\left( \omega ^{2}/v_{s}^{2}+%
\mathbf{q}^{2}+\xi ^{-2}\right) ^{-1},  \label{k25b}
\end{equation}%
where $\xi $ is a correlation length characterizing proximity to the
antiferromagnetic transition point and $v_{s}$ is the velocity of
antiferromagnetic excitations. The term $S_{\mathrm{int}}\left[ \chi \right]
$ describes the interaction between the fermions of the region $1$ and $2.$

We concentrate on more energetically favorable singlet electron-hole
pairings between the regions $1$ and $2$. This allows one to replace the
action $S_{\mathrm{int}}\left[ \chi \right] $ by the following effective
action
\begin{equation}
S_{\mathrm{int}}\left[ \chi ,\chi ^{+}\right] \rightarrow S_{\mathrm{int}%
}^{\left( \mathrm{current}\right) }\left[ \chi ,\chi ^{+}\right] +S_{\mathrm{%
int}}^{\left( \mathrm{density}\right) }\left[ \chi ,\chi ^{+}\right] ,
\label{k26}
\end{equation}%
where
\begin{eqnarray}
&&S_{\mathrm{int}}^{\left( \mathrm{current}\right) }\left[ \chi ,\chi ^{+}%
\right] =-\frac{3\lambda ^{2}}{8}\int D_{\mathrm{0}}\left( X-X^{\prime
}\right)  \notag \\
&&\times \left( \chi ^{+}\left( X^{\prime }\right) \Sigma _{2}\chi \left(
X\right) \right) \left( \chi ^{+}\left( X\right) \Sigma _{2}\chi \left(
X^{\prime }\right) \right) dXdX^{\prime },  \notag \\
&&  \label{k27}
\end{eqnarray}%
\begin{eqnarray}
&&S_{\mathrm{int}}^{\left( \mathrm{density}\right) }\left[ \chi ,\chi ^{+}%
\right] =\frac{3\lambda ^{2}}{8}\int D_{\mathrm{0}}\left( X-X^{\prime
}\right)  \notag \\
&&\times \left( \chi ^{+}\left( X^{\prime }\right) \Sigma _{1}\chi \left(
X\right) \right) \left( \chi ^{+}\left( X\right) \Sigma _{1}\chi \left(
X^{\prime }\right) \right) dXdX^{\prime }.  \notag \\
&&  \label{k28}
\end{eqnarray}%
In addition to the interaction via antiferromagnetic paramagnons, a $\mathbf{%
Q}_{AF}$-component term $S_{\mathrm{c}}\left[ \chi ,\chi ^{+}\right] $ of
the Coulomb interaction has been added in equation (\ref{k24})
\begin{eqnarray}
&&S_{\mathrm{c}}\left[ \chi ,\chi ^{+}\right] =\frac{1}{2}\int V_{\mathrm{c}%
}\left( X-X^{\prime }\right)  \notag \\
&&\times \left( \chi ^{+}\left( X\right) \Sigma _{1}\chi \left( X\right)
\right) \left( \chi ^{+}\left( X^{\prime }\right) \Sigma _{1}\chi \left(
X^{\prime }\right) \right) dXdX^{\prime },  \notag \\
&&  \label{k30}
\end{eqnarray}%
which is very similar to $S_{\mathrm{int}}^{\left( \mathrm{density}\right) }%
\left[ \chi \right] $ and does not give a contribution of the form of $S_{%
\mathrm{int}}^{\left( \mathrm{current}\right) }\left[ \chi \right] $, Eq. (%
\ref{k27}). Actually, Eq. (\ref{k27}) describes an attraction of the loop
currents, while Eqs. (\ref{k28}, \ref{k30}) stand for a repulsion of the $%
\left( \pi ,\pi \right) $ component of the charges.

The limit of the overlapping hot spots allows a variety of electron-hole as
well as superconducting pairings \cite{volkov3}. In the model described by
Eqs. (\ref{k26}, \ref{k30}) several different phases have been identified
within mean field schemes. This includes, depending on the parameters of the
model, d-wave superconductivity, Pomeranchuk deformation of the Fermi
surface, d-formfactor charge-density waves with modulation vectors parallel
to the bonds and d-density wave (DDW) state (loop currents). It is relevant
to emphasize that all relevant energies like gaps in the spectrum, the Fermi
energy, chemical potential, energy of the Pomeranchuk deformation of the
Fermi surface, etc., are of the same order of magnitude. This corresponds to
the experiments showing that all relevant energies of the various states are
of order of several hundreds Kelvin, which is much smaller than usual
electronic energies of the order of $1eV.$

Taking into account only the attraction term, Eq. (\ref{k27}), one obtains
\cite{volkov3} the DDW state with the static currents represented in Fig. %
\ref{fig:hotspots}b. The loop currents obtained in SFMOHS flow along the
bonds of the square lattice and their direction is shown by arrows. The
period of the oscillations equals the double period of the lattice. Within
this picture one can speak of static magnetic moments oscillating in space.

Although using SFMOHS with the interaction specified by Eqs. (\ref{k25a}-\ref%
{k30}) one can explicitly make calculations of various physical quantities
for static order parameters, it is worth further simplifying the form of the
interaction when investigating the possibility of formation of TQTC. This
allows one to avoid unnecessary complications in calculations but, after
all, the main issue of the present work is to demonstrate that TQTC is a
general phenomenon, and keeping a detailed form of the interaction is not
helpful for achieving this goal.

One comes to the simplified version of the model replacing the interactions $%
D_{\mathrm{0}}\left( X-X^{\prime }\right) $ and $V_{\mathrm{c}}\left(
X-X^{\prime }\right) $ by $\delta $-functions in both space and imaginary
time
\begin{eqnarray}
3\lambda ^{2}D_{\mathrm{0}}\left( X-X^{\prime }\right) &\rightarrow &2U_{%
\mathrm{0}}\delta \left( X-X^{\prime }\right) ,\;  \label{k31} \\
2V_{\mathrm{c}}\left( x-X^{\prime }\right) &\rightarrow &U_{\mathrm{c}%
}\delta \left( X-X^{\prime }\right) .  \notag
\end{eqnarray}%
The main results of this paper are derived using a scheme equivalent to a
mean field approximation. Therefore, we start with a simplified model
already adopted for using this scheme (the two-particle interaction contains
squares of the sums over momenta and spins). We write such a simplified
action $S\left[ \chi ,\chi ^{+}\right] $ in the form%
\begin{eqnarray}
&&S\left[ \chi ,\chi ^{+}\right] =S_{\mathrm{0}}\left[ \chi ,\chi ^{+}\right]
\label{e0} \\
&&-\frac{U_{\mathrm{0}}}{4V}\int_{0}^{1/T}\Big(\sum_{p}\chi _{p}^{+}\left(
\tau \right) \Sigma _{2}\chi _{p}\left( \tau \right) \Big)^{2}d\tau  \notag
\\
&&+\frac{\tilde{U}_{\mathrm{0}}}{4V}\int_{0}^{1/T}\Big(\sum_{p}\chi
_{p}^{+}\left( \tau \right) \Sigma _{1}\chi _{p}\left( \tau \right) \Big)%
^{2}d\tau ,  \notag
\end{eqnarray}%
where

\begin{eqnarray}
&&S_{\mathrm{0}}\left[ \chi ,\chi ^{+}\right]  \label{e2} \\
&=&\int_{0}^{1/T}\chi _{p}^{+}\left( \tau \right) \left[ \partial _{\tau
}+\varepsilon ^{+}\left( \mathbf{p}\right) +\Sigma _{3}\varepsilon
^{-}\left( \mathbf{p}\right) \text{ }\right] \chi _{p}\left( \tau \right)
d\tau .  \notag
\end{eqnarray}%
and
\begin{equation}
\tilde{U}_{\mathrm{0}}=U_{\mathrm{0}}+U_{\mathrm{c}}>U_{\mathrm{0}}
\label{k1b}
\end{equation}%
It is assumed that $U_{\mathrm{0}}>0$, $p=\left\{ \mathbf{p,}\alpha \right\}
$ stands for the momentum $\mathbf{\ p}$ and spin $\alpha $, and $V$ is the
volume of the system.

The interaction terms correspond to a long-range coherence of pairs composed
of electrons and holes belonging to the different bands but the interaction
between the electrons is a short-range one.

The combination $\sum_{p}\chi _{p}^{+}\left( \tau \right) \Sigma _{1}\chi
_{p}\left( \tau \right) $ stands for the $\left( \pi ,\pi \right) $ charge
oscillation, while $\sum_{p}\chi _{p}^{+}\left( \tau \right) \Sigma _{2}\chi
_{p}\left( \tau \right) $ describes loop currents, Fig. \ref{fig:hotspots}b.

Action Eq. (\ref{e0}) written for the electron-hole pairs is similar to the
one in the Bardeen-Cooper-Schrieffer (BCS) model for Cooper pairs \cite{bcs}%
. It contains an inter-band attraction (term with matrix $\Sigma _{2}$) and
repulsion (term with $\Sigma _{1}$). Taking into account only the term with
the attraction one obtains an order parameter $B$ corresponding in the
language of SFMOHS to spontaneous static loop currents, Fig. \ref%
{fig:hotspots}b. In order to obtain the new interesting state with a
time-dependent order parameter one should consider both the interactions. It
is crucial that the term with $\Sigma _{1}$ in Eq. (\ref{e0}) describing the
interaction of charges is repulsive. The correspondence of action, Eq. (\ref%
{e0}), and the BCS model could be achieved formally putting $\tilde{U}_{%
\mathrm{0}}=-U_{\mathrm{0}}$ but this would contradict to the assumption (%
\ref{k1b}). This is the reason why the results obtained in the present paper
cannot be applied to the BCS superconductors.

\section{\label{sec:HS}Partition function and equations for the minimum of a
free energy functional of boson fields.}

\subsection{General formulas for the partition function.}

The order parameter of the model determined by the action (\ref{e0}) can be
either static or oscillating both in real and imaginary time. Although the
properties of the TQTC, Eqs. (\ref{i3}-\ref{i4}), are expected to follow
from the real-time dependence of the order parameter, thermodynamics is
determined by its imaginary-time behavior.

Now we write the partition function $Z$, Eq. (\ref{k24}), as
\begin{equation}
Z=\int \exp \left[ -S\left[ \chi ,\chi ^{+}\right] \right] D\chi ,
\label{e3}
\end{equation}%
where the action $S\left[ \chi ,\chi ^{+}\right] $ is specified by Eqs. (\ref%
{e0}, \ref{e2}), and following the standard mean field theory introduce
order parameters $b\left( \tau \right) $ and $b_{1}\left( \tau \right) .$
Making a rotation of the fields $\chi $ in the space of numbers $1$ and $2$
of the bands%
\begin{equation}
\chi _{p}\left( \tau \right) =\mathcal{U}_{0}\eta _{p}\left( \tau \right)
,\quad \mathcal{U}_{0}=\frac{1}{\sqrt{2}}\left(
\begin{array}{cc}
1 & i \\
i & 1%
\end{array}%
\right)  \label{e6}
\end{equation}%
and using the relations
\begin{equation}
\mathcal{U}_{0}^{+}\Sigma _{2}\mathcal{U}_{0}=\Sigma _{3},\;\mathcal{U}%
_{0}^{+}\Sigma _{3}\mathcal{U}_{0}=-\Sigma _{2},\quad \mathcal{U}%
_{0}^{+}\Sigma _{1}\mathcal{U}_{0}=\Sigma _{1},  \label{e7}
\end{equation}%
we write the action $S\left[ \chi ,\chi ^{+}\right] $ in terms of the
anticommuting variables in the form%
\begin{equation}
S\left[ \eta \right] =S_{\mathrm{0}}\left[ \eta \right] +S_{\mathrm{int}}%
\left[ \eta \right] ,  \label{e5}
\end{equation}%
where%
\begin{equation}
S_{\mathrm{0}}\left[ \eta \right] =\int_{0}^{1/T}\eta _{p}^{+}\left( \tau
\right) \left( \partial _{\tau }+\varepsilon ^{+}\left( \mathbf{p}\right)
-\varepsilon ^{-}\left( \mathbf{p}\right) \Sigma _{2}\right) \eta _{p}\left(
\tau \right) d\tau ,  \label{e5a}
\end{equation}%
and%
\begin{eqnarray}
&&S_{\mathrm{int}}\left[ \eta \right] =-\frac{1}{4V}\int_{0}^{1/T}\Big[U_{%
\mathrm{0}}\Big(\sum_{p}\eta _{p}^{+}\left( \tau \right) \Sigma _{3}\eta
_{p}\left( \tau \right) \Big)^{2}  \notag \\
&&-\tilde{U}_{\mathrm{0}}\Big(\sum_{p}\eta _{p}^{+}\left( \tau \right)
\Sigma _{1}\eta _{p}\left( \tau \right) \Big)^{2}\Big]d\tau .  \label{e4}
\end{eqnarray}%
Now we use a method of integration over auxiliary bosonic fields $b\left(
\tau \right) $ and $b_{1}\left( \tau \right) $ (Hubbard-Stratonovich
transformation) to decouple the interaction in $S_{\mathrm{int}}\left( \eta
\right) $, Eq. (\ref{e4})
\begin{eqnarray}
&&\exp \left[ -S_{\mathrm{int}}\left[ \eta \right] \right] =\int \exp \left[
-\int_{0}^{1/T}\left[ \frac{b^{2}\left( \tau \right) }{U_{\mathrm{0}}}+\frac{%
b_{1}^{2}\left( \tau \right) }{\tilde{U}_{\mathrm{0}}}\right] d\tau \right]
\notag \\
&&\times Z_{b}^{-1}\exp \left[ \int_{0}^{1/T}\eta _{p}^{+}\left( b\left(
\tau \right) \Sigma _{3}+ib_{1}\left( \tau \right) \Sigma _{1}\right) d\tau %
\right] DbDb_{1},  \notag \\
&&  \label{l1}
\end{eqnarray}%
where
\begin{equation*}
Z_{b}=\int DbDb_{1}\exp \left[ -\int_{0}^{1/T}\left[ \frac{b^{2}\left( \tau
\right) }{U_{\mathrm{0}}}+\frac{b_{1}^{2}\left( \tau \right) }{\tilde{U}_{%
\mathrm{0}}}\right] d\tau \right] .
\end{equation*}%
The fields $b\left( \tau \right) $ and $b_{1}\left( \tau \right) $ must obey
the bosonic boundary conditions
\begin{equation}
b\left( \tau \right) =b\left( \tau +1/T\right) ,\quad b_{1}\left( \tau
\right) =b_{1}\left( \tau +1/T\right) .  \label{k5a}
\end{equation}%
Substituting Eq. (\ref{l1}) into Eqs. (\ref{e4}, \ref{e5}) we obtain an
effective action quadratic in $\eta ,$ which allows us to integrate exactly
over $\eta ,\eta ^{+}.$ As a result, we represent the free energy $F$ in the
form
\begin{equation}
F=-T\ln \left[ \int \exp \left[ -\frac{\mathcal{F}\left[ b,b_{1}\right] }{T}%
\right] DbDb_{1}\right] .  \label{k7}
\end{equation}%
Herein, the free energy functional $\mathcal{F}\left[ b,b_{1}\right] $ equals

\begin{eqnarray}
\frac{\mathcal{F}\left[ b,b_{1}\right] }{T} &=&\int_{0}^{1/T}\Big[-2\sum_{%
\mathbf{p}}\mathrm{tr}\left[ \ln \left( h\left( \tau ,\mathbf{p}\right)
-ib_{1}\left( \tau \right) \Sigma _{1}\right) \right] _{\tau ,\tau }  \notag
\\
&&+V\left( \frac{b^{2}\left( \tau \right) }{U_{\mathrm{0}}}+\frac{%
b_{1}^{2}\left( \tau \right) }{\tilde{U}_{\mathrm{0}}}\right) \Big]d\tau ,
\label{k10}
\end{eqnarray}%
where
\begin{equation}
h\left( \tau ,\mathbf{p}\right) =\partial _{\tau }+\varepsilon ^{+}\left(
\mathbf{p}\right) -\varepsilon ^{-}\left( \mathbf{p}\right) \Sigma
_{2}-b\left( \tau \right) \Sigma _{3},  \label{e8b}
\end{equation}%
and symbol `$\mathrm{tr}$' means trace in the space of the bands $1,2$.

Both the terms in Eq. (\ref{k10}) are proportional to the volume $V$ of the
system, and therefore the saddle-point method becomes exact for calculation
of the functional integral over the fields $b\left( \tau \right) $ and $%
b_{1}\left( \tau \right) $ in the limit $V\rightarrow \infty $. In other
words, the fields $b\left( \tau \right) $ and $b_{1}\left( \tau \right) $
should be found from the condition of the minimum of the free energy
functional $\mathcal{F}\left[ b,b_{1}\right] $, Eq. (\ref{k10}).
Substitution of the functions obtained in this way into Eqs. (\ref{k10}, \ref%
{k7}) gives the free energy $F.$

\subsection{Equations for the minimum of the free energy functional $%
\mathcal{F}\left[ b,b_{1}\right] $ and their solutions at $b_{1}\left(
\protect\tau \right) =0.$}

Minimization of the free energy functional $\mathcal{F}\left[ b,b_{1}\right]
,$ Eq. (\ref{k10}), leads to the following equations for $b\left( \tau
\right) $ and $b_{1}\left( \tau \right) $

\begin{equation}
b\left( \tau \right) =-U_{\mathrm{0}}\mathrm{tr}\int \Sigma _{3}\left[
H^{-1}\left( \tau ,\mathbf{p}\right) \right] _{\tau ,\tau }\frac{d\mathbf{p}%
}{\left( 2\pi \right) ^{2}},  \label{e14}
\end{equation}%
\begin{equation}
b_{1}\left( \tau \right) =-i\tilde{U}_{\mathrm{0}}\mathrm{tr}\int \Sigma _{1}%
\left[ H^{-1}\left( \tau ,\mathbf{p}\right) \right] _{\tau ,\tau }\frac{d%
\mathbf{p}}{\left( 2\pi \right) ^{2}},  \label{e15}
\end{equation}%
where%
\begin{equation*}
H\left( \tau ,\mathbf{p}\right) =h\left( \tau ,\mathbf{p}\right)
-ib_{1}\left( \tau \right) \Sigma _{1}.
\end{equation*}%
Equations (\ref{e14}-\ref{e15}) have a time-independent solution \cite%
{volkov3}
\begin{equation}
b_{1}\left( \tau \right) =0,\;b\left( \tau \right) =\gamma ,  \label{e16}
\end{equation}%
where $\gamma $ determines the gap in the spectrum and can be bound from the
equation
\begin{equation}
1=\frac{U_{\mathrm{0}}}{2}\int \frac{\tanh \frac{\kappa _{\mathbf{p}%
}^{\left( 0\right) }+\varepsilon ^{+}\left( \mathbf{p}\right) /T}{2}+\tanh
\frac{\kappa _{\mathbf{p}}^{\left( 0\right) }-\varepsilon ^{+}\left( \mathbf{%
p}\right) /T}{2}}{\sqrt{\left( \varepsilon ^{-}\left( \mathbf{p}\right)
\right) ^{2}+\gamma ^{2}}}\frac{d\mathbf{p}}{\left( 2\pi \right) ^{2}},
\label{e17}
\end{equation}%
and
\begin{equation}
\kappa _{\mathbf{p}}^{\left( 0\right) }=\frac{\sqrt{\left( \varepsilon
^{-}\left( \mathbf{p}\right) \right) ^{2}+\gamma ^{2}}}{T}.  \label{e18}
\end{equation}%
Of course, the trivial solution $b_{1}\left( \tau \right) =0,\;b\left( \tau
\right) =0$ also exists but we are interested in the region of parameters of
the model where non-zero solutions appear.

Provided $\varepsilon ^{+}\left( \mathbf{p}\right) $ is not very large, Eq. (%
\ref{e17}) simplifies at low temperatures to the form%
\begin{equation}
1=U_{\mathrm{0}}\int \frac{1}{\sqrt{\left( \varepsilon ^{-}\left( \mathbf{p}%
\right) \right) ^{2}+\gamma ^{2}}}\frac{d\mathbf{p}}{\left( 2\pi \right) ^{2}%
}.  \label{e19}
\end{equation}

However, more sophisticated time-dependent solutions of Eqs. (\ref{e14}, \ref%
{e15}) also exist even at $\tilde{U}_{\mathrm{0}}=0.$ Although it was
assumed in the beginning that $\tilde{U}_{\mathrm{0}}>0$ (\ref{k1b}) and
this is the most interesting case, we start with formally considering the
limit $\tilde{U}_{\mathrm{0}}=0$ because it helps to understand the
structure of the solutions at arbitrary $\tilde{U}_{\mathrm{0}}$.

Setting $\tilde{U}_{\mathrm{0}}=0$ in Eq. (\ref{e15}) and, hence $%
b_{1}\left( \tau \right) =0$, we come to the following equation for $b\left(
\tau \right) $

\begin{equation}
b\left( \tau \right) =-U_{\mathrm{0}}\mathrm{tr}\int \Sigma _{3}\left[
h^{-1}\left( \tau ,\mathbf{p}\right) \right] _{\tau ,\tau }\frac{d\mathbf{p}%
}{\left( 2\pi \right) ^{2}},  \label{e20}
\end{equation}%
with the operator $h\left( \tau ,\mathbf{p}\right) $ specified by Eq. (\ref%
{e8b}).

\begin{figure*}[htp]
\centering
\begin{subfigure}[b] {0.49\linewidth}
  \includegraphics[width=\linewidth]{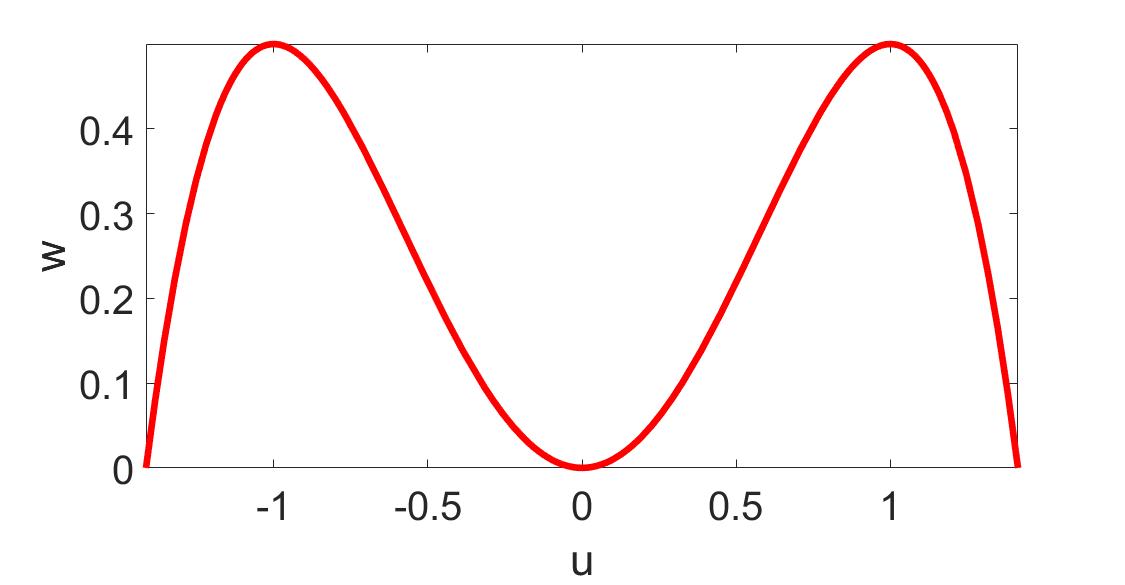}
 \caption{Potential $w$ at  $k=0.9999$.}
 \end{subfigure} 
\begin{subfigure}[b]{0.49\linewidth}
  \includegraphics[width=\linewidth]{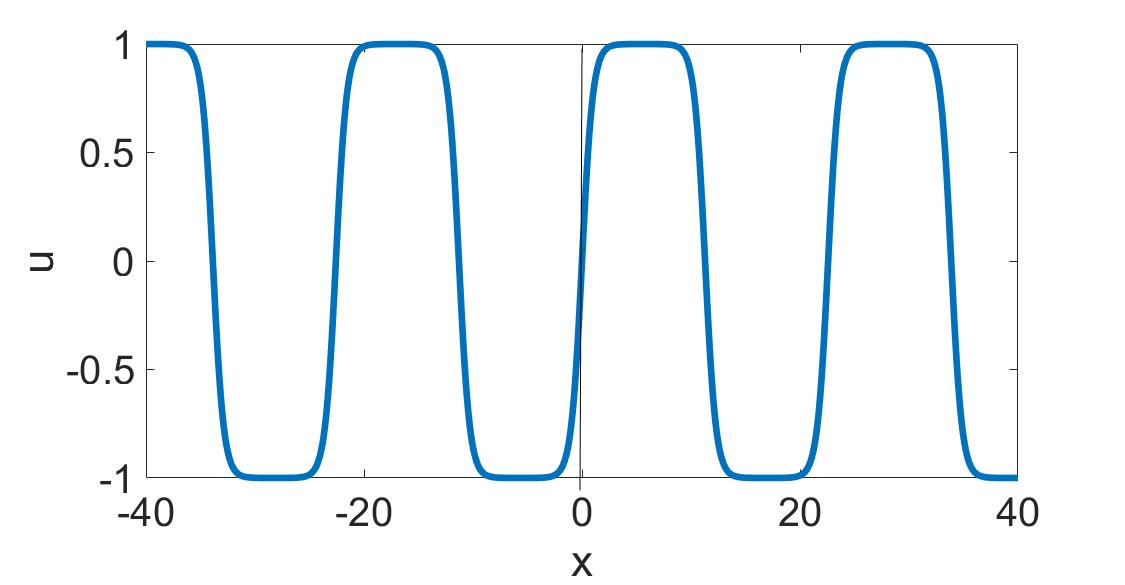}
 \caption{Solution $u$ as a function of $x$ at $k=0.9999$}
 \end{subfigure}
\caption{(Color online.) Jacobi elliptic function.}
\label{fig:jacobi}
\end{figure*}

Although Eq. (\ref{e20}) is quite non-trivial due to a possible dependence
of $b\left( \tau \right) $ on $\tau $, solutions $b_{0}\left( \tau \right) $
of this equations can be written exactly in terms of a Jacobi double
periodic elliptic function $\mathrm{sn}\left( x|k\right) $,
\begin{equation}
b_{0}\left( \tau \right) =k\gamma \mathrm{sn}\left( \gamma \left( \tau -\tau
_{0}\right) |k\right) ,  \label{e21}
\end{equation}%
where $0<k<1$ is the modulus, $\gamma $ is an energy, and $\tau _{0}$ is an
arbitrary shift of the imaginary time in the interval $0<\tau _{0}<1/T$ (see
also Fig. \ref{fig:jacobi}). The period of the oscillations for an arbitrary
$k$ equals $4K\left( k\right) /\gamma ,$ where $K\left( k\right) $ is the
elliptic integral of the first kind, and therefore the condition
\begin{equation}
\gamma =4K\left( k\right) mT  \label{e22}
\end{equation}%
with integer $m>0$ must be satisfied to fulfill Eqs. (\ref{k5a}).

One can visualize the function $u=k\mathrm{sn}\left( x|k\right) $ satisfying
the equation
\begin{equation}
\left( \frac{du}{dx}\right) ^{2}=u^{4}-\left( 1+k^{2}\right) u^{2}+k^{2}
\label{e23}
\end{equation}%
by taking into account the fact that it describes motion of a classical
particle with unit mass and energy $k^{2}/2$ in the potential $w=\frac{1}{2}%
\left( u^{2}\left( 1+k^{2}\right) -u^{4}\right) $. In the limit $%
k\rightarrow 1$, the particle starts moving from the top a hill in Fig. \ref%
{fig:jacobi}a, stops on the other hill, and then moves back forming an
instanton-anti-instanton pair (IAP). In this limit, the function $u$ has a
simple form of alternating functions $\tanh x$ and -$\tanh x$ (instantons
and anti-instantons), and is represented in Fig. \ref{fig:jacobi}b. In the
limit of small $k$, the Jacobi elliptic function has the asymptotic behavior
$\mathrm{sn}\left( x|k\right) \rightarrow \sin x$ corresponding to a
harmonic oscillation of the classical particle near a minimum of $w$.

At small $1-k,$ the period $4K\left( k\right) /\gamma $ of $b_{0}\left( \tau
\right) $ grows logarithmically as $-\ln $ $\left( 1-k\right) $ and the
solution $b_{0}\left( \tau \right) $ consists of $2m$ well separated
alternating instantons and anti-instantons. It is important that the
integral over the period of the oscillations in Eq. (\ref{e21}) equals zero.
The average over the position $\tau _{0}$ equals zero as well
\begin{equation}
\overline{b_{0}\left( \tau \right) }=0,  \label{k12b}
\end{equation}%
where bar stands for such an averaging.

The solution (\ref{e21}) of Eq. (\ref{e20}) satisfying the boundary
condition, Eq. (\ref{e22}), has been used previously in Refs. \cite%
{mukhin,mukhin1} starting from a different model. A similar solution for a
mean field equation arising in 1D models of polymers (depending on
coordinates but not on the imaginary time) has been discovered long ago \cite%
{brazovskii} and later used also in Refs. \cite{mertsching,machida}.

One can demonstrate that the function $b_{0}\left( \tau \right) $, Eq. (\ref%
{e21}), is really the solution of Eq. (\ref{e20}) by finding the
eigenfunctions $\Psi _{s\mathbf{p}}^{\left( 0\right) }\left( \tau \right) $
and eigenenergies $\epsilon _{s\mathbf{p}}^{\left( 0\right) }$ of the
operator $h\left( \tau ,\mathbf{p}\right) ,$ Eq. (\ref{e8b}), with $b\left(
\tau \right) =b_{0}\left( \tau \right) ,$ where $b_{0}\left( \tau \right) $
satisfies equation
\begin{equation}
\dot{b}_{0}^{2}\left( \tau \right) =b_{0}^{4}\left( \tau \right) -\gamma
^{2}\left( 1+k^{2}\right) b_{0}^{2}\left( \tau \right) +\gamma ^{4}k^{2},
\label{k46}
\end{equation}%
and making a spectral expansion of the operator $h^{-1}\left( t,\mathbf{p}%
\right) .$

It is instructive to give details of the calculations because an extension
of this formalism will be used for non-zero $b_{1}\left( \tau \right) $. At $%
b_{1}\left( \tau \right) =0$ we have equations
\begin{eqnarray}
&&h\left( \tau ,\mathbf{p}\right) \Psi _{s\mathbf{p}}^{\left( 0\right)
}\left( \tau \right) =\epsilon _{s\mathbf{p}}^{\left( 0\right) }\Psi _{s%
\mathbf{p}}^{\left( 0\right) }\left( \tau \right) ,  \notag \\
&&\bar{\Psi}_{s\mathbf{p}}^{\left( 0\right) }\left( \tau \right) \bar{h}%
\left( \tau ,\mathbf{p}\right) =\epsilon _{s\mathbf{p}}^{\left( 0\right) }%
\bar{\Psi}_{s\mathbf{p}}^{\left( 0\right) }\left( \tau \right) .  \label{k47}
\end{eqnarray}%
In Eqs. (\ref{k47}) operator $h\left( \tau ,\mathbf{p}\right) $ has been
defined in Eq. (\ref{e8b}) and $\bar{h}\left( \tau ,\mathbf{p}\right) $
equals%
\begin{equation}
\bar{h}\left( \tau ,\mathbf{p}\right) =-\overleftarrow{\partial }_{\tau
}+\varepsilon ^{+}\left( \mathbf{p}\right) -\varepsilon ^{-}\left( \mathbf{p}%
\right) \Sigma _{2}-b_{0}\left( \tau \right) \Sigma _{3},  \label{e8c}
\end{equation}%
where the derivative $\overleftarrow{\partial }_{\tau }$ acts on the left.
The eigenfunctions $\bar{\Psi}_{s\mathbf{p}}^{\left( 0\right) }\left( \tau
\right) $ and $\Psi _{s\mathbf{p}}^{\left( 0\right) }\left( \tau \right) $
have to obey the antiperiodic boundary conditions%
\begin{equation}
\Psi _{s\mathbf{p}}^{\left( 0\right) }\left( \tau \right) =-\Psi _{s\mathbf{p%
}}^{\left( 0\right) }\left( \tau +1/T\right) ,\;\bar{\Psi}_{s\mathbf{p}%
}^{\left( 0\right) }\left( \tau \right) =-\bar{\Psi}_{s\mathbf{p}}^{\left(
0\right) }\left( \tau +1/T\right)  \label{e25}
\end{equation}

Solutions $\Psi _{s\mathbf{p}}^{\left( 0\right) }\left( \tau \right) $, $%
\bar{\Psi}_{s\mathbf{p}}^{\left( 0\right) }\left( \tau \right) $ of Eqs. (%
\ref{k47}) satisfying Eqs. (\ref{e25}) can be sought in the form
\begin{eqnarray}
\Psi _{s\mathbf{p}}^{\left( 0\right) }\left( \tau \right) &=&\mathcal{N}_{%
\mathbf{p}}\Upsilon _{l\mathbf{p}}\left( \tau \right) e^{-i\pi \left(
2n+1\right) \tau T}e^{-\left( -1\right) ^{l}\kappa _{\mathbf{p}}T\tau },
\label{k48} \\
\bar{\Psi}_{s\mathbf{p}}^{\left( 0\right) }\left( \tau \right) &=&\left(
-1\right) ^{l+1}\mathcal{N}_{\mathbf{p}}\bar{\Upsilon}_{l\mathbf{p}}\left(
\tau \right) e^{i\pi \left( 2n+1\right) \tau T}e^{\left( -1\right)
^{l}\kappa _{\mathbf{p}}T\tau },  \notag
\end{eqnarray}%
where $s=\left\{ l,n\right\} $, $l=1,2$, $n=0,\pm 1,\pm 2...$ and $\mathcal{N%
}_{\mathbf{p}}$ is a normalization factor.

Functions $\Upsilon _{l\mathbf{p}}\left( \tau \right) $ equal
\begin{equation}
\Upsilon _{l\mathbf{p}}\left( \tau \right) =\left(
\begin{array}{c}
1 \\
\left( i\varepsilon ^{-}\left( \mathbf{p}\right) \right) ^{-1}\left(
-\partial _{\tau }+b_{0}\left( \tau \right) \right)%
\end{array}%
\right) Z_{l\mathbf{p}}\left( \tau \right) ,  \label{k49}
\end{equation}

\begin{equation}
\bar{\Upsilon}_{l\mathbf{p}}\left( \tau \right) =\left(
\begin{array}{cc}
\left( i\varepsilon ^{-}\left( \mathbf{p}\right) \right) ^{-1}\left(
\partial _{\tau }-b_{0}\left( \tau \right) \right) & 1%
\end{array}%
\right) \bar{Z}_{l\mathbf{p}}\left( \tau \right) ,  \label{k50}
\end{equation}%
where
\begin{equation}
\bar{Z}_{l}\left( \tau \right) =\left\{
\begin{array}{cc}
X_{\mathbf{p}}\left( \tau \right) , & l=1 \\
Y_{\mathbf{p}}\left( \tau \right) , & l=2%
\end{array}%
\right. ,\quad Z_{l}\left( \tau \right) =\left\{
\begin{array}{cc}
Y_{\mathbf{p}}\left( \tau \right) , & l=1 \\
X_{\mathbf{p}}\left( \tau \right) , & l=2%
\end{array}%
\right. ,  \label{k51}
\end{equation}%
and the functions $X_{\mathbf{p}}\left( \tau \right) $ and $Y_{\mathbf{p}%
}\left( \tau \right) $ are growing and decaying with $\tau $ solutions of
the same equation

\begin{eqnarray}
\left[ -\partial _{\tau }^{2}+\left( \varepsilon ^{-}\left( \mathbf{p}%
\right) \right) ^{2}+Q\left( \tau \right) \right] Y_{\mathbf{p}}\left( \tau
\right) &=&0,  \label{k52} \\
\left[ -\partial _{\tau }^{2}+\left( \varepsilon ^{-}\left( \mathbf{p}%
\right) \right) ^{2}+Q\left( \tau \right) \right] X_{\mathbf{p}}\left( \tau
\right) &=&0,  \notag
\end{eqnarray}%
with
\begin{equation}
Q\left( \tau \right) =b_{0}^{2}\left( \tau \right) +\dot{b}_{0}\left( \tau
\right) .  \label{e24}
\end{equation}

Although the functions $\Psi _{s\mathbf{p}}^{\left( 0\right) }\left( \tau
\right) $, $\bar{\Psi}_{s\mathbf{p}}^{\left( 0\right) }\left( \tau \right) $
obey the antiperiodicity conditions (\ref{e25}), the solutions $X_{\mathbf{p}%
}\left( \tau \right) $ and $Y_{\mathbf{p}}\left( \tau \right) $ cannot be
periodic. We assume that they change over the period $1/T$ as
\begin{equation}
X_{\mathbf{p}}\left( \tau +1/T\right) =e^{\kappa _{\mathbf{p}}}X_{\mathbf{p}%
}\left( \tau \right) ,\;Y_{\mathbf{p}}\left( \tau +1/T\right) =e^{-\kappa _{%
\mathbf{p}}}Y_{\mathbf{p}}\left( \tau \right) ,  \label{e26}
\end{equation}%
where $\kappa _{\mathbf{p}}$ is a function of $\mathbf{p}$ only. The
antiperiodicity of the eigenfunctions $\Psi _{s\mathbf{p}}^{\left( 0\right)
}\left( \tau \right) $, $\bar{\Psi}_{s\mathbf{p}}^{\left( 0\right) }\left(
\tau \right) $ is guaranteed by the presence in Eq. (\ref{k48}) of the
exponentials containing $\kappa _{\mathbf{p}}$. The solutions $X_{\mathbf{p}%
}\left( \tau \right) $ and $Y_{\mathbf{p}}\left( \tau \right) $ of Eqs. (\ref%
{k52}) are related to each other as
\begin{equation}
\dot{X}_{\mathbf{p}}\left( \tau \right) Y_{\mathbf{p}}\left( \tau \right)
-X_{\mathbf{p}}\left( \tau \right) \dot{Y}_{\mathbf{p}}\left( \tau \right)
=C_{\mathbf{p}},  \label{k53}
\end{equation}%
where $C_{\mathbf{p}}$ is a time-independent function of $\mathbf{p.}$

Eqs. (\ref{k48}-\ref{k53}) are sufficient to prove the orthogonality of the
eigenfunctions $\Psi _{s\mathbf{p}}^{\left( 0\right) }\left( \tau \right) $
\begin{equation}
\left( \bar{\Psi}_{s\mathbf{p}},\Psi _{s^{\prime }\mathbf{p}}\right) \equiv
T\int_{0}^{1/T}\bar{\Psi}_{s\mathbf{p}}\left( \tau \right) \Psi _{s^{\prime }%
\mathbf{p}}\left( \tau \right) d\tau =\delta _{ss^{\prime }}.  \label{e27}
\end{equation}%
and write the normalization $\mathcal{N}_{\mathbf{p}}$ in the form
\begin{equation}
\mathcal{N}_{\mathbf{p}}^{2}=\frac{i\varepsilon ^{-}\left( \mathbf{p}\right)
T}{C_{\mathbf{p}}}.  \label{k54}
\end{equation}%
Substituting Eqs. (\ref{k48}-\ref{k51}) into Eq. (\ref{k47}) one finds the
eigenenergies $\epsilon _{s\mathbf{p}}^{\left( 0\right) }$
\begin{equation}
\epsilon _{s\mathbf{p}}^{\left( 0\right) }=\varepsilon _{\mathbf{p}%
}^{+}\left( \mathbf{p}\right) -i\left( 2n+1\right) \pi T+\left( -1\right)
^{l+1}\kappa _{\mathbf{p}}T.  \label{e28}
\end{equation}
The last step to be done for writing the eigenfunctions $\Psi _{s\mathbf{p}%
}^{\left( 0\right) }\left( \tau \right) $, $\bar{\Psi}_{s\mathbf{p}}^{\left(
0\right) }\left( \tau \right) $ explicitly is to solve equations (\ref{k52}%
). The final solution for $X_{\mathbf{p}}\left( \tau \right) $ and $Y_{%
\mathbf{p}}\left( \tau \right) $ can be written in the form
\begin{equation}
X_{\mathbf{p}}\left( \tau \right) =w_{\mathbf{p}}\left( \tau \right) \exp %
\left[ \int_{0}^{\tau }\frac{\Omega _{\mathbf{p}}}{w_{\mathbf{p}}^{2}\left(
\tau ^{\prime }\right) }d\tau ^{\prime }\right] ,  \label{k55}
\end{equation}%
\begin{equation}
Y_{\mathbf{p}}\left( \tau \right) =w_{\mathbf{p}}\left( \tau \right) \exp %
\left[ -\int_{0}^{\tau }\frac{\Omega _{\mathbf{p}}}{w_{\mathbf{p}}^{2}\left(
\tau ^{\prime }\right) }d\tau ^{\prime }\right] ,  \label{k56}
\end{equation}%
where
\begin{eqnarray}
&&\Omega _{\mathbf{p}}=\left\vert \varepsilon ^{-}\left( \mathbf{p}\right)
\right\vert ^{3}  \label{k57} \\
&&\times \sqrt{\left( 1+\frac{\gamma ^{2}}{4}\frac{\left( 1-k\right) ^{2}}{%
\left( \varepsilon ^{-}\left( \mathbf{p}\right) \right) ^{2}}\right) \left(
1+\frac{\gamma ^{2}}{4}\frac{\left( 1+k\right) ^{2}}{\left( \varepsilon
^{-}\left( \mathbf{p}\right) \right) ^{2}}\right) },  \notag
\end{eqnarray}%
and
\begin{equation}
w_{\mathbf{p}}\left( \tau \right) =\left( \left( \varepsilon ^{-}\left(
\mathbf{p}\right) \right) ^{2}+\frac{1+k^{2}}{4}-\frac{Q\left( \tau \right)
}{2}\right) ^{1/2}.  \label{k58}
\end{equation}%
These solutions can be checked substituting equations (\ref{k55}, \ref{k56})
into Eqs. (\ref{k52}) and using Eq. (\ref{k46}). One can also see from
equation (\ref{k53}) that
\begin{equation}
C_{\mathbf{p}}=2\Omega _{\mathbf{p}}.  \label{k59}
\end{equation}%
and it is clear that the parameter $\kappa _{\mathbf{p}}$ equals
\begin{equation}
\kappa _{\mathbf{p}}=\int_{0}^{1/T}\frac{\Omega _{\mathbf{p}}}{w_{\mathbf{p}%
}^{2}\left( \tau \right) }d\tau .  \label{k59a}
\end{equation}%
One can derive from equations (\ref{k49}-\ref{k51}) and (\ref{k55}-\ref{k59}%
) the following useful relations
\begin{equation}
\left( \bar{\Upsilon}_{l\mathbf{p}}\left( \tau \right) \Sigma _{3}\Upsilon
_{l\mathbf{p}}\left( \tau \right) \right) =2ib_{0}\left( \tau \right)
\varepsilon ^{-}\left( \mathbf{p}\right) ,\;  \label{k60}
\end{equation}%
\begin{equation}
\left( \bar{\Upsilon}_{l\mathbf{p}}\left( \tau \right) \Sigma _{1}\Upsilon
_{l\mathbf{p}}\left( \tau \right) \right) =-\dot{b}_{0}\left( \tau \right) .
\label{k61}
\end{equation}
Using the spectral expansion of the function $h^{-1}\left( \tau ,\mathbf{p}%
\right) $ one can bring Eq. (\ref{e20}) to the form

\begin{equation}
b\left( \tau \right) =-U_{0}\sum_{s}\frac{\bar{\Psi}_{s\mathbf{p}}^{\left(
0\right) }\left( \tau \right) \Sigma _{3}\Psi _{s\mathbf{p}}^{\left(
0\right) }\left( \tau \right) }{\epsilon _{s\mathbf{p}}^{\left( 0\right) }}.
\label{k40}
\end{equation}%
Substituting eigenfunctions $\Psi _{s\mathbf{p}}^{\left( 0\right) }\left(
\tau \right) ,$ Eqs. (\ref{k48}-\ref{k54}), and eigenenergies $\epsilon _{s%
\mathbf{p}}^{\left( 0\right) }$, Eq. (\ref{e28}), into Eq. (\ref{k40}) we
obtain finally the mean field equation for $b_{0}\left( \tau \right) $
\begin{eqnarray}
&&b_{0}\left( \tau \right) =\frac{U_{0}}{2}\int \Big[\tanh \frac{\kappa _{%
\mathbf{p}}+\varepsilon _{\mathbf{p}}^{+}/T}{2}+\tanh \frac{\kappa _{\mathbf{%
p}}-\varepsilon _{\mathbf{p}}^{+}/T}{2}\Big]  \notag \\
&&\times \frac{b_{0}\left( \tau \right) \left\vert \varepsilon ^{-}\left(
\mathbf{p}\right) \right\vert }{\sqrt{\left( \left( \varepsilon ^{-}\left(
\mathbf{p}\right) \right) ^{2}+\gamma ^{2}\frac{\left( 1-k\right) ^{2}}{4}%
\right) \left( \left( \varepsilon ^{-}\left( \mathbf{p}\right) \right)
^{2}+\gamma ^{2}\frac{\left( 1+k\right) ^{2}}{4}\right) }}\frac{d\mathbf{p}}{%
\left( 2\pi \right) ^{2}}.  \notag \\
&&  \label{e29}
\end{eqnarray}%
We see that $b_{0}\left( \tau \right) $ drops out from Eq. (\ref{e29}), and
the latter is algebraic, which means that $b_{0}\left( \tau \right) $ is an
exact solution of Eq. (\ref{e20}). Eq. (\ref{e29}) is valid for arbitrary
temperature $T$ and modulus $k.$ It should be solved together with Eq. (\ref%
{e22}) and one should find the solution for a given number $m$ of IAP. One
has to calculate also the integral providing the value $\kappa _{\mathbf{p}%
}, $ Eq. (\ref{k59a}). Actually, it can be expressed in terms of the
elliptic integrals of the first $K\left( \tilde{k}\right) $ and third $\Pi
\left( \tilde{k}\right) $ kinds.

\begin{equation}
\kappa _{\mathbf{p}}=\frac{\left\vert \varepsilon ^{-}\left( \mathbf{p}%
\right) \right\vert }{T}\sqrt{\frac{\left( \varepsilon ^{-}\left( \mathbf{p}%
\right) \right) ^{2}+\frac{\gamma ^{2}}{4}\left( 1-k\right) ^{2}}{\left(
\varepsilon ^{-}\left( \mathbf{p}\right) \right) ^{2}+\frac{\gamma ^{2}}{4}%
\left( 1+k\right) ^{2}}}\frac{\Pi \left( n,\tilde{k}\right) }{K\left( \tilde{%
k}\right) }  \label{e30}
\end{equation}%
where%
\begin{equation}
n=\frac{\gamma ^{2}k}{\left( \varepsilon ^{-}\left( \mathbf{p}\right)
\right) ^{2}+\frac{\gamma ^{2}}{4}\left( 1+k\right) ^{2}},\;\tilde{k}=\frac{2%
\sqrt{k}}{1+k}.  \label{e31}
\end{equation}%
Equations (\ref{e29}-\ref{e31}) overlap with results of previous
publications \cite{mukhin,mukhin1}.

It is clear that at low temperatures there can be many solutions of Eq. (\ref%
{e29}) because it contains two unknown parameters $\gamma $ and $k.$ The
proper solution should correspond to the minimum of the free energy $%
\mathcal{F}\left[ b,b_{1}\right] $, Eqs. (\ref{k10}, \ref{e8b}). We will see
that in the absence of the field $b_{1}\left( \tau \right) ,$ which is the
case when the interaction $\tilde{U}_{\mathrm{0}}$ in Eq. (\ref{e15}) is
formally put to zero, the time-independent solution, Eq. (\ref{e16}), is
most favorable. At non-zero $\tilde{U}_{\mathrm{0}}$ one has to solve Eqs. (%
\ref{e14}, \ref{e15}) together. Of course, this will modify the solution $%
b\left( \tau \right) $ but the solution alone cannot help answering the
question whether configurations with a finite number of IAP can be
energetically favorable or not. In other words, one has to calculate the
free energy for a given number $m$ of IAP and check whether this energy can
be lower than that with $m=0$ or not.

The calculations to be done look rather involved. Fortunately, the limit of
a diluted system of IAP corresponding to a large period ($1-k\ll 1$) is most
favorable for obtaining a non-zero imaginary-time-dependent order parameter,
and calculations become considerably simpler in this limit.

It is worth mentioning that, in the limit $1-k\ll 1,$ Eq. (\ref{e29}),
reduces to Eq. (\ref{e17}), and many thermodynamic properties of the diluted
system of IAP can be very close to those of DDW state. At the same time,
behavior of physical quantities like loop currents determined directly by
the order parameter can be drastically different.

\section{\label{sec:Free}Free energy.}

\subsection{General scheme of the calculations.}

In principle, the free energy $F$ should be calculated by first solving Eqs.
(\ref{e14}, \ref{e15}) for $b\left( \tau \right) ,$ $b_{1}\left( \tau
\right) $ for a given number $m$ of IAP and substituting the solution into
the free energy functional $\mathcal{F}\left[ b,b_{1}\right] $, Eqs. (\ref%
{k10}, \ref{e8b}). This would give a free energy $F\left[ m\right] $ as a
function of $m.$ The minimum of this function is the free energy of the
system.

This procedure resembles calculation of the free energy of vortices in a
superconductor subjected to a magnetic $B$ field. In order to investigate
the possibility of entering vortices into the superconductor one calculates
the energy of a single vortex and the energy of its interaction with the
magnetic field. The total energy is positive in the Meissner state, while it
is negative in the mixed Abrikosov state. The energy equals zero at the
critical field $B_{c1}$. Even if it is not easy to find the precise form of
the single vortex, the existence of the transition can be established in
this way taking an approximate solution for the vortex.

Solving Eqs. (\ref{e14}, \ref{e15}) exactly or investigating the problem
numerically is not an easy task, and it is beyond the scope of the present
work. Instead, we develop here an approximate scheme that provides
physically plausible results. We proceed by introducing eigenfunctions $\Psi
_{s\mathbf{p}}\left( \tau \right) $, its conjugates $\bar{\Psi}_{s\mathbf{p}%
}\left( \tau \right) $ and eigenenergies $\epsilon _{s\mathbf{p}},$
satisfying equations
\begin{eqnarray}
\left( h\left( \tau ,\mathbf{p}\right) -ib_{1}\left( \tau \right) \Sigma
_{1}\right) \Psi _{s\mathbf{p}}\left( \tau \right) &=&\epsilon _{s\mathbf{p}%
}\Psi _{s\mathbf{p}}\left( \tau \right) ,  \label{k32} \\
\bar{\Psi}_{s\mathbf{p}}\left( \tau \right) \left( \bar{h}\left( \tau ,%
\mathbf{p}\right) -ib_{1}\left( \tau \right) \Sigma _{1}\right) &=&\epsilon
_{s\mathbf{p}}\bar{\Psi}_{s\mathbf{p}}\left( \tau \right) ,  \notag
\end{eqnarray}%
and antiperiodicity conditions
\begin{equation}
\Psi _{s\mathbf{p}}\left( \tau +1/T\right) =-\Psi _{s\mathbf{p}}\left( \tau
\right) ,\quad \bar{\Psi}_{s\mathbf{p}}\left( \tau +1/T\right) =-\bar{\Psi}%
_{s\mathbf{p}}\left( \tau \right) .  \label{k34}
\end{equation}%
Operators $h\left( \tau ,\mathbf{p}\right) $ and $\bar{h}\left( \tau ,%
\mathbf{p}\right) $ are specified in Eqs. (\ref{e8b}, \ref{e8c}). Functions $%
\Psi _{s\mathbf{p}}\left( \tau \right) $, $\bar{\Psi}_{s\mathbf{p}}\left(
\tau \right) $ form an orthogonal basis of the eigenfunctions
\begin{equation}
\left( \bar{\Psi}_{s\mathbf{p}},\Psi _{s^{\prime }\mathbf{p}}\right) \equiv
T\int_{0}^{1/T}\bar{\Psi}_{s\mathbf{p}}\left( \tau \right) \Psi _{s^{\prime }%
\mathbf{p}}\left( \tau \right) d\tau =\delta _{ss^{\prime }}.  \label{k37}
\end{equation}%
Setting formally $b_{1}\left( \tau \right) =0$ one comes back to the
eigenfunctions $\Psi _{s\mathbf{p}}^{\left( 0\right) }\left( \tau \right) $,
$\bar{\Psi}_{s\mathbf{p}}^{\left( 0\right) }\left( \tau \right) $ and
eigenenergies $\epsilon _{s\mathbf{p}}^{\left( 0\right) },$ Eqs. (\ref{k47}-%
\ref{k59a}). Then, one can write the `electronic'\ part $\mathcal{F}_{%
\mathrm{el}}$ (first term in the integrand in Eq. (\ref{k10})) in the form
\begin{eqnarray}
&&\frac{\mathcal{F}_{\mathrm{el}}\left[ b\left( \tau \right) ,b_{1}\left(
\tau \right) \right] }{VT}  \label{k38} \\
&=&2\int_{0}^{1/T}\sum_{\mathbf{p}}\mathrm{tr}\left[ \ln \left( h\left( \tau
,\mathbf{p}\right) -ib_{1}\left( \tau \right) \Sigma _{1}\right) \right]
_{\tau ,\tau }d\tau  \notag \\
&=&-2\sum_{s}\int \ln \frac{\epsilon _{s\mathbf{p}}}{T}\frac{d\mathbf{p}}{%
\left( 2\pi \right) ^{2}}.  \notag
\end{eqnarray}

One should keep in mind that the eigenvalues $\epsilon _{s\mathbf{p}}$ are
functionals of the functions $b\left( \tau \right) $ and $b_{1}\left( \tau
\right) .$ The fact that the functional $\mathcal{F}_{\mathrm{el}}\left[
b\left( \tau \right) ,b_{1}\left( \tau \right) \right] $ can be expressed in
terms of only the eigenvalues simplifies calculations. We cannot find $%
\epsilon _{s\mathbf{p}}$ and $\Psi _{s\mathbf{p}}\left( \tau \right) $
exactly for arbitrary $b_{1}\left( \tau \right) $ and simply use a
perturbation theory for the eigenvalues $\epsilon _{s\mathbf{p}}$. In the
zeroth approximation one puts $b_{1}\left( \tau \right) =0$ and obtains the
eigenvalues $\epsilon _{s\mathbf{p}}^{\left( 0\right) }$, Eqs. (\ref{e28}, %
\ref{e30}, \ref{e31}).

As the next step, we assume non-zero $b_{1}\left( \tau \right) $ and write
\begin{equation}
b\left( \tau \right) =b_{0}\left( \tau \right) +\delta b\left( \tau \right) ,
\label{k41}
\end{equation}%
where $b_{0}\left( \tau \right) $ is given by Eq. (\ref{e21}). Then, we
expand $\mathcal{F}_{\mathrm{el}}\left[ b\left( \tau \right) ,b_{1}\left(
\tau \right) \right] ,$ Eq. (\ref{k38}), in $b_{1}\left( \tau \right) $ and $%
\delta b\left( \tau \right) $ up to the second order in these variables.
This will allows us to obtain an interaction between the fields $b\left(
\tau \right) $ and $b_{1}\left( \tau \right) $ and take into account a
screening of this interaction.

Although putting $b_{1}\left( \tau \right) =0$ leads to the correct static
solution for $b,$ it is generally not a good assumption because, as we will
see, the field $b_{1}\left( \tau \right) $ linearly couples to the time
derivative $\dot{b}_{0}\left( \tau \right) $ generating an additional term
in the free energy of IAP
\begin{equation}
\frac{\mathcal{F}_{\mathrm{int}}\left[ b_{1}\right] }{VT}=-\frac{J}{2}%
\int_{0}^{1/T}\dot{b}_{0}\left( \tau \right) b_{1}\left( \tau \right) d\tau ,
\label{e32}
\end{equation}%
where $J$ is a constant.

Fluctuations of $b_{1}\left( \tau \right) $ generate an effective attraction
between the instantons and anti-instantons and favor formation of $\tau $
-dependent structures. Formally replacing $\tau $ by a space coordinate one
can see that the mechanism of the attraction is similar to the one of the
electron-phonon interaction in solids. The field $b_{1}\left( \tau \right) $
plays in this picture the role of phonons and its fluctuations may result in
a sufficiently strong attraction of instantons and anti-instantons and,
eventually, lead to a function $b\left( \tau \right) $ oscillating in the
imaginary time $\tau .$

The calculation of the free energy functional $\mathcal{F}_{\mathrm{el}}%
\left[ b\left( \tau \right) ,b_{1}\left( \tau \right) \right] $ is done by
substituting
\begin{equation}
\epsilon _{s\mathbf{p}}=\epsilon _{s\mathbf{p}}^{\left( 0\right) }+\epsilon
_{s\mathbf{p}}^{\left( 1\right) }+\epsilon _{s\mathbf{p}}^{\left( 2\right) },
\label{k42}
\end{equation}%
into Eq. (\ref{k38}) and calculating $\epsilon _{s\mathbf{p}}^{\left(
1\right) }$ and $\epsilon _{s\mathbf{p}}^{\left( 2\right) }$ with the help
of standard quantum-mechanical formulas%
\begin{equation}
\epsilon _{s\mathbf{p}}^{\left( 1\right) }=-\int_{0}^{1/T}\Pi _{ss}\left(
\tau ,\mathbf{p}\right) d\tau ,  \label{k43}
\end{equation}%
\begin{equation}
\epsilon _{s\mathbf{p}}^{\left( 2\right) }=\sum_{s^{\prime }\neq
s}\int_{0}^{1/T}\frac{\Pi _{ss^{\prime }}\left( \tau ,\mathbf{p}\right) \Pi
_{s^{\prime }s}\left( \tau ,\mathbf{p}\right) }{\epsilon _{s\mathbf{p}%
}^{\left( 0\right) }-\epsilon _{s^{\prime }\mathbf{p}}^{\left( 0\right) }}%
d\tau ,  \label{k43a}
\end{equation}%
where
\begin{equation*}
\Pi _{ss^{\prime }}\left( \tau ,\mathbf{p}\right) =\bar{\Psi}_{s\mathbf{p}%
}^{\left( 0\right) }\left( \tau \right) \left( ib_{1}\left( \tau \right)
\Sigma _{1}+\delta b\left( \tau \right) \Sigma _{3}\right) \Psi _{s^{\prime }%
\mathbf{p}}^{\left( 0\right) }\left( \tau \right) .
\end{equation*}

As soon as the electronic part is calculated, one should minimize $\mathcal{F%
}\left[ b\left( \tau \right) ,b_{1}\left( \tau \right) \right] ,$ Eq. (\ref%
{k10}, \ref{e8b}), with respect to $b_{1}\left( \tau \right) $ and $\delta
b\left( \tau \right) ,$ and calculate the free energy in terms of the
solution $b_{0}\left( \tau \right) ,$ Eq. (\ref{e21}).

\subsection{Free energy for an arbitrary spectrum $\protect\varepsilon %
^{-}\left( \mathbf{p}\right) $.}

We consider here the most interesting limit of small $1-k\ll 1$ assuming
that temperatures are low, $T\ll \gamma $. Taking $b_{0}\left( \tau \right) $
in the form of Eq. (\ref{e21}) one can determine the parameter $\gamma $
from Eq. (\ref{e19}) which is, at the same time, the mean field equation for
the time-independent order parameter (gap in the electron spectrum) of the
DDW state \cite{volkov3}. The modulus $k$ and, hence, the number of IAP
drops out from Eq. (\ref{e19}) and can be determined only from the condition
of the minimum of free energy. In the absence of the field $b_{1}\left( \tau
\right) ,$ the minimum is always reached at the time-independent solution $%
b=\gamma $. The fields $b_{1}\left( \tau \right) $ couple to $\dot{b}%
_{0}\left( \tau \right) $, Eq. (\ref{e32}), and a finite number $m$ of the
instantons-antiinstantons can provide the absolute minimum of the free
energy. This statement can be checked by calculating the free energy for
finite $m$ in the linear approximation in this number. Negative values of
the difference $\Delta F$ between this free energy and the free energy of
the DDW will indicate the possibility of the imaginary-time-dependent state.

In the limit $1-k\ll 1,$ one can write the parameter $\kappa _{\mathbf{p}}$ (%
\ref{e30}, \ref{e31}) in a simplified form
\begin{equation}
\kappa _{\mathbf{p}}=\kappa _{\mathbf{p}}^{\left( DDW\right) }+\kappa _{%
\mathbf{p}}^{\left( inst\right) },  \label{k64a}
\end{equation}%
where%
\begin{equation}
\kappa _{\mathbf{p}}^{\left( DDW\right) }=\frac{\sqrt{\left( \varepsilon
^{-}\left( \mathbf{p}\right) \right) ^{2}+\gamma ^{2}}}{T}  \label{k64}
\end{equation}%
is the function $\kappa _{\mathbf{p}}$ for the DDW state, and
\begin{equation}
\kappa _{\mathbf{p}}^{\left( inst\right) }=-m\ln \frac{1+\frac{\gamma }{%
\sqrt{\left( \varepsilon ^{-}\left( \mathbf{p}\right) \right) ^{2}+\gamma
^{2}}}}{1-\frac{\gamma }{\sqrt{\left( \varepsilon ^{-}\left( \mathbf{p}%
\right) \right) ^{2}+\gamma ^{2}}}},  \label{k64b}
\end{equation}%
is the contribution of the instantons and antiinstantons. The term $\kappa _{%
\mathbf{p}}^{\left( inst\right) }$ is smaller than $\kappa _{\mathbf{p}%
}^{\left( DDW\right) }$ at low temperatures but still can be large at large $%
m.$

Eqs. (\ref{k64a}-\ref{k64b}) are used for calculation of $\epsilon _{s%
\mathbf{p}}^{\left( 0\right) }$, Eq. (\ref{e28}). Substituting Eqs. (\ref%
{k64a}-\ref{k64b}, \ref{e28}) into (\ref{k38}), performing summation over $n$%
, and adding a contribution coming from the $b^{2}\left( \tau \right) $-term
in Eq. (\ref{k10}) we obtain with the help of the self-consistency equation (%
\ref{e19}) the energy of $F_{\mathrm{inst}}$ of `non-interacting'\
instantons and antiinstantons
\begin{eqnarray}
&&\frac{F_{\mathrm{inst}}}{2mVT}  \label{k66} \\
&=&\int \left[ \ln \frac{1+\frac{\gamma }{\sqrt{\left( \varepsilon
^{-}\left( \mathbf{p}\right) \right) ^{2}+\gamma ^{2}}}}{1-\frac{\gamma }{%
\sqrt{\left( \varepsilon ^{-}\left( \mathbf{p}\right) \right) ^{2}+\gamma
^{2}}}}-\frac{2\gamma }{\sqrt{\left( \varepsilon ^{-}\left( \mathbf{p}%
\right) \right) ^{2}+\gamma ^{2}}}\right] \frac{d\mathbf{p}}{\left( 2\pi
\right) ^{2}}.  \notag
\end{eqnarray}%
The energy $F_{\mathrm{inst}}$, Eq. (\ref{k66}), is always positive. The
effective attraction arises due to the interaction of the instantons with
the field $b_{1}\left( \tau \right) $, Eq. (\ref{e32}), and can be obtained
calculating the first order $\epsilon _{s\mathbf{p}}^{\left( 1\right) }$,
Eq. (\ref{k43}). Using Eq. (\ref{k61}) one can easily carry out summation
over $n$ to obtain Eq. (\ref{e32}) with the constant $J,$%
\begin{equation}
J=\frac{1}{2}\int \frac{sgn\left( \varepsilon ^{-}\left( \mathbf{p}\right)
\right) }{\sqrt{\left( \left( \varepsilon ^{-}\left( \mathbf{p}\right)
\right) ^{2}+\gamma ^{2}\frac{\left( 1-k\right) ^{2}}{4}\right) \left(
\left( \varepsilon ^{-}\left( \mathbf{p}\right) \right) ^{2}+\gamma
^{2}\right) }}\frac{d\mathbf{p}}{\left( 2\pi \right) ^{2}}.  \label{k67}
\end{equation}%
At small $1-k$ the integral $J$, Eq. (\ref{k67}), can be very large and the
attraction very strong due to a large contribution coming from the region of
small $\varepsilon ^{-}\left( \mathbf{p}\right) $. Therefore, the second
order $\epsilon _{s\mathbf{p}}^{\left( 2\right) },$ Eq. (\ref{k43a}), is
also important because it leads to a screening of the interaction (\ref{e32}%
, \ref{k67}) and partially cuts the singularity at $\varepsilon ^{-}\left(
\mathbf{p}\right) =0$ arising in the integrand in (\ref{k67}) in the limit $%
k\rightarrow 1.$

As a result, one can write the difference $\Delta \mathcal{F}\left[
b_{1},\delta b\right] $ between the free energy of the system with $m$
alternating instantons and antiinstantons and that of the system without
instantons as
\begin{equation}
\Delta \mathcal{F}\left[ b_{1},\delta b\right] =F_{\mathrm{inst}}+\mathcal{F}%
_{\mathrm{int}}\left[ b_{1}\right] +\mathcal{F}_{\mathrm{2}}\left[
b_{1},\delta b\right]  \label{k67b}
\end{equation}%
with $F_{\mathrm{inst}}$ given by Eq. (\ref{k66}), $\mathcal{F}_{\mathrm{int}%
}\left[ b_{1}\right] $ specified by Eqs. (\ref{e32}, \ref{k67}) and a
quadratic form $\mathcal{F}_{\mathrm{2}}\left[ b_{1},\delta b\right] $ of $%
b_{1}\left( \tau \right) $ and $\delta b\left( \tau \right) $
\begin{eqnarray}
&&\frac{\mathcal{F}_{\mathrm{2}}\left[ b_{1},\delta b\right] }{VT}
\label{k67a} \\
&=&\int_{0}^{1/T}\Big[\left( \mathcal{A}_{0}+\frac{1}{4}\left( \mathcal{A}%
_{1}-\frac{\mathcal{C}^{2}}{\mathcal{B}}\right) \dot{b}_{0}^{2}\left( \tau
\right) \right) b_{1}^{2}\left( \tau \right) +  \notag \\
&&+\mathcal{B}b_{0}^{2}\left( \tau \right) \left( \delta b\left( \tau
\right) -\frac{\mathcal{C}\dot{b}_{0}\left( \tau \right) }{2\mathcal{B}%
b_{0}\left( \tau \right) }b_{1}\left( \tau \right) \right) ^{2}\Big]d\tau ,
\notag
\end{eqnarray}%
where the constants $\mathcal{A}_{0}$, $\mathcal{A}_{1}$, $\mathcal{B}$ and $%
\mathcal{C}$ equal
\begin{equation}
\mathcal{A}_{0}\mathcal{=}\left( 1+\frac{U_{0}}{\tilde{U}_{0}}\right) \int
\frac{1}{\sqrt{\left( \varepsilon ^{-}\left( \mathbf{p}\right) \right)
^{2}+\gamma ^{2}}}\frac{d\mathbf{p}}{\left( 2\pi \right) ^{2}},  \label{k69}
\end{equation}%
\begin{equation}
\mathcal{A}_{1}=\int \frac{1}{\left( \left( \varepsilon ^{-}\left( \mathbf{p}%
\right) \right) ^{2}+\gamma ^{2}\right) ^{3/2}\left( \left( \varepsilon
^{-}\left( \mathbf{p}\right) \right) ^{2}+\gamma ^{2}\frac{\left( 1-k\right)
^{2}}{4}\right) }\frac{d\mathbf{p}}{\left( 2\pi \right) ^{2}},  \label{k70}
\end{equation}

\begin{equation}
\mathcal{B=}\frac{1}{\left( \left( \varepsilon ^{-}\left( \mathbf{p}\right)
\right) ^{2}+\gamma ^{2}\right) ^{3/2}}\frac{d\mathbf{p}}{\left( 2\pi
\right) ^{2}},  \label{k71}
\end{equation}%
\begin{equation}
\mathcal{C=}\int \frac{\varepsilon ^{-}\left( \mathbf{p}\right) }{\left(
\left( \varepsilon ^{-}\left( \mathbf{p}\right) \right) ^{2}+\gamma
^{2}\right) ^{3/2}\left( \left( \varepsilon ^{-}\left( \mathbf{p}\right)
\right) ^{2}+\frac{\gamma ^{2}\left( 1-k\right) ^{2}}{4}\right) }\frac{d%
\mathbf{p}}{\left( 2\pi \right) ^{2}}.  \label{k72}
\end{equation}%
The minimum of $\mathcal{F}_{\mathrm{2}}\left[ b_{1},\delta b\right] $ with
respect to $\delta b\left( \tau \right) $ is achieved at%
\begin{equation}
\delta b\left( \tau \right) =\frac{\mathcal{C}\dot{b}_{0}\left( \tau \right)
}{2\mathcal{B}b_{0}\left( \tau \right) }b_{1}\left( \tau \right) .
\label{k67c}
\end{equation}%
Then, one finds the minimum value of $\Delta \mathcal{F}\left[ b_{1},\delta b%
\right] ,$ Eq. (\ref{k67b}) giving the free energy $\Delta F$
\begin{eqnarray}
&&\frac{\Delta F}{VT}=\frac{F_{\mathrm{inst}}}{VT}  \label{k68} \\
&&-J^{2}\int_{0}^{1/T}\left[ \mathcal{A}_{0}+\frac{1}{4}\left( \mathcal{A}%
_{1}\mathcal{-}\frac{\mathcal{C}^{2}}{\mathcal{B}_{0}}\right) \dot{b}%
_{0}^{2}\left( \tau \right) \right] ^{-1}\dot{b}_{0}^{2}\left( \tau \right)
d\tau ,  \notag
\end{eqnarray}%
where $\Delta F$ is the difference between the total free energy $F$ and the
free energy $F_{\hom }$ of the system with the order parameter homogeneous
in the imaginary time.

At first glance, the energy $\Delta F$ is proportional to $T$, which is
small at low temperatures. However, in the limit of small $k,$ when the
period of the oscillations is large, the energy $\Delta F$ is proportional
to the number $m,$ that can be large and proportional to $1/T.$ The energy $%
\Delta F$ is proportional to the volume as well, and therefore the
contribution of the IAP into thermodynamical quantities will exceed those
coming from fluctuations. The case $\Delta F/\left( 2mTV\right) >0$
corresponds to the state with the static order parameter, while in the
region of parameters where $\Delta F/V\left( 2mT\right) <0$ one expects a
chain of alternating instantons and anti-instantons. A more accurate
calculations are necessary to determine the number $m$ as a function of
temperature and parameters of the model. Here we restrict ourselves by
investigating the stability against formation the chain of the alternating
instanton and antiinstantons.

The main contribution to the integral, Eq. (\ref{k68}), comes from the
vicinity of zeros of $b_{0}\left( \tau \right) ,$ where the derivative $\dot{%
b}_{0}\left( \tau \right) $ is essentially non-zero. Therefore, the integral
is proportional to $2m$ (as well as $F_{\mathrm{inst}},$ Eq. (\ref{k66})),
and the integration is reduced to the integration over the half period of
the function $b_{0}\left( \tau \right) .$ The free energy $\Delta F$ is also
proportional to $2m,$ and one can calculate the energy per one instanton
replacing $b_{0}\left( \tau \right) $ by $\gamma \tanh \gamma \tau $ and
then integrating over $\tau $ from $-\infty $ to $\infty $ because the
distance between the instantons and antiinstantons is very large in the
limit $k\rightarrow 1.$ Equation (\ref{k68}) can be further simplified
introducing a new variable of integration $v=\gamma \tanh \gamma \tau .$

In order to compute the energy $\Delta F$ explicitly one should choose a
specific form of the electron spectrum. Having in mind SFMOHS \cite%
{volkov1,volkov2,volkov3} we write the spectrum as%
\begin{equation}
\varepsilon _{1}\left( \mathbf{p}\right) =\alpha p_{x}^{2}-\beta
p_{y}^{2}+P-\mu ,\quad \varepsilon _{2}\left( \mathbf{p}\right) =\alpha
p_{y}^{2}-\beta p_{x}^{2}-P-\mu ,  \label{k73}
\end{equation}%
where $\mu $ is the chemical potential and $P$ is a Pomeranchuk order
parameter that may appear in the model under consideration. The spectrum
displayed in Eq. (\ref{k73}) corresponds at $P=0$ to the Fermi surface
depicted in Fig. \ref{fig:hotspots}a. In principle, the Pomeranchuk order
parameter can compete with the DDW and the state with the instantons and
antiinstantons, and one should then consider all these phases together. In
order to avoid too complicated formulas here, we neglect the inverse effect
of the instantons of the value of $P$ and consider the latter as an
independent parameter. We introduce also an energy cutoff $\Lambda $
limiting the areas of the hot spots as%
\begin{equation}
\frac{\alpha +\beta }{2}p^{2}<\Lambda ,  \label{k74}
\end{equation}%
with $p^{2}=p_{x}^{2}+p_{y}^{2}.$

\begin{figure*}[htp]
\centering
\begin{subfigure}[b] {0.3\linewidth}
  \includegraphics[width=\linewidth]{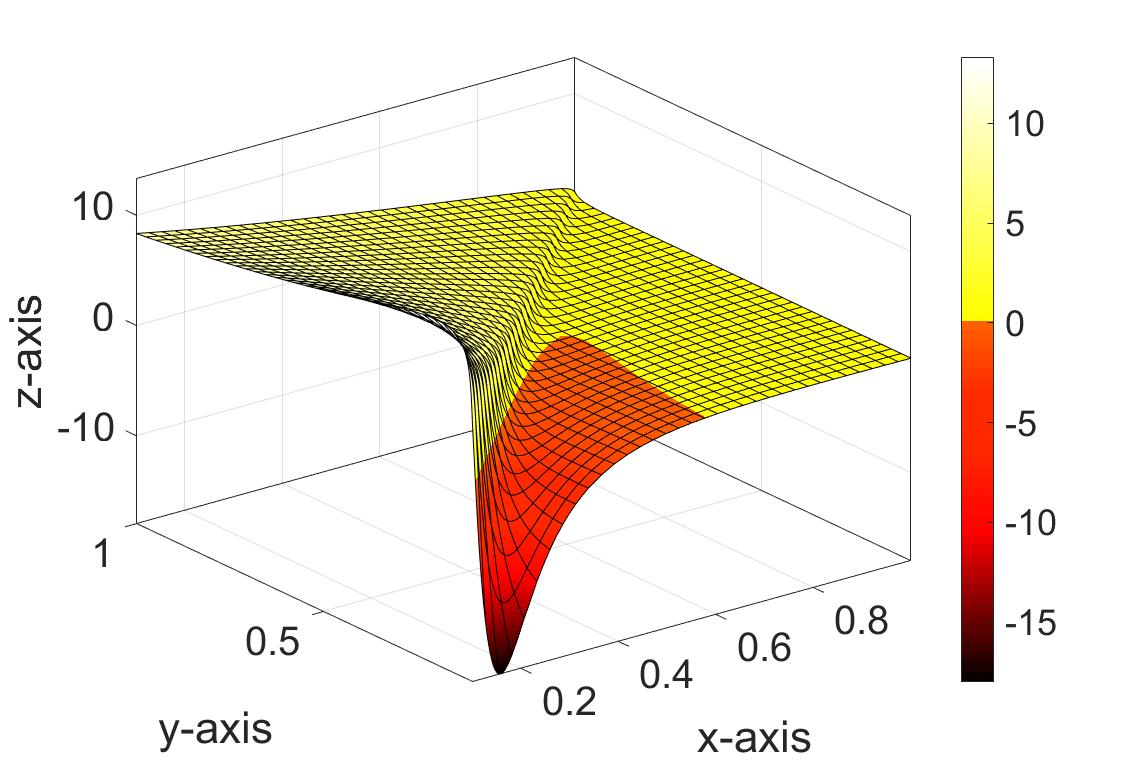}
 \caption{$k=0.99$, $a=0$}
 \end{subfigure}
\begin{subfigure}[b]{0.3\linewidth}
 \includegraphics[width=\linewidth]{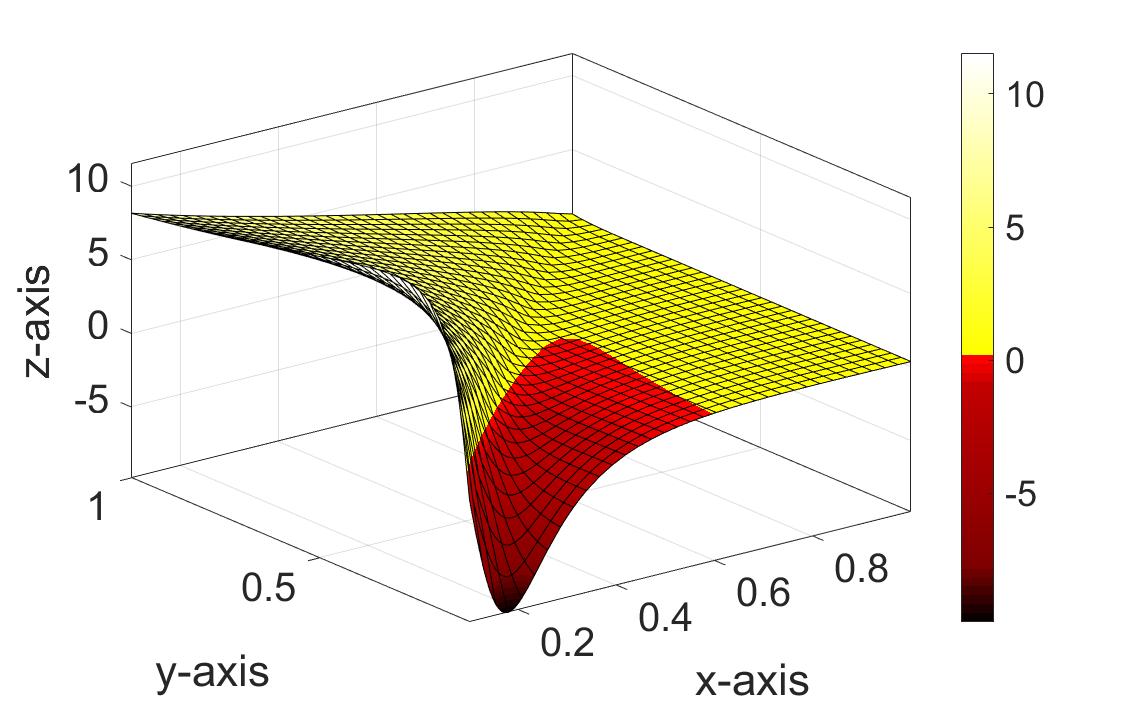}
 \caption{$k=0.90$, $a=0$}
 \end{subfigure}
\begin{subfigure}[b]{0.3\linewidth}
 \includegraphics[width=\linewidth]{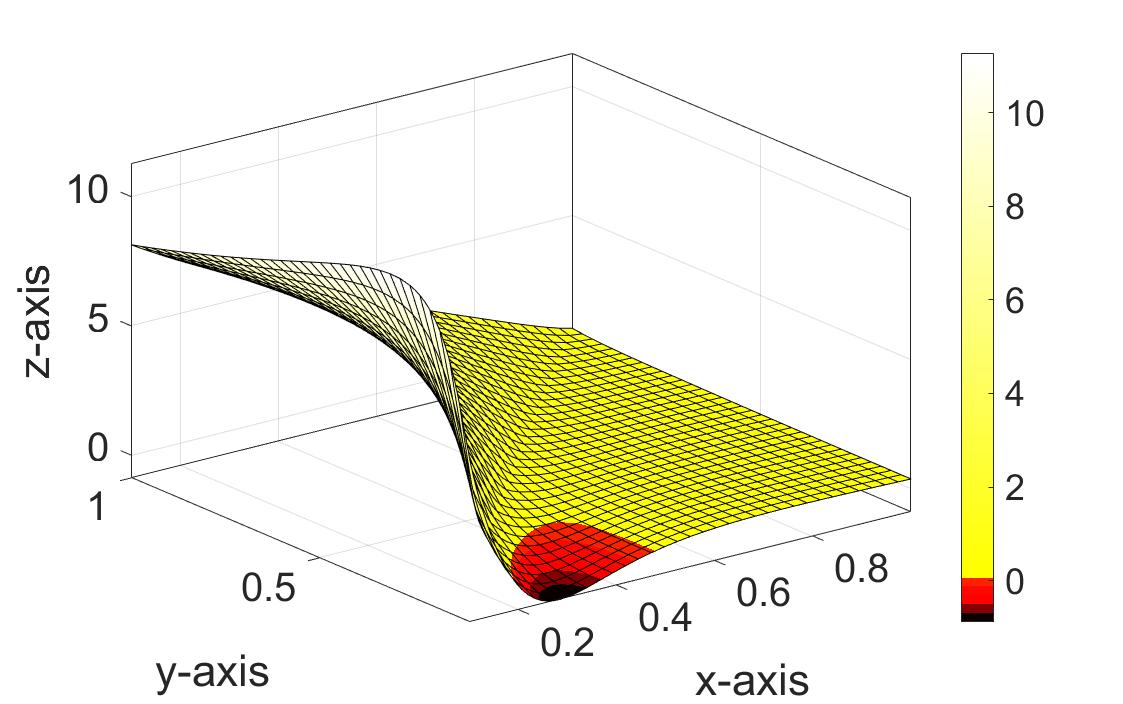}
 \caption{$k=0.70$, $a=0$}
 \end{subfigure}
\par
\begin{subfigure}[b]{0.3\linewidth}
 \includegraphics[width=\linewidth]{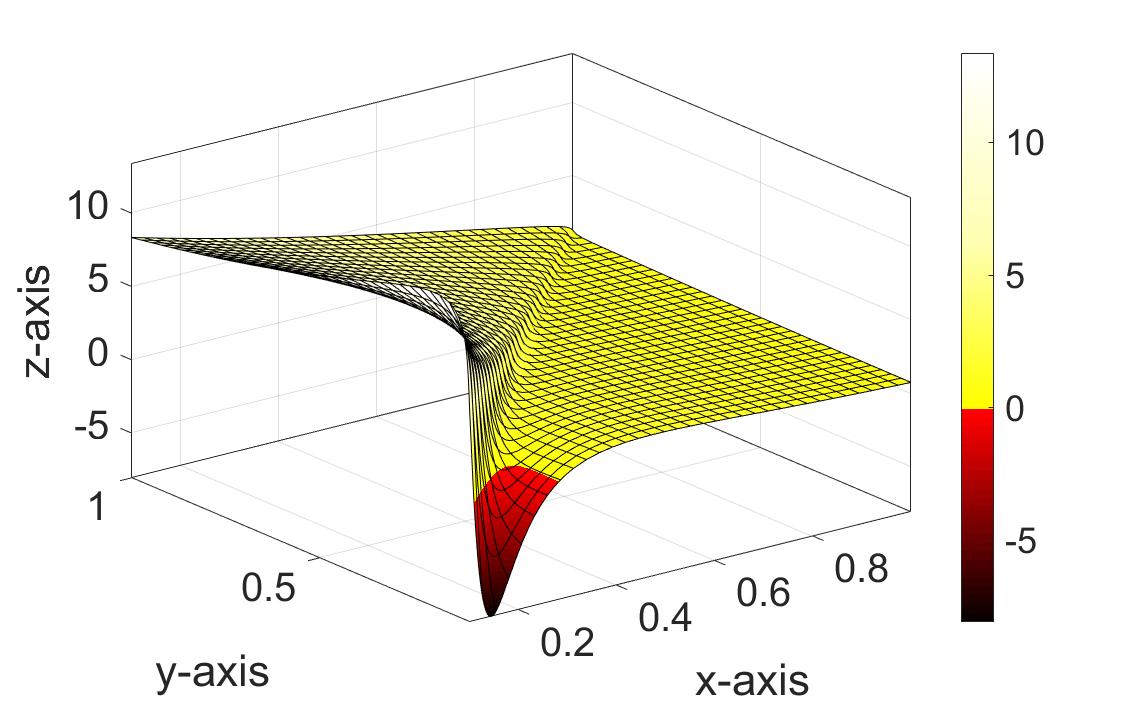}
 \caption{$k=0.99$, $a=1$}
 \end{subfigure}
\begin{subfigure}[b]{0.3\linewidth}
 \includegraphics[width=\linewidth]{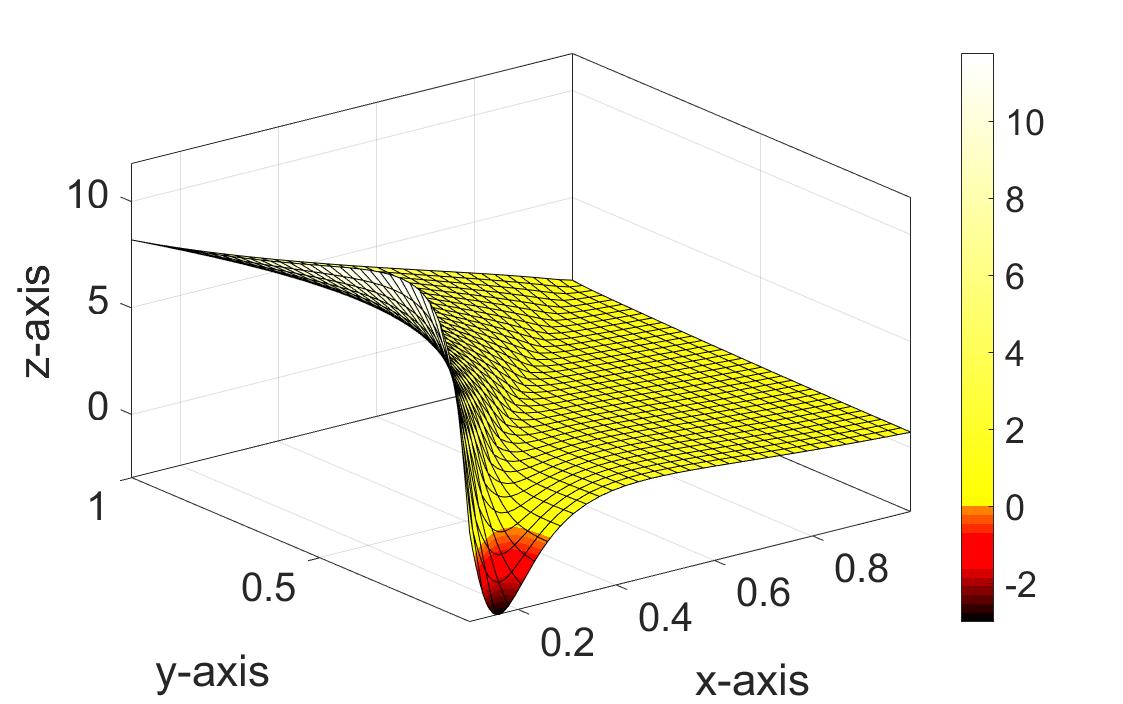}
 \caption{$k=0.90$, $a=1$}
 \end{subfigure}
\begin{subfigure}[b]{0.3\linewidth}
 \includegraphics[width=\linewidth]{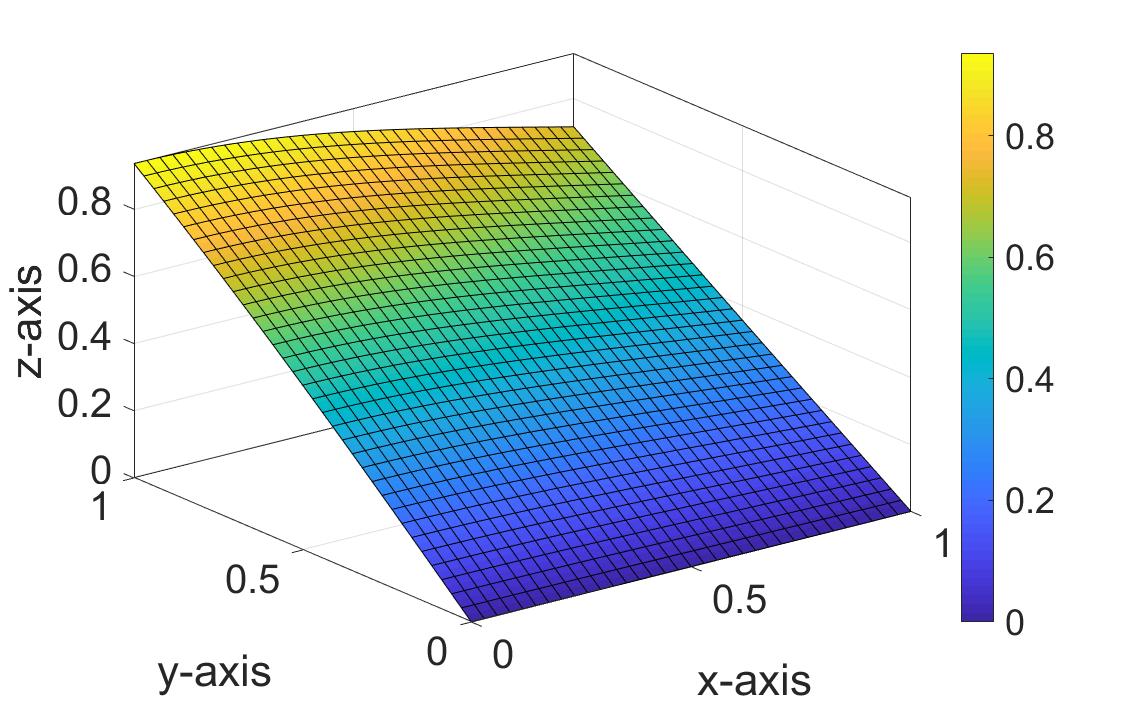}
 \caption{$z=g_{0}^{-1}(x,y)$}
 \end{subfigure}
\caption{(Color online.) Free energy of instanton-antiinstanton pairs. }
\label{fig:energy}
\end{figure*}

The dependence of a function $S=\Delta F/\left( 2mTV\right) $ on parameters
characterizing the energy spectrum in the SFMOHS is represented in Fig. \ref%
{fig:energy}. We use the following notations
\begin{equation*}
a=U_{0}/\tilde{U}_{0}\text{, }g_{0}=\frac{1}{2\pi }\frac{U_{0}}{\alpha
+\beta }.
\end{equation*}%
Figs. \ref{fig:energy}(a-e) represent dependence of $z=2\pi ^{2}\left(
\alpha +\beta \right) S/\Lambda $ on $x=P/\gamma $ and $y=\Lambda /\gamma $
for several values of $k$ and $a$. Fig. \ref{fig:energy}f describes the
solution of the self-consistency equation (\ref{e19}) for arbitrary $\Lambda
$ and $P$. More information about the computations can be found in the
Appendix.

We see from Figs. \ref{fig:energy} that there is a certain region of
parameters where the free energy $\Delta F$ is negative, which indicates the
instability of the imaginary-time-independent state (DDW). The region of
small $1-k$ and $a$ is most favorable for the formation of the lattice of
IAP. As we consider here structures periodic in space (oscillations with
vector $\mathbf{Q}_{AF}$ connecting the bands $1$ and $2$, Fig. \ref%
{fig:hotspots}b), the periodic in $\tau $ order parameter $b\left( \tau
\right) $ providing the minimum of the free energy is at the same time the
amplitude of the periodic oscillations in space.

The present calculations do not determine the number $m$ of IAP as a
functions of temperature. Leaving this problem for future investigations, we
will calculate in the next sections physical quantities without specifying
the value of $m$ or, alternatively, $k$ related to the latter by Eq. (\ref%
{e22}).

\section{\label{sec:Time_Crystal}Thermodynamic quantum time-space crystal.}

In the previous sections we have shown that the state with the order
parameter $b\left( \tau \right) $ represented by a chain of alternating
instantons and antiinstantons in the imaginary time can be more favorable
energetically than the one with the static order parameter. Now we discuss
properties of this state concerning its behavior in real time.

The thermodynamical quantities have been calculated using imaginary time $%
\tau $ and Matsubara frequencies. Methods of calculations of dynamic
quantities for `conventional systems' are well developed. Linear response
functions are calculated by an analytical continuation from imaginary
Matsubara frequencies to real ones \cite{agd}. This method is based on an
assumption that the response function is analytical at $\infty $ on the
complex plane of the frequencies. However, due to periodicity in both real $%
t $ and imaginary time $\tau $ of the function $b_{0}\left( \tau \right) $,
Eq. (\ref{e21}), one obtains a Fourier series rather than a Fourier integral
in the complex plane of the frequencies and the standard analytical
continuation cannot be applied for this essentially thermodynamic problem.

Fortunately, one can use now the fact that the function $b_{0}\left( \tau
\right) ,$ Eq. (\ref{e21}), is analytical in the complex plane of $\tau $
(except poles). Equation (\ref{k46}) as well as equations (\ref{k49}-\ref%
{k52}) can be used everywhere on this plane including the axis of real time $%
t=-i\tau .$

At zero temperature $T=0$, one can rather easily represent real-time
correlation functions in terms of a functional integral in real time. This
is the standard field-theoretical formulation of quantum mechanics in terms
of path integrals. Following this method one represents the current-current
correlation function $N\left( t_{1}-t_{2}\right) $ in the form
\begin{eqnarray}
&&N\left( t_{1}-t_{2}\right)  \label{m1} \\
&=&\frac{U_{0}^{2}}{V^{2}}\sum_{\mathbf{p,p}^{\prime }\mathbf{,}\alpha
,\alpha ^{\prime }}\left\langle \left( \eta _{p}^{+}\left( t_{1}\right)
\Sigma _{3}\eta _{p}\left( t_{1}\right) \right) \left( \eta _{p^{\prime
}}^{+}\left( t_{2}\right) \Sigma _{3}\eta _{p^{\prime }}\left( t_{2}\right)
\right) \right\rangle _{\tilde{S}}.  \notag
\end{eqnarray}%
In Eq. (\ref{m1}) the same fields $\eta ,$ $\eta ^{+}$ as in the previous
sections are used. However, they are functions of real time, and the angle
brackets stand for averaging with the action written in real time. The
function $N\left( t_{1}-t_{2}\right) $ is proportional to a function
entering the scattering cross-sections. The overall coefficient in Eq. (\ref%
{m1}) is chosen for simplicity.

The equivalent functional integration is based on averaging with action $%
\tilde{S}$ written in real time%
\begin{equation}
\left\langle ...\right\rangle _{\tilde{S}}=\frac{\int \left( ...\right) e^{-i%
\tilde{S}\left[ \eta \right] }D\eta }{\int e^{-i\tilde{S}\left[ \eta \right]
}D\eta }.  \label{m2}
\end{equation}%
In Eq. (\ref{m2}), the action $\tilde{S}\left[ \eta \right] $ has the
following form
\begin{equation}
\tilde{S}_{\mathrm{0}}\left[ \eta \right] =\tilde{S}_{\mathrm{0}}\left[ \eta %
\right] +\tilde{S}_{\mathrm{int}}\left[ \eta \right] ,  \label{m3}
\end{equation}%
where
\begin{equation}
\tilde{S}_{\mathrm{0}}\left[ \eta \right] =\sum_{p}\int_{-\infty }^{\infty
}\eta _{p}^{+}\left( t\right) \left( -i\partial _{t}+\varepsilon ^{+}\left(
\mathbf{p}\right) -\varepsilon ^{-}\left( \mathbf{p}\right) \Sigma
_{2}\right) \eta _{p}\left( t\right) dt,  \label{m4}
\end{equation}%
and%
\begin{eqnarray}
&&\tilde{S}_{\mathrm{int}}\left[ \chi \right] =-\frac{1}{4V}\int_{-\infty
}^{\infty }\Big[U_{\mathrm{0}}\Big(\sum_{p}\eta _{p}^{+}\left( t\right)
\Sigma _{3}\eta _{p}\left( t\right) \Big)^{2}  \notag \\
&&-\tilde{U}_{\mathrm{0}}\Big(\sum_{p}\eta _{p}^{+}\left( t\right) \Sigma
_{1}\eta _{p}\left( t\right) \Big)^{2}\Big]dt.  \label{m6}
\end{eqnarray}%
The action $\tilde{S}\left[ \eta \right] $ can easily be obtained from Eqs. (%
\ref{e5}-\ref{e4}) making the Wick rotation $\tau \rightarrow it.$ Again,
the evaluation of the functional integrals can be performed decoupling the
interaction by Gaussian integration over auxiliary fields $B\left( t\right) $
and $B_{1}\left( t\right) $. This leads us to the electron part of action $%
\mathcal{S}\left[ \eta ,\,\eta ^{+},B,B_{1}\left( t\right) \right] $
containing both fermion $\eta ,\eta ^{+}$ and boson $B\left( t\right) ,$ $%
B_{1}\left( t\right) $ fields%
\begin{equation}
\mathcal{S}\left[ \eta \,\eta ^{+},B,B_{1}\left( t\right) \right]
=\int_{-\infty }^{\infty }\eta _{p}^{+}\left( t\right) \mathcal{H}\left( t,%
\mathbf{p}\right) \eta _{p}\left( t\right) dt  \label{m7}
\end{equation}%
with the operator $\mathcal{H}\left( t,\mathbf{p}\right) $ equal to
\begin{equation}
\mathcal{H}\left( t,\mathbf{p}\right) =-i\partial _{t}+\varepsilon
^{+}\left( \mathbf{p}\right) -\mathcal{H}_{0}\left( t,\mathbf{p}\right) ,
\label{m8}
\end{equation}%
where
\begin{equation}
\mathcal{H}_{0}\left( t,\mathbf{p}\right) =\varepsilon ^{-}\left( \mathbf{p}%
\right) \Sigma _{2}+i\left( B\left( t\right) \Sigma _{3}+B_{1}\left(
t\right) \Sigma _{1}\right) .  \label{m8a}
\end{equation}%
Then, we integrate over $\eta ,\eta ^{+}$ and reduce the full action $S$ to
the form

\begin{equation}
S=-\ln \left[ \int \exp \left[ -i\mathcal{S}\left[ B,B_{1}\right] \right]
DBDB_{1}\right] ,  \label{m9}
\end{equation}%
where the action $\mathcal{S}\left[ B,B_{1}\right] $ equals
\begin{eqnarray}
\mathcal{S}\left[ B,B_{1}\right] &=&\int_{-\infty -}^{\infty }\Big[-2\sum_{%
\mathbf{p}}\mathrm{tr}\left[ \ln \left( \mathcal{H}\left( t,\mathbf{p}%
\right) \right) \right] _{t,t}  \label{m10} \\
&&-V\left( \frac{B^{2}\left( t\right) }{U_{\mathrm{0}}}-\frac{%
B_{1}^{2}\left( t\right) }{\tilde{U}_{\mathrm{0}}}\right) \Big]dt,  \notag
\end{eqnarray}%
Minimizing $\mathcal{S}\left[ B,B_{1}\right] $ with respect to $B\left(
t\right) $ and $B_{1}\left( t\right) $ we come to equations
\begin{eqnarray}
B\left( t\right) &=&-iU_{\mathrm{0}}\mathrm{tr}\int \Sigma _{3}\mathcal{G}_{%
\mathbf{p}}\left( t,t^{\prime }\right) \frac{d\mathbf{p}}{\left( 2\pi
\right) ^{2}},  \label{m11} \\
B_{1}\left( t\right) &=&i\tilde{U}_{\mathrm{0}}\mathrm{tr}\int \Sigma _{1}%
\mathcal{G}_{\mathbf{p}}\left( t,t^{\prime }\right) \frac{d\mathbf{p}}{%
\left( 2\pi \right) ^{2}},  \label{m12}
\end{eqnarray}%
where the matrix Green function $\mathcal{G}_{\mathbf{p}}\left( t,t^{\prime
}\right) $ satisfies the equations%
\begin{eqnarray}
\mathcal{H}\left( t,\mathbf{p}\right) \mathcal{G}_{\mathbf{p}}\left(
t,t^{\prime }\right) &=&-\delta \left( t-t^{\prime }\right) ,  \label{m12c}
\\
\mathcal{G}_{\mathbf{p}}\left( t,t^{\prime }\right) \mathcal{\bar{H}}\left(
t^{\prime },\mathbf{p}\right) &=&-\delta \left( t-t^{\prime }\right) .
\label{m12ca}
\end{eqnarray}%
In Eq. (\ref{m12d}) the operator $\mathcal{\bar{H}}\left( t,\mathbf{p}%
\right) $ equals
\begin{equation}
\mathcal{\bar{H}}\left( t,\mathbf{p}\right) =i\overleftarrow{\partial }%
_{t}+\varepsilon ^{+}\left( \mathbf{p}\right) -\mathcal{H}_{0}\left( t,%
\mathbf{p}\right) .  \label{m12ea}
\end{equation}%
We will see that the solutions $B\left( t\right) $, $B_{1}\left( t\right) $
are periodic. The integrands in Eqs. (\ref{m11}, \ref{m12}) can be
calculated using spectral expansions similar to those performed in Sections (%
\ref{sec:HS}, \ref{sec:Free}). Indeed, one can introduce eigenfunctions $%
\psi _{s\mathbf{p}}\left( t,\mathbf{p}\right) $ and eigenvalues $E_{s\mathbf{%
p}}$ of the operator $\mathcal{H}\left( t,\mathbf{p}\right) $, Eq. (\ref{m8}%
), as%
\begin{eqnarray}
\mathcal{H}\left( t,\mathbf{p}\right) \psi _{s\mathbf{p}}\left( t\right)
&=&E_{s\mathbf{p}}\psi _{s\mathbf{p}}\left( t\right) ,  \label{m12a} \\
\bar{\psi}_{s\mathbf{p}}\left( t\right) \mathcal{\bar{H}}\left( t,\mathbf{p}%
\right) &=&E_{s\mathbf{p}}\bar{\psi}_{s\mathbf{p}}\left( t\right) .  \notag
\end{eqnarray}%
Equations (\ref{m11}, \ref{m12}), and (\ref{m12a}) are very similar to Eqs. (%
\ref{e14}, \ref{e15}) and (\ref{k32}), respectively, and can be obtained
from those setting $T=0$ and making the rotation $\tau \rightarrow it$.

It can rather easily be demonstrated that there are real periodic solutions $%
B\left( t\right) $ and $B_{1}\left( t\right) $ of Eqs. (\ref{m11}, \ref{m12}%
) determining the minimum of the action $\mathcal{S}\left[ B,B_{1}\right] ,$
Eq. (\ref{m10}). Indeed, subtracting Eqs. (\ref{m12c}, \ref{m12ca}) from
each other one obtains putting $t^{\prime }=t$ the following equation
\begin{equation}
-i\partial _{t}\left( \mathcal{G}_{\mathbf{p}}\left( t,t\right) \right) -%
\left[ \mathcal{H}_{0}\left( t,\mathbf{p}\right) ,\mathcal{G}_{\mathbf{p}%
}\left( t,t\right) \right] =0,  \label{m12d}
\end{equation}%
where $\left[ ..,..\right] $ stands for the commutator.

Representing the solution $\mathcal{G}_{\mathbf{p}}\left( t,t\right) $ in
the form
\begin{equation}
\mathcal{G}_{\mathbf{p}}\left( t,t\right) =iS_{\mathbf{p}}^{1}\left(
t\right) \Sigma _{1}+S_{\mathbf{p}}^{2}\left( t\right) \Sigma _{2}+iS_{%
\mathbf{p}}^{3}\Sigma _{3},  \label{m12e}
\end{equation}%
one can transform Eq. (\ref{m12e}) as
\begin{equation}
\partial _{t}\mathbf{S}_{\mathbf{p}}\left( t\right) =2\mathbf{B}_{\mathbf{p}%
}\left( t\right) \times \mathbf{S}_{\mathbf{p}}\left( t\right) ,
\label{m12f}
\end{equation}%
where the vectors $\mathbf{S}_{\mathbf{p}}\left( t\right) $ and $\mathbf{B}_{%
\mathbf{p}}\left( t\right) $ equal
\begin{eqnarray}
\mathbf{S}_{\mathbf{p}}\left( t\right) &=&\left( S_{\mathbf{p}}^{1}\left(
t\right) ,S_{\mathbf{p}}^{2}\left( t\right) ,S_{\mathbf{p}}^{3}\left(
t\right) \right) ,\quad  \label{m12g} \\
\mathbf{B}_{\mathbf{p}}\left( t\right) &=&\left( B_{1}\left( t\right)
,\varepsilon ^{-}\left( \mathbf{p}\right) ,B\left( t\right) \right) .  \notag
\end{eqnarray}%
The vector product in Eq. (\ref{m12f}) is defined for arbitrary $3$%
-component vectors $\mathbf{L}$ and $\mathbf{M}$ as
\begin{equation}
\left[ \mathbf{L}\times \mathbf{M}\right] ^{k}=e_{ijk}g_{kk}L^{i}L^{j},
\label{m12m}
\end{equation}%
where $e_{ijk}$ is the antisymmetric tensor ($e_{123}=1$), and $g=\mathrm{%
diag}\left( -1,1,-1\right) $ is metric. Accordingly, a scalar product of the
vectors $\mathbf{L}$ and $\mathbf{M}$ equals%
\begin{equation}
\left( \mathbf{LM}\right)
=\sum_{k}g_{kk}L^{k}M^{k}=-L^{1}M^{1}+L^{2}M^{2}-L^{3}M^{3}.  \label{m12n}
\end{equation}

Using the definitions (\ref{m12m}, \ref{m12n}) we multiply both the sides of
Eq. (\ref{m12f}) by the vector $\mathbf{S}_{\mathbf{p}}\left( t\right) $ to
obtain
\begin{equation}
\partial _{t}\mathbf{S}_{\mathbf{p}}^{2}\left( t\right) =0,\quad \mathbf{S}_{%
\mathbf{p}}^{2}\left( t\right) =q_{\mathbf{p}}^{2}.  \label{m12h}
\end{equation}%
In Eq. (\ref{m12h}), $q_{\mathbf{p}}$ is a time-independent function of $%
\mathbf{p.}$ Further, Eqs. (\ref{m11}, \ref{m12}) can be rewritten in the
form
\begin{eqnarray}
B\left( t\right) &=&2U_{\mathrm{0}}\int S_{\mathbf{p}}^{3}\left( t\right)
\frac{d\mathbf{p}}{\left( 2\pi \right) ^{2}},  \label{m12k} \\
B_{1}\left( t\right) &=&-2\tilde{U}_{\mathrm{0}}\int S_{\mathbf{p}%
}^{1}\left( t\right) \frac{d\mathbf{p}}{\left( 2\pi \right) ^{2}}.
\label{m12l}
\end{eqnarray}%
Assuming that the functions $B\left( t\right) $ and $B_{1}\left( t\right) $
are periodic in time with a period $T_{0}$ one can see using Eqs. (\ref{m12f}%
, \ref{m12h}) that $\mathbf{S}_{\mathbf{p}}\left( t\right) $ is also
periodic with the same period.

Indeed, one writes in this case the equation for $\mathbf{S}_{\mathbf{p}%
}\left( t+T_{0}\right) $ as
\begin{equation}
\partial _{t}\mathbf{S}_{\mathbf{p}}\left( t+T_{0}\right) =2\mathbf{S}_{%
\mathbf{p}}\left( t+T_{0}\right) \mathbf{\times B}_{\mathbf{p}}\left(
t\right) ,  \label{m13a}
\end{equation}%
and one can write the solution in the form%
\begin{equation}
\mathbf{S}_{\mathbf{p}}\left( t+T_{0}\right) =w_{\mathbf{p}}\mathbf{S}_{%
\mathbf{p}}\left( t\right) ,  \label{m13b}
\end{equation}%
where $w_{\mathbf{p}}$ is a time-independent constant. Using Eq. (\ref{m12h}%
, \ref{m12k}, \ref{m12l}) one comes immediately to conclusion that $w_{%
\mathbf{p}}=1,$ and the solution $\mathbf{S}_{\mathbf{p}}\left( t\right) $
is also periodic. The periodicity property can be written as
\begin{equation}
\mathbf{B}\left( t+T_{0}\right) =\mathbf{B}\left( t\right) ,\quad \mathbf{S}%
_{\mathbf{p}}\left( t\right) =\mathbf{S}_{\mathbf{p}}\left( t+T_{0}\right) .
\label{m13c}
\end{equation}

So, one can expect periodic functions $B\left( t\right) $, $\,B_{1}\left(
t\right) $ providing the minimum of the action $\mathcal{S}\left[ B,B_{1}%
\right] ,$ Eq. (\ref{m10}). Equations (\ref{m12f}-\ref{m12l}) allow one to
fix symmetries of the solutions
\begin{eqnarray}
B\left( t\right) &=&-B\left( -t\right) ,\quad B_{1}\left( t\right)
=B_{1}\left( -t\right) ,  \notag \\
S^{3}\left( t\right) &=&-S^{3}\left( -t\right) ,\quad S^{1}\left( t\right)
=S^{1}\left( -t\right) ,\quad S^{2}\left( t\right) =S^{2}\left( -t\right) .
\notag \\
&&  \label{m13d}
\end{eqnarray}

Comparing Eqs. (\ref{m11}, \ref{m12}) with Eqs. (\ref{e14}, \ref{e15})
derived when minimizing the free energy functional $\mathcal{F}\left[ b,b_{1}%
\right] $, Eqs. (\ref{k10}, \ref{e8b}), we conclude that
\begin{equation}
iB\left( t\right) =b\left( it\right) ,\quad B_{1}\left( t\right)
=b_{1}\left( it\right) .  \label{m13}
\end{equation}%
It is very important that if $B\left( t\right) $ and $B_{1}\left( t\right) $
are solutions of Eqs. (\ref{m11}, \ref{m12}), then $B\left( t-t_{0}\right) $
and $B_{1}\left( t-t_{0}\right) $ are also solutions at an arbitrary $t_{0}$%
. It is clear that there can be many solutions even at a fixed $t_{0}$. For
example, for $B_{1}\left( t\right) =0$ one comes in the limit $k\rightarrow
1 $ to Eq. (\ref{e19}) for any period of the function $B\left( t\right) $.
The relation (\ref{m13}) allows one to obtain proper $B\left( t\right) $ and
$B_{1}\left( t\right) $ as soon as $b\left( \tau \right) $ and $b_{1}\left(
\tau \right) $ are obtained from the condition for the minimum of the free
energy functional $\mathcal{F}\left[ b,b_{1}\right] $, Eq. (\ref{k10}).

Now, using Eq. (\ref{m1}, \ref{m8}) we can integrate over the fermion fields
$\eta ,\eta ^{+}$ in Eq. (\ref{m1}) to obtain
\begin{widetext}
\begin{equation}
N\left( t_{1}-t_{2}\right) =-U_{\mathrm{0}}^{2}\overline{\int \mathrm{tr}%
\left[ \Sigma _{3}\mathcal{G}\left( t_{1}-t_{0},\mathbf{p}_{1}\right) %
\right] _{t_{1}t_{1}}\frac{d\mathbf{p}_{1}}{\left( 2\pi \right) ^{2}}\int
\mathrm{tr}\left[ \Sigma _{3}\mathcal{G}\left( t_{2}-t_{0},\mathbf{p}%
_{2}\right) \right] _{t_{2}t_{2}}\frac{d\mathbf{p}_{2}}{\left( 2\pi \right)
^{2}}},  \label{m14}
\end{equation}%
\end{widetext}where the bar stands for the averaging in $t_{0}$ over the
period of the structure. Integration over $t_{0}$ is absolutely necessary
because the extremum of the action functional is degenerate with the respect
to the time shifts, and one should integrate over all the extremum states.
Finally, using Eq. (\ref{m11}) we write the correlation function $N\left(
t\right) $ in the form
\begin{equation}
N\left( t_{1}-t_{2}\right) =\overline{B\left( t_{1}-t_{0}\right) B\left(
t_{2}-t_{0}\right) },  \label{m15}
\end{equation}%
where the function $B\left( t\right) $ is the exact periodic solution of
Eqs. (\ref{m11}-\ref{m12ea}) (or Eq. (\ref{m12f}-\ref{m12l})).

The loop current around the elementary cell is proportional to
\begin{equation}
iU_{\mathrm{0}}\overline{\mathrm{tr}\left[ \Sigma _{3}\mathcal{G}\left(
t-t_{0},\mathbf{p}_{1}\right) \right] }=\overline{B\left( t-t_{0}\right) }.
\label{m16}
\end{equation}%
The periodicity of the exact solution guarantees a non-decaying oscillating
behavior of the function $N\left( t_{1}-t_{2}\right) $, Eq. (\ref{m15}). As
it is difficult to find the exact solution of Eqs. (\ref{m12f}-\ref{m12l}),
we use the approximate solution $B_{0}\left( t\right) $ writing it with the
help of Eq. (\ref{m13}) as
\begin{equation}
B_{0}\left( t\right) =-ib_{0}\left( it\right) ,  \label{m16a}
\end{equation}%
where $b_{0}\left( \tau \right) $ is the Jacobi elliptic function introduced
in Eq. (\ref{k46}).

The Jacobi elliptic function $\mathrm{sn}\left( iu,k\right) $ of an
imaginary argument $iu$ is related to an antisymmetric elliptic function $%
\mathrm{sc}\left( u|k\right) $ with the period $2K\left( k\right) $ as \cite%
{as}
\begin{equation}
\mathrm{sn}\left( iu|k\right) =i\mathrm{sc}\left( u|k^{\prime }\right)
,\;k^{2}+k^{\prime 2}=1,  \label{m17}
\end{equation}%
and one can write the order parameter $B\left( t-t_{0}\right) $ in real time
in the form
\begin{equation}
B_{0}\left( t-t_{0}\right) =-ib_{0}\left( i\left( t-t_{0}\right) \right)
=\gamma k\mathrm{sc}\left( \gamma \left( t-t_{0}\right) |k^{\prime }\right) .
\label{m18}
\end{equation}%
In Eq. (\ref{m16}), $t_{0}$ is arbitrary and integrating over the degeneracy
of the extremum one obtains immediately
\begin{equation}
\overline{B\left( t-t_{0}\right) }=0,  \label{m19}
\end{equation}%
Comparing Eq. (\ref{m19}) with Eq. (\ref{m16}) we conclude that the loop
currents equal zero at any time $t,$ which means that there is no radiation,
and the energy is conserved. As concerns the correlation function $N\left(
t_{1}-t_{2}\right) $, one should insert $B\left( t-t_{0}\right) $ from Eq. (%
\ref{m18}) into Eq. (\ref{m15}) and average over $t_{0}.$

The Fourier transform of the function $\mathrm{sc}\left( u|k^{\prime
}\right) $ is well known \cite{as}
\begin{eqnarray}
&&\mathrm{sc}\left( u|k^{\prime }\right) =\frac{\pi }{2kK\left( k^{\prime
}\right) }\tan \frac{\pi u}{2K\left( k^{\prime }\right) }  \label{m20} \\
&&+\frac{2\pi }{kK\left( k^{\prime }\right) }\sum_{n=1}^{\infty }\left(
-1\right) ^{n}\frac{\exp \left( -\frac{2\pi nK\left( k\right) }{K\left(
k^{\prime }\right) }\right) }{1+\exp \left( -\frac{2\pi nK\left( k\right) }{%
K\left( k^{\prime }\right) }\right) }\sin \left( \frac{n\pi u}{K\left(
k^{\prime }\right) }\right) ,  \notag
\end{eqnarray}%
where $K\left( k\right) $ is the elliptic integral of the first kind.

The oscillating behavior of this function can, again, be understood from
Fig. \ref{fig:jacobi}a. Indeed, the oscillation in the imaginary time $\tau $
could be visualized as classical motion of a particle in the potential $w=%
\frac{1}{2}\left( u^{2}\left( 1+k^{2}\right) -u^{4}\right) .$ Therefore, the
motion in real time should be described by motion in the potential $\tilde{w}%
\left( u\right) =-$ $w\left( u\right) $ and is periodic.

Using the Fourier series for $\tan u$
\begin{equation*}
\tan u=-2\sum_{n=1}^{\infty }\left( -1\right) ^{n}\sin \left( 2nu\right) ,
\end{equation*}%
we write the function $\mathrm{sc}\left( u|k^{\prime }\right) $ in a more
compact form

\begin{equation}
\mathrm{sc}\left( u|k^{\prime }\right) =-\frac{\pi }{kK\left( k^{\prime
}\right) }\sum_{n=1}^{\infty }\left( -1\right) ^{n}\tanh \frac{\pi nK\left(
k\right) }{K\left( k^{\prime }\right) }\sin \frac{n\pi u}{K\left( k^{\prime
}\right) }.  \label{m21}
\end{equation}%
Writing in Eq. (\ref{m21}) $u=\gamma \left( t-t_{0}\right) $ and
substituting it into Eq. (\ref{m18}, \ref{m15}) we write the correlation
function $N\left( t\right) $ as
\begin{equation}
N\left( t\right) =2\gamma ^{2}\sum_{n=1}^{\infty }f_{n}^{2}\cos \left(
n\omega _{0}t\right) ,  \label{m22}
\end{equation}%
where
\begin{equation}
f_{n}=\frac{\pi }{2K\left( k^{\prime }\right) }\tanh \frac{\pi nK\left(
k\right) }{K\left( k^{\prime }\right) },\;\omega _{0}=\frac{\pi \gamma }{%
K\left( k^{\prime }\right) }.  \label{m23}
\end{equation}%
In the limit $k\rightarrow 1$ one has the following asymptotic behavior for
the elliptic integral
\begin{equation}
K\left( k\right) \varpropto \frac{1}{2}\ln \frac{8}{1-k},\quad K\left(
k^{\prime }\right) \varpropto \frac{\pi }{2},  \label{m24}
\end{equation}%
and Eq. (\ref{m22}, \ref{m23}) simplifies to the following form
\begin{equation}
N\left( t\right) \approx 2\gamma ^{2}\sum_{n=1}^{\infty }\Big[1-\left( \frac{%
1-k}{8}\right) ^{2n}\Big]\cos \left( 2\gamma nt\right) .  \label{m25}
\end{equation}

In the limit $k\rightarrow 1$, the correlation function $N\left( t\right) $
shows an oscillating behavior with the frequencies $2\gamma n$ (we put
everywhere $\hbar =1$). The energy $2\gamma $ is the energy of the breaking
of electron-hole pairs and one can interpret the result (\ref{m25}) as
oscillations between the static order and normal state. The oscillating form
of $N\left( t_{1}-t_{2}\right) $ resembles oscillations of the order
parameter in the non-equilibrium superconductors \cite%
{vk,spivak,barankov,altshuler,altshuler1,dzero,moor} but, in contrast to
those, the function $N\left( t\right) $ does not decay in time.
Non-perturbative quantum dynamic effects have been studied in Ref. \cite%
{galitski} using the imaginary time representation. One obtains the
non-decaying behavior because now a thermodynamically stable state is
considered. The contribution of high harmonics $n$ does not decay with $n,$
which originates from the existence of the poles in the first term in Eq. (%
\ref{m20}). Apparently, this is a consequence of using the function $%
B_{0}\left( t\right) $, Eq. (\ref{m16a}), instead of the exact solution $%
B\left( t\right) $ in Eq. (\ref{m15}), and more accurate calculations would
give decaying amplitudes of high harmonics.

At the same time, the non-decaying form the function $N\left( t\right) $ is
guaranteed by the periodicity of the solution $B\left( t\right) $, Eqs. (\ref%
{m13c}), and the corresponding possibility of expanding this solution in
Fourier series like those written in Eq. (\ref{m20}).

\section{\label{sec:Operator}Operator order parameter.}

The correlation function $N\left( t_{1}-t_{2}\right) $ of the functions $%
B\left( t\right) $ was calculated by averaging the product $B\left(
t_{1}-t_{0}\right) B\left( t_{2}-t_{0}\right) $ over the position $t_{0}.$
The same results for the correlation functions can be obtained using an
alternative description based on the notion of an `operator order parameter'
$\hat{B}.$ In order to describe the oscillating behavior of the time crystal
one can formally introduce a Hamiltonian $\hat{H}_{TC}$ \textrm{\ }of a
harmonic oscillator%
\begin{equation}
\hat{H}_{TC}=\left( a^{+}a+\frac{1}{2}\right) \omega _{0},  \label{e40}
\end{equation}%
where the energy $\omega _{0}$ has been introduced in Eqs. (\ref{m22}, \ref%
{m23}), and $a^{+},$ $a$ are boson creation and annihilation operators
satisfying the commutation relations
\begin{equation}
aa^{+}-aa^{+}=1.  \label{e41}
\end{equation}

Instead of averaging over $t_{0}$ we represent now correlation functions a
form of quantum mechanical averages with the Hamiltonian $\hat{H}_{TC}$. For
this purpose we write for any $t$ and $t_{0}$ the following identity
\begin{equation}
\exp \left( in\omega _{0}\left( t+t_{0}\right) \right) \left\vert
n\right\rangle =\frac{\left( e^{i\omega _{0}\left( t+t_{0}\right)
}a^{+}\right) ^{n}}{\sqrt{n!}}\left\vert 0\right\rangle ,  \label{e42}
\end{equation}%
where $\left\vert n\right\rangle $ means $n$-th state of the Hamiltonian $%
\hat{H}_{TC}$, Eq. (\ref{e40}).

Further, we have a standard relation
\begin{equation}
a^{+}\left( t\right) =e^{i\hat{H}_{TC}t}ae^{-i\hat{H}_{TC}t}=ae^{i\omega
_{0}t},  \label{e43}
\end{equation}%
which allows us to write
\begin{eqnarray}
&&e^{in\omega _{0}\left( t+t_{0}\right) }\left\vert n\right\rangle
\label{e44} \\
&=&\frac{1}{\sqrt{n!}}e^{i\hat{H}_{TC}\left( t+t_{0}\right) }\left(
a^{+}\right) ^{n}e^{-i\hat{H}_{TC}\left( t+t_{0}\right) }\left\vert
0\right\rangle .  \notag
\end{eqnarray}%
Introducing operator $A$
\begin{equation}
A=\sum_{n=1}^{\infty }f_{n}\frac{a^{n}}{\sqrt{n!}},\;A\left( t\right) =e^{i%
\hat{H}_{TC}t}Ae^{-i\hat{H}_{TC}t},  \label{e45}
\end{equation}%
where $f_{n}$ is given by Eq. (\ref{m23}), and its Hermitian conjugate $%
A^{+},$ one obtains easily

\begin{eqnarray}
A\left( t\right) \left\vert 0\right\rangle &=&0,\;  \label{e47} \\
A^{+}\left( t\right) \left\vert 0\right\rangle &=&\sum_{n=1}^{\infty }f_{n}%
\frac{\left( a^{+}\left( t\right) \right) ^{n}}{\sqrt{n!}}\left\vert
0\right\rangle =\sum_{n=1}^{\infty }f_{n}e^{in\omega _{0}t}\left\vert
n\right\rangle .  \notag
\end{eqnarray}

Now we consider two different methods of calculation of correlation
functions. Following the first method we calculate the correlation functions
by averaging over $t_{0},$%
\begin{equation}
\sum_{n=1}^{\infty }f_{n}\overline{e^{in\omega _{0}\left( t+t_{0}\right) }}%
=0,  \label{e49}
\end{equation}
\begin{equation}
\sum_{n_{1},n_{2}=1}^{\infty }f_{n_{1}}f_{n_{2}}\overline{e^{in_{1}\omega
_{0}\left( t+t_{0}\right) }e^{in_{2}\omega _{0}t_{0}}}=0,  \label{e50}
\end{equation}%
and%
\begin{equation}
\sum_{n_{1},n_{2}=1}^{\infty }f_{n_{1}}f_{n_{2}}\overline{e^{in_{1}\omega
_{0}\left( t+t_{0}\right) }e^{-i\omega _{0}n_{2}t_{0}}}=\sum_{n=1}^{\infty
}f_{n}^{2}\exp \left( i\omega _{0}nt\right) ,  \label{e51}
\end{equation}%
where the bar means averaging over $t_{0}$.

On the other hand, we can write using Eqs. (\ref{e47}) and the normalization
of the states $\left\vert n\right\rangle $ the following relations
\begin{equation}
\sum_{n=1}^{\infty }f_{n}\overline{e^{in\omega _{0}\left( t+t_{0}\right) }}%
=\left\langle 0\left\vert A^{+}\left( t+t_{0}\right) \right\vert
0\right\rangle =0,  \label{e52}
\end{equation}%
\begin{equation}
\sum_{n_{1},n_{2}=1}^{\infty }f_{n_{1}}f_{n_{2}}\overline{e^{in_{1}\omega
_{0}\left( t+t_{0}\right) }e^{-in_{2}\omega _{0}t_{0}}}=\left\langle
0\left\vert A\left( 0\right) A^{+}\left( t\right) \right\vert 0\right\rangle
,  \label{e52a}
\end{equation}%
and
\begin{equation}
\sum_{n_{1},n_{2}=1}^{\infty }f_{n_{1}}f_{n_{2}}\overline{e^{-in_{1}\omega
_{0}\left( t+t_{0}\right) }e^{in_{2}\omega _{0}t_{0}}}=\left\langle
0\left\vert A\left( t\right) A^{+}\left( 0\right) \right\vert 0\right\rangle
.  \label{e53}
\end{equation}%
In the language of the quantized order parameter $A$, one can replace with
the help of Eqs. (\ref{e52}-\ref{e53}) the averaging over the phase by a
quantum mechanical averaging and write
\begin{equation}
N\left( t\right) =\gamma ^{2}\left( \left\langle 0\left\vert A\left(
t\right) A^{+}\left( 0\right) \right\vert 0\right\rangle +\left\langle
0\left\vert A\left( 0\right) A^{+}\left( t\right) \right\vert 0\right\rangle
\right) .  \label{e54}
\end{equation}%
where $\left\vert 0\right\rangle $ stands for the wave function of the
ground state of the Hamiltonian $\hat{H}_{TC},$ Eq. (\ref{e40}). At the same
time, quantum averages of the operators $A$ and $A^{+}$ vanish
\begin{equation}
\left\langle 0\left\vert A\left( t\right) \right\vert 0\right\rangle
=\left\langle 0\left\vert A^{+}\left( t\right) \right\vert 0\right\rangle =0.
\label{e55}
\end{equation}%
Actually, using this representation one can calculate multi-time correlation
functions%
\begin{eqnarray}
&&N_{2p}\left( t_{1},t_{2}....,t_{2p}\right)  \label{e56} \\
&=&\overline{B\left( t_{1}\right) B\left( t_{2}\right) ...B\left(
t_{p}\right) B\left( t_{p+1}\right) B\left( t_{p+2}\right) ...B\left(
t_{2p}\right) },  \notag
\end{eqnarray}%
with $B\left( t\right) $ determined by Eq. (\ref{m18}). Writing $\sin (n\pi
u/K(k^{\prime }))$ in Eq. (\ref{m21}) as the sum of two exponentials one can
multiply all $B\left( t_{l}\right) $ in Eq. (\ref{e56}) and average over $%
t_{0}$ term by term. It can be checked that each average can be written as a
quantum-mechanical average of products of operators $A\left( t\right) $ and $%
A^{+}\left( t\right) $.

As a result, the correlation function $N_{2p}\left(
t_{1},t_{2}....,t_{2p}\right) $ takes the form
\begin{eqnarray}
&&N_{2p}\left( t_{1},t_{2}....,t_{2p}\right)  \label{e57} \\
&=&\frac{1}{\left( p!\right) ^{2}}\sum_{P}\left\langle 0\left\vert
\prod_{0\leq l\leq p}A\left( t_{l}\right) A^{+}\left( t_{p+l}\right)
\right\vert 0\right\rangle .  \notag
\end{eqnarray}%
where, the symbol $\sum_{P}$ means the sum of all permutations of the
operators $A\left( t_{l}\right) $ and $A^{+}\left( t_{l}\right) $ in the
product. Correlation functions of odd number of times are equal to zero. At $%
p=1$ one obtains the correlation function $N\left( t_{1}-t_{2}\right) ,$ Eq.
(\ref{e54}). In the limit $k\rightarrow 1,$ Eq. (\ref{e54}) simplifies to (%
\ref{m25}).

Equations (\ref{m15}, \ref{e54}) and (\ref{e56}, \ref{e57}) demonstrate
equivalence between the averaging of classical order parameters over the
positions in time and the quantum-mechanical averaging of operator order
parameters. This resembles the equivalence between the coherent states and
the number states descriptions in quantum optics \cite{glauber,loudon}.

One can interpret the operator $A$ as an operator order parameter. This type
of the order parameters extends the variety of conventional order parameters
like scalars, vectors, matrices used in theoretical physics. As the quantum
mechanical average of the operators $A$ and $A^{+}$ vanishes, one cannot
expect any loss of energy due to e.g. emission of light. At the same time,
already two-time correlation functions do show oscillating behavior and this
does not mean any loss of energy. The system described by the Hamiltonian $%
\hat{H}_{TC},$ Eq. (\ref{e40}), remains in the ground state, and the
oscillations are due to virtual transitions between the states. Remarkably,
the distance between the energy levels does not decay in the limit of
infinite volume, $V\rightarrow \infty ,$ which demonstrates a coherence all
over the sample.

The two-times correlation functions of type (\ref{m25}, \ref{e54}) describe
inelastic quantum mechanical scattering and corresponding experiments can be
used for observing the time crystals. The non-decaying time oscillations can
be an important property for designing qubits but, for making devices, one
should identify physically relevant systems described by Eqs. (\ref{e0}-\ref%
{k1b}).

\section{\label{sec:Mean_Field}Spontaneous breaking of the time-translation
symmetry: time-dependent wave functions of equilibrium states and their
meaning.}

The oscillating behavior of the correlation function $N\left( t\right) $,
Eqs. (\ref{m22}, \ref{m25}), has been demonstrated in the preceding sections
using methods of functional integration. Within this method one starts
writing physical quantities in a form of a functional integral over
anticommuting fermion fields. These integrals are transformed to functional
integrals over boson fields, and one can calculate these integrals using the
saddle-point method. The oscillating in both imaginary and real time order
parameters appears as a result of minimizing an effective free energy
(action) functional of boson fields. As a result, correlation functions
oscillating in real time arise in the thermodynamically stable state. This
is the oscillating in time order parameter $B\left( t\right) $ that is
responsible for this striking effect. Using a time-independent order
parameter in Eqs. (\ref{m14}, \ref{m15}) one would obtain a standard
time-independent long-range order in space only.

However, previous discussions in the literature on the time-crystals were
based on the more traditional Hamiltonian formalism (see, e.g. Ref. \cite%
{watanabe}). Assuming that the starting Hamiltonian does not depend on time
it is less straightforward to understand an oscillating behavior of
correlation functions at large times. Therefore, we sketch in this section
the derivation of the main results within the Hamiltonian approach and
clarify why the results do not agree with the conclusions of Ref. \cite%
{watanabe}.\qquad\

\subsection{Hamiltonian approach.}

We start this subsection with writing a Hamiltonian $\hat{H}$ corresponding
to the action $S\left[ \chi ,\chi ^{+}\right] ,$ Eq. (\ref{e0}). It
describes the system of interacting electrons in two bands $1$ and $2$
\begin{eqnarray}
\hat{H} &=&\sum_{p}c_{p}^{+}\left( \varepsilon ^{+}\left( \mathbf{p}\right)
+\varepsilon ^{-}\left( \mathbf{p}\right) \Sigma _{3}\right) c_{p}
\label{g1} \\
&&+\frac{1}{4V}\Big[\tilde{U}_{\mathrm{0}}\Big(\sum_{p}c_{p}^{+}\Sigma
_{1}c_{p}\Big)^{2}-U_{\mathrm{0}}\Big(\sum_{p}c_{p}^{+}\Sigma _{2}c_{p}\Big)%
^{2}\Big].  \notag
\end{eqnarray}%
Two-component vectors $c_{p}=\left\{ c_{p}^{1},c_{p}^{2}\right\} ,$ contain
creation and destruction operators $c_{p}^{1}$ and $c_{p}^{2}$ for the
fermions of the bands $1$ and $2$, $p=\left\{ \alpha ,\mathbf{p}\right\} ,$
where $\alpha $ stands for spin (the spin variable $\alpha $ is not very
important here). Hamiltonian (\ref{g1}) corresponds to the action $S\left[
\chi \right] ,$ Eqs. (\ref{e0}-\ref{k1b}), and the rest of the notations is
the same. It resembles the BCS \cite{bcs} Hamiltonian in theory of
superconductivity specially designed for describing superconductivity. The
electron-electron interaction in Eq. (\ref{g1}) is short-ranged but the
interaction terms have a somewhat simplified separable form containing
summation over two momenta only. Equation (\ref{g1}) describes a Hamiltonian
of the grand canonical ensemble written at fixed chemical potential $\mu $
(see also Eq. (\ref{k25c})). Considering grand canonical ensemble is
standard in study of macroscopic systems of interacting electrons. Solving
the model with the Hamiltonian $\hat{H}$, Eq. (\ref{g1}), is equivalent to
solving models with a more general electron-electron interaction in the mean
field approximation. In the limit of a large volume $V\rightarrow \infty $,
one can replace the sum over the momenta by integrals using the standard
replacement
\begin{equation}
\sum_{p}\left( ...\right) \rightarrow V\int \left( ...\right) \frac{d\mathbf{%
p}}{\left( 2\pi \right) ^{d}},  \label{g1a}
\end{equation}%
($d$ is dimension) and see that $\hat{H}$ is proportional to the volume $V,$
as it should be. The form of the interaction in Eq. (\ref{g1}) corresponds
the `infinite range' interaction of effective electron-hole pairs (in the
BCS Hamiltonian one writes interaction of electron-electron pairs), and that
is why the mean field theory coincides with the exact solution for the
Hamiltonian $\hat{H}$, Eq. (\ref{g1}). It is well known that, generally, the
mean field approximation is not necessarily good in one and two dimensions
but it works very well in three or quasi-two dimensional models at least
qualitatively. Therefore, the Hamiltonian $\hat{H}$ is a good starting point
for studying ordered phases of the electron systems in 3D or quasi-2D unless
they are close to a phase transition.

Of course, time is not present in Eq. (\ref{g1}). It is very important for
the results obtained in the present work that the coupling constant $\tilde{U%
}_{\mathrm{0}}$ is positive and sufficiently large. In a recent publication
\cite{mukhin2019}, similar models were considered with $\tilde{U}_{\mathrm{0}%
}=0$ and $\tilde{U}_{\mathrm{0}}=-U_{\mathrm{0}}$, and it was demonstrated
that the instanton lattice had a higher energy than the homogeneous state.
The same result would be obtained if we used negative or not sufficiently
large positive $\tilde{U}_{\mathrm{0}}.$

Having in mind that the mean field approximation is exact for the model
described by the Hamiltonian $\hat{H}$ we use this method for explicit
calculations. One starts replacing $\hat{H}$, Eq. (\ref{g1}), by a mean
field Hamiltonian containing an effective potential instead of the
interparticle interaction. It can depend on the imaginary time $\tau $ when
thermodynamic properties are studied or one can introduce a real-time
dependence for studying real-time correlation functions. Introducing a mean
field Hamiltonian is equivalent to decoupling the interaction by integration
over auxiliary fields as it has been done in the previous sections. The
effective free energy functional obtained in this way depends on the
auxiliary field and, minimizing this functional, one obtains equations that
are equivalent to the mean field equations. Of course, there can be many
solutions of the mean field equations for the effective potential, and one
should use the one corresponding to the minimum of the free energy. The
difference between the free energies of different states is proportional to
the volume $V,$ and in the limit $V\rightarrow \infty $ one should find the
one with the lowest energy.

\subsection{Time-dependent wave functions.}

Usually, one uses in quantum mechanics wave functions $\tilde{\Psi}\left(
t,X\right) $ ($X$ are coordinates of particles) that are solutions of the
Schr\"{o}dinger equation
\begin{equation}
i\frac{\partial \tilde{\Psi}\left( t,X\right) }{\partial t}=\hat{H}\tilde{%
\Psi}\left( t,X\right) .  \label{f1}
\end{equation}%
The time-dependent solutions $\tilde{\Psi}\left( t,X\right) $ can be written
in a form of a superposition
\begin{equation}
\tilde{\Psi}\left( t,X\right) =\sum_{n}c_{n}\tilde{\Psi}_{n}\left(
t,X\right) ,  \label{f2}
\end{equation}%
where the time-dependent eigenfunctions $\tilde{\Psi}_{n}\left( t,X\right) $
equal
\begin{equation}
\tilde{\Psi}_{n}\left( t,X\right) =\Psi _{n}\left( X\right) \exp \left(
-iE_{n}t\right) ,  \label{f2a}
\end{equation}%
and the time-independent wave functions $\bar{\Psi}_{n}\left( X\right) $
satisfy the stationary Schr\"{o}dinger equation%
\begin{equation}
\hat{H}\Psi _{n}\left( X\right) =E_{n}\Psi _{n}\left( X\right) .  \label{f3}
\end{equation}%
The time-dependent wave functions $\tilde{\Psi}_{n}\left( t,X\right) $ obey
orthogonality relations
\begin{eqnarray}
&&\int \tilde{\Psi}_{n}^{\ast }\left( t,X\right) \tilde{\Psi}_{m}\left(
t,X\right) dX  \label{f3a} \\
&=&\int \exp \left( i\left( E_{n}-E_{m}\right) t\right) \Psi _{n}^{\ast
}\left( X\right) \Psi _{m}\left( X\right) dX=\delta _{mn}.  \notag
\end{eqnarray}

Solutions of the Schr\"{o}dinger equation (\ref{f1}) are generally
time-dependent but can be represented in a form of Eq. (\ref{f2}, \ref{f2a})
with the help of the time-independent eigenfunctions $\Psi _{n}\left(
X\right) $. The scalar product of two time-dependent wave functions in Eq. (%
\ref{f3a}) contains integration over the coordinates $X$ but not over time $%
t $.

As a spontaneous breaking of the symmetry is expected, the symmetries of the
Hamiltonian may differ below the transition point from those of the original
Hamiltonian $\hat{H}$. There are plenty of examples of such a behavior. One
of the examples, most close to the present case, is formation of a charge
density wave (CDW) in a model invariant with respect to space translations.
Below the transition, a periodic in space order parameter appears, which
leads to an additional periodic dependence of electron wave functions on
coordinates (Bloch theorem). Thermodynamic quantities depend directly on the
amplitude of CDW but not on the modulation vector $\mathbf{Q,}$ although the
latter enters explicitly the mean field Hamiltonian. At the same time, the
space modulation can be observed in two-point correlation functions. One
cannot calculate physical quantities like, e.g., correlation functions at
different space points analytically without calculating first the order
parameter. Mean field theory serves usually as an efficient tool for
studying how the original symmetries of the model are broken. Fluctuations
can in some cases be important in low dimensions but in 3D or in quasi-2D
this procedure is in most cases reliable. One can say that the CDW appears
as a result of the breaking of the space-translation symmetry of the
original space-translation invariant Hamiltonian.

The scenario concerning formation of the thermodynamic quantum time crystal
considered in this paper is similar. The main difference with respect to the
formation of the CDW is the breaking of the time-translation symmetry of the
time-translation invariant (static) Hamiltonian. Being at the moment more
interested in the correlation functions of real time, we will replace the
Hamiltonian $\hat{H}$, Eq. (\ref{g1}), by a mean-field Hamiltonian $\hat{H}^{%
\mathrm{mf}}\left( t\right) $ quadratic in the operators $c_{p},$ $%
c_{p}^{+}, $ which simplifies the calculations. The Hamiltonian $\hat{H}^{%
\mathrm{mf}}\left( t\right) $ contains periodic in time $t$ order parameters
$B\left( t\right) $ and $B_{1}\left( t\right) $ that have to be found from
mean-field equations. It is clear that the wave functions of $H^{\mathrm{mf}%
}\left( t\right) $ will depend on time in a non-trivial way different from
Eqs. (\ref{f2a}, \ref{f3}). At first glance, one might think that such a
mean field is simply unreasonable because it violates the exact form of the
wave functions written in Eqs. (\ref{f2a}, \ref{f3}).

In order to clarify this discrepancy we adopt now the scheme of the BCS
theory \cite{bcs,bdg} to the present situation. Although we have to find
solutions of the Schr\"{o}dinger equation with the Hamiltonian $\hat{H},$
Eq. (\ref{g1}), written for the grand canonical ensemble, it is instructive
to start with a system of a fixed number of electrons.

A trial antisymmerized time-dependent wave function $\tilde{\Psi}_{N}\left(
t\right) $ of $N$ electrons and $N$ holes (spin indices are omitted for
simplicity) can be written as

\begin{equation}
\tilde{\Psi}_{N}\left( t\right) =Ae^{-iE_{N}t}\prod_{j=1}^{N}\psi \left(
\mathbf{r}_{ej}-\mathbf{r}_{hj}\right)  \label{f100}
\end{equation}%
where $\mathbf{r}_{ej}$ and $\mathbf{r}_{hj},$ $j=1,2,,,,N$ numerate
coordinates of electrons and holes. Functions $\psi \left( \mathbf{r}_{ej}-%
\mathbf{r}_{hj}\right) $ satisfy a stationary Schr\"{o}dinger equation for a
single electron-hole pair, the operator $A$ antisymmetrizes the product of
these functions, and the energy $E_{N}$ is the total energy of $N$ isolated
pairs. The Hamiltonian of a single electron-hole pair is taken in the form
\begin{equation}
\hat{h}_{\mathrm{eh}}\left( \mathbf{r}_{1},\mathbf{r}_{2}\right)
=\varepsilon _{e}\left( -i\nabla _{\mathbf{r}_{1}}\right) +\varepsilon
_{h}\left( -i\nabla _{\mathbf{r}_{2}}\right) +V\left( \mathbf{r}_{1}-\mathbf{%
r}_{2}\right) ,  \label{f102}
\end{equation}%
where $\varepsilon _{e}\left( -i\nabla _{\mathbf{r}}\right) $ and $%
\varepsilon _{h}\left( -i\nabla _{\mathbf{r}}\right) $ are operators of
kinetic energy of the electron and hole, and $V\left( \mathbf{r}_{1}-\mathbf{%
r}_{2}\right) $ is an interaction. Putting the pair in the center of mass we
find the eigenvalue $\varepsilon _{0}$ and the eigenfunction $\psi \left(
\mathbf{r}_{1}-\mathbf{r}_{2}\right) $ from the equation
\begin{equation}
\hat{h}_{\mathrm{eh}}\left( \mathbf{r}_{1},\mathbf{r}_{2}\right) \psi \left(
\mathbf{r}_{1}-\mathbf{r}_{2}\right) =\varepsilon _{0}\psi \left( \mathbf{r}%
_{1}-\mathbf{r}_{2}\right) .  \label{f103}
\end{equation}%
It is clear that the energy $E_{N}$ of non-interacting electron-hole pairs
equals
\begin{equation}
E_{N}=N\varepsilon _{0}.  \label{f106}
\end{equation}%
In principle, taking into account an effective interaction with other
electron-hole pairs can change the value $\varepsilon _{0}.$ In this case,
one should use an energy $\varepsilon _{0N}$ depending on $N$ instead of $%
\varepsilon _{0}$ and write
\begin{equation}
E_{N}=N\varepsilon _{0N}  \label{f106a}
\end{equation}%
Introducing Fourier-transform of the function $\psi \left( \mathbf{r}\right)
$%
\begin{equation}
\psi \left( \mathbf{r}\right) =\int g_{\mathbf{p}}e^{i\mathbf{p}r}\frac{d%
\mathbf{p}}{\left( 2\pi \right) ^{d}},  \label{f103a}
\end{equation}%
we write the function $\tilde{\Psi}_{N}\left( t\right) $ in the form%
\begin{eqnarray}
\tilde{\Psi}_{N}\left( t\right) &=&A\prod_{j=1}^{N}\left( \sum_{\mathbf{p}%
_{j}}g_{\mathbf{p}_{j}}e^{i\mathbf{p}_{j}\left( \mathbf{r}_{ej}-\mathbf{r}%
_{hj}\right) }e^{-i\varepsilon _{0N}t}\right)  \notag \\
&=&\prod_{j=1}^{N}\left( \sum_{\mathbf{p}_{j}}g_{\mathbf{p}_{j}}c_{1\mathbf{p%
}_{j}}^{+}c_{2\mathbf{p}_{j}}e^{-i\varepsilon _{0N}t}\right) \left\vert
0\right\rangle .  \label{f107}
\end{eqnarray}%
In Eq. (\ref{f107}) the symbol $\left\vert 0\right\rangle $ stands for the
ground state, and $c_{1\mathbf{p}_{j}},$ $c_{2\mathbf{p}_{j}}$ and $c_{1%
\mathbf{p}_{j}}^{+},$ $c_{2\mathbf{p}_{j}}^{+}$ are destruction and creation
operators for the state with momentum $\mathbf{p}$ of the $j$-th pair. All
the transformations, Eqs. (\ref{f100}-\ref{f107}) have been exact so far,
and the time dependence of the function $\tilde{\Psi}_{N}\left( t\right) $
has the form given by Eqs. (\ref{f2a}-\ref{f3}) usual for static
Hamiltonians.

The trial function $\tilde{\Psi}_{N}\left( t\right) $ is still too
complicated and is not helpful for solving the Schr\"{o}dinger equation with
the Hamiltonian $\hat{H},$ Eq. (\ref{g1}). Moreover, it is written for the
fixed number of electrons-hole pairs $N,$ while the Hamiltonian $\hat{H}$
has been introduced for the grand canonical ensemble. Therefore, let us
write a trial function $\tilde{\Phi}$ for the grand canonical ensemble in
the form of a superposition
\begin{equation}
\tilde{\Phi}\left( t\right) =\sum_{N=1}^{\infty }\lambda _{N}\tilde{\Psi}%
_{N}\left( t\right) .  \label{f108}
\end{equation}%
The coefficients $\lambda _{N}$ have to be chosen in a form that would make
it possible to find exact time-dependent solutions for the Hamiltonian $\hat{%
H},$ Eq. (\ref{g1}). This task is achieved writing the function $\tilde{\Phi}%
\left( t\right) $ as

\begin{equation}
\tilde{\Phi}\left( t\right) =\prod_{p}\left( u_{\mathbf{p}}\left( t\right)
+v_{\mathbf{p}}\left( t\right) c_{p}^{1+}c_{p}^{2}\right) \left\vert
0\right\rangle  \label{g5}
\end{equation}

Equation (\ref{g5}) is written for $\varepsilon _{1}\left( \mathbf{p}\right)
>\varepsilon _{2}\left( \mathbf{p}\right) $ (at $\varepsilon _{1}\left(
\mathbf{p}\right) <\varepsilon \left( \mathbf{p}_{2}\right) $ one should
exchange the bands, $1\rightleftarrows 2$ ), and $\left\vert 0\right\rangle $
is the state of the Hamiltonian of non-interacting particles (Fermi step
function). Equation (\ref{g5}) is very similar to the trial function in the
BCS theory \cite{bcs} but the functions $u_{\mathbf{p}}\left( t\right) ,$ $%
v_{\mathbf{p}}\left( t\right) $ depend on time, and the products $%
c_{p}^{1+}c_{p}^{2}$ create electron-hole pairs instead of the Cooper pairs.
The superconducting analog of the function $\tilde{\Phi}\left( t\right) $,
Eq. (\ref{g5}), has been used in Ref. \cite{spivak} for studying
non-equilibrium states in superconductors. The dependence of $\lambda _{N}$
on $N$ is sharp and $\left\langle N^{2}\right\rangle -\left\langle
N\right\rangle ^{2}\sim \left\langle N\right\rangle $, where $\left\langle
N\right\rangle $ is the average number of electrons at a fixed chemical
potential. At the same time, $\left\langle N\right\rangle \gg 1$, and the
variation of $\lambda _{N}$ is negligible when several electron-hole pairs
enter or leave the sample.

We emphasize that the function $\tilde{\Phi}\left( t\right) ,$ Eq. (\ref{g5}%
), is the most general form of the exact solution of the non-stationary Schr%
\"{o}dinger equation with the Hamiltonian $\hat{H}$, Eq. (\ref{g1}), and
this form is different from Eqs. (\ref{f2a}, \ref{f3}). Of course, Eq. (\ref%
{g5}) could be introduced as a guess without the preliminary discussion
resulting in Eqs. (\ref{f100}-\ref{f108}). In this case, one would have to
just check that the function $\tilde{\Phi}\left( t\right) $ is the solution
to the Hamiltonian $\hat{H}.$ However, this qualitative discussion may be
helpful for understanding the origin of the form of the wave functions $%
\tilde{\Phi}\left( t\right) ,$ Eq. (\ref{g5}). Taking time-independent
coefficients $u_{\mathbf{p}}$ and $v_{\mathbf{p}}$ in the trial function $%
\tilde{\Phi}$, Eq. (\ref{g5}), corresponds to the standard mean-field
approximation used in Ref. \cite{volkov3}.

Following the proposed mean field procedure we replace the Hamiltonian $\hat{%
H},$ Eq. (\ref{g1}), by the mean field Hamiltonian $\hat{H}^{\mathrm{mf}%
}\left( t\right) ,$

\begin{eqnarray}
\hat{H} &\rightarrow &\hat{H}^{\mathrm{mf}}\left( t\right)
=\sum_{p}c_{p}^{+}M_{\mathbf{p}}^{\mathrm{mf}}\left( t\right) c_{p}
\label{g2} \\
&&-V\left( \frac{B^{2}\left( t\right) }{U_{\mathrm{0}}}-\frac{%
B_{1}^{2}\left( t\right) }{\tilde{U}_{\mathrm{0}}}\right) ,  \notag
\end{eqnarray}%
where
\begin{equation}
\hat{M}_{\mathbf{p}}^{\mathrm{mf}}\left( t\right) =\varepsilon ^{+}\left(
\mathbf{p}\right) +\varepsilon ^{-}\left( \mathbf{p}\right) \Sigma
_{3}-i\left( B\left( t\right) \Sigma _{2}+B_{1}\left( t\right) \Sigma
_{1}\right) ,  \label{g3}
\end{equation}%
and functions $B\left( t\right) $ and $B_{1}\left( t\right) $ play the role
of the order parameter. They are periodic in time with a period $T_{0}$
(these are the same functions as those introduced in Eqs. (\ref{m7}-\ref{m12}%
)). Their explicit form of the time dependence can be obtained from
self-consistency equations%
\begin{eqnarray}
\frac{U_{\mathrm{0}}}{2V}\sum_{p}\left\langle c_{p}^{+}\Sigma
_{2}c_{p}\right\rangle _{\mathrm{mf}} &=&B\left( t\right) ,  \label{f4} \\
\frac{\tilde{U}_{\mathrm{0}}}{2V}\sum_{p}\left\langle c_{p}^{+}\Sigma
_{1}c_{p}\right\rangle _{\mathrm{mf}} &=&-B_{1}\left( t\right) .  \label{f5}
\end{eqnarray}%
In Eqs. (\ref{f4}, \ref{f5}), the angle brackets $\left\langle
...\right\rangle _{\mathrm{mf}}$ stand for the quantum mechanical averaging
with the Hamiltonian $\hat{H}^{\mathrm{mf}}\left( t\right) $, Eq. (\ref{g2}%
). Equations (\ref{f4}, \ref{f5}) are identical to Eqs. (\ref{m11}, \ref{m12}%
).

In order to find wave functions $\Phi $, one should solve the effective Schr%
\"{o}dinger equation%
\begin{equation}
i\frac{\partial \tilde{\Phi}\left( t\right) }{\partial t}=H^{\mathrm{mf}%
}\left( t\right) \tilde{\Phi}\left( t\right) .  \label{f6}
\end{equation}%
The solutions $\tilde{\Phi}\left( t\right) $ of Eq. (\ref{f6}) are
time-dependent but the expansion, Eqs. (\ref{f2}-\ref{f3}), is no longer
valid when the functions $B\left( t\right) $, $B_{1}\left( t\right) $ are
time-dependent. As we have seen, this does not lead to any contradiction
because Eqs. (\ref{f2}-\ref{f3}) have been written for a system with a fixed
number of the electrons, while the solutions $\tilde{\Phi}\left( t\right) $
has been obtained for the grand canonical ensemble.

Using Eqs. (\ref{f6}, \ref{g3}) one comes to Bogolyubov-de Gennes-like
equations \cite{bdg} for the functions $u_{\mathbf{p}}\left( t\right) $ and $%
v_{\mathbf{p}}\left( t\right) $
\begin{equation}
\left( -i\frac{\partial }{\partial t}+\hat{M}_{\mathbf{p}}^{\mathrm{mf}%
}\left( t\right) \right) \tilde{\phi}_{\mathbf{p}}\left( t\right) =0,\quad
\label{g7}
\end{equation}%
where
\begin{equation}
\tilde{\phi}_{\mathbf{p}}\left( t\right) =\left(
\begin{array}{c}
u_{\mathbf{p}}\left( t\right) \\
v_{\mathbf{p}}\left( t\right)%
\end{array}%
\right) .  \label{g8}
\end{equation}

It is very important that, if the periodic order parameters $B\left(
t\right) $ and $B_{1}\left( t\right) $ are solutions of self-consistency
equations, the functions $B\left( t-t_{0}\right) $ and $B_{1}\left(
t-t_{0}\right) $ are also solutions. This means that the state is
degenerate, and one should integrate over $t_{0}$ at the end of calculations
to take into account the degeneracy. This procedure contrast calculations
performed when studying non-equilibrium phenomena where a certain fixed time
$t_{0}$ is always present marking the beginning of a process.

In the functional integral formulation of the preceding sections, the
necessity of integration over $t_{0}$ simply followed from the degeneracy of
the minimum of the effective action against the shift of time by $t_{0}$. In
mean field theories, one should first guess a mean field solution and then
check the self-consistency. If one finds a solution for any shift of $t_{0}$
(which is the case in the present situation), one should average physical
quantities over $t_{0}$ at the end of calculations. From the mathematical
point of view, the necessity of the integration over $t_{0}$ is more
transparent in the functional integral formulation developed previously.

Of course, there is also a region of parameters of the model where a
time-independent order parameter
\begin{equation}
B=-i\gamma ,\quad B_{1}=0,  \label{g4}
\end{equation}%
($\gamma $ is the gap in the spectrum) corresponding to the DDW state \cite%
{volkov3} is more favorable but the time-dependent solutions are of the
major interest now and we concentrate on those.

A non-trivial form of the time dependence of the wave functions $\tilde{\Phi}%
_{n}\left( t\right) $ (different from $\exp \left( -iE_{n}t\right) ,$ where $%
E_{n}$ is an eigenenergy) originates from the time-dependence of the order
parameters $B\left( t\right) $ and $B_{1}\left( t\right) $ appearing below a
critical temperature, and this is an unusual feature. However, using the
fact that the functions $B\left( t\right) $ and $B_{1}\left( t\right) $ are
periodic one can construct an expansion alternative to Eq. (\ref{f2}) with
the help of the Floquet theorem \cite{floquet} by writing a solution $\phi _{%
\mathbf{p}n}\left( t\right) $ of Eq. (\ref{g7}) as%
\begin{equation}
\tilde{\phi}_{\mathbf{p}n}\left( t\right) =\exp \left( -i\mathcal{E}_{%
\mathbf{p}n}t\right) \phi _{\mathbf{p}n}\left( t\right) ,  \label{g9}
\end{equation}%
where $\phi _{\mathbf{p}n}\left( t\right) $ is a periodic function with the
period $T_{0},$%
\begin{equation}
\phi _{\mathbf{p}n}\left( t\right) =\phi _{\mathbf{p}n}\left( t+T_{0}\right)
,  \label{g10}
\end{equation}%
and $\mathcal{E}_{\mathbf{p}n}$ is quasienergy. The functions $\phi _{%
\mathbf{p}n}\left( t\right) $ and the quasienergies $\mathcal{E}_{\mathbf{p}%
n}$ obey the equation
\begin{equation}
\left( -i\frac{\partial }{\partial t}+\hat{M}_{\mathbf{p}}^{\mathrm{mf}%
}\left( t\right) \right) \phi _{\mathbf{p}n}\left( t\right) =\mathcal{E}_{%
\mathbf{p}n}\phi _{\mathbf{p}n}\left( t\right) .  \label{g11}
\end{equation}%
Comparing Eq. (\ref{g11}) with Eq. (\ref{m12a}) we see that these are the
same equations for the same wave functions.

Introducing the scalar product one proves in the standard way that the
functions $\phi _{\mathbf{p}n}\left( t\right) $ form the orthonormal basis,
\begin{equation}
\left\{ \phi _{\mathbf{p}n}^{+}\left( t\right) ,\phi \left( t\right) _{%
\mathbf{p}n^{\prime }}\right\} \equiv T_{0}^{-1}\int_{0}^{T_{0}}\left( \phi
_{\mathbf{p}n}^{+}\left( t\right) ,\phi _{\mathbf{p}n^{\prime }}\left(
t\right) \right) dt=\delta _{nn^{\prime }}.  \label{g12}
\end{equation}%
The figure brackets are used here to emphasize that, in the Floquet theory,
time integration is included in the scalar product. This contrasts the
scalar product, Eq. (\ref{f3a}), used in the conventional quantum mechanics
with static Hamiltonians. For time-independent order parameters $B\left(
t\right) $ and $B_{1}\left( t\right) $, the functions $\phi _{\mathbf{p}%
n}\left( t\right) $ are constants. The periodic dependence of the functions $%
\phi _{\mathbf{p}n}\left( t\right) $ on time contrasts the traditional
time-independence of the functions $\Psi _{n}$, Eq. (\ref{f3}). As the mean
field scheme is exact in the model with the Hamiltonian $\hat{H},$ Eq. (\ref%
{g1}), the time-dependent functions $\tilde{\Phi}_{\mathbf{p}n}\left(
t\right) $, Eq. (\ref{g5}-\ref{g11}), are exact eigenfunctions of the Schr%
\"{o}dinger equation with the Hamiltonian $\hat{H},$ Eq. (\ref{g1}).

At first glance, it looks as if the time dependence of $\tilde{\phi}_{%
\mathbf{p}n}\left( t\right) $ violated basic principle of thermodynamic
because the conventional expression for the partition function $Z$,%
\begin{equation}
Z=\sum_{\mathrm{states}}\exp \left( -E_{n}/T\right) ,  \label{f7}
\end{equation}%
where the sum is as usual performed over all states, does not contain time.

However, it is still possible to formulate a proper definition of the
partition function $Z$ using the Floquet states (\ref{g9}, \ref{g10}). It is
worth emphasizing that we do not attempt here constructing new quantum
mechanics. As the main results have already been obtained in the previous
sections using the functional integrals, the scheme suggested in this
Section serves merely as an illustration.

Although time $t$ is explicitly present in Eqs. (\ref{g2}-\ref{g11}), the
quantum mechanical average of an operator $\hat{A}\left( t\right) $
containing time due to dependence on the periodic functions $B\left(
t\right) $ and $B_{1}\left( t\right) $ is time independent. Indeed, using
Eqs. (\ref{g9}, \ref{g10}), recalling that $B\left( t-t_{0}\right) ,$ $%
B_{1}\left( t-t_{0}\right) $ and $\phi _{\mathbf{p}n}\left( t-t_{0}\right) $
are also solutions of Eqs. (\ref{f4}, \ref{f5}, \ref{g7}), and therefore
integrating over $t_{0},$ the average of the operator $\hat{A}$ can be
written as%
\begin{eqnarray}
A_{\mathbf{p},nn} &=&T_{0}^{-1}\int_{0}^{T_{0}}\tilde{\phi}_{\mathbf{p}%
n}^{+}\left( t-t_{0}\right) \hat{A}\left( t-t_{0}\right) \tilde{\phi}_{%
\mathbf{p}n}\left( t-t_{0}\right) dt_{0}  \notag \\
&=&T_{0}^{-1}\int_{0}^{T_{0}}\phi _{\mathbf{p}n}^{+}\left( t-t_{0}\right)
\hat{A}\left( t-t_{0}\right) \phi _{\mathbf{p}n}\left( t-t_{0}\right) dt_{0}
\notag \\
&=&T_{0}^{-1}\int_{0}^{T_{0}}\phi _{\mathbf{p}n}^{+}\left( t\right) \hat{A}%
\left( t\right) \phi _{\mathbf{p}n}\left( t\right) dt  \label{g13}
\end{eqnarray}%
Equation (\ref{g13}) shows that $\hat{A}_{\mathbf{p},nn}$ is
time-independent.

Actually, the Hamiltonian formalism presented in this section is a natural
generalization of the conventional one. Although, the wave functions and
order parameters depend on time, one can construct thermodynamics in an
almost standard way. Usually, calculating the partition function $Z$ one
starts with the Gibbs formula%
\begin{eqnarray}
Z &=&\mathrm{Tr\exp }\left( -\hat{H}/T\right)  \label{g14} \\
&=&\sum_{n}\left( \tilde{\Psi}_{n}^{+}\left( t\right) ,\tilde{\Psi}%
_{n}\left( t\right) \right) \exp \left( -E_{n}/T\right)  \notag \\
&=&\sum_{n}\left( \Psi _{n}^{+},\Psi _{n}\right) \exp \left( -E_{n}/T\right)
,  \notag
\end{eqnarray}%
with the functions $\tilde{\Psi}_{n}\left( t\right) $ and $\Psi _{n}$
introduced in Eqs. (\ref{f2a}, \ref{f3}). Eq. (\ref{g14}) contains
explicitly the sum over all states. This is the standard formalism that can
be used for describing the normal metal phase or the DDW phase with the
order parameter given by Eq. (\ref{g4}).

Now we solve Eq. (\ref{f6}) using the Hamiltonian $\hat{H}^{\mathrm{mf}%
}\left( t-t_{0}\right) $, Eq. (\ref{g2}). As the Hamiltonian $\hat{H}^{%
\mathrm{mf}}\left( t-t_{0}\right) $ is periodic in time, we use the Floquet
theorem and write the function $\tilde{\Phi}\left( t\right) $ satisfying Eq.
(\ref{f6}) in the form
\begin{equation}
\tilde{\Phi}_{n}\left( t\right) =\exp \left( -i\mathcal{E}_{n}t\right) \Phi
_{n}\left( t\right) ,  \label{g19c}
\end{equation}%
where $\Phi _{n}\left( t\right) $ are periodic functions with the period $%
T_{0}$ satisfying the equation
\begin{equation}
\left( \hat{H}^{\mathrm{mf}}\left( t\right) -i\frac{\partial }{\partial t}%
\right) \Phi _{n}\left( t\right) =\mathcal{E}_{n}\Phi _{n}\left( t\right) ,
\label{g19d}
\end{equation}%
and $\mathcal{E}_{n}$ is quasienergy.

It looks natural to replace $\Psi _{n}\left( t\right) $ and $E_{n}$ in Eq. (%
\ref{g14}) by $\Phi _{n}\left( t\right) $ and $\mathcal{E}_{n}$ and write
taking into account the degeneracy with respect to the shift of time $%
t\rightarrow t-t_{0}$ the following formula for the partition function%
\begin{eqnarray}
Z &=&T_{0}^{-1}\sum_{n}\int_{0}^{T_{0}}dt_{0}  \label{g19b} \\
&&\times \left( \Phi _{n}^{+}\left( t-t_{0}\right) ,\Phi _{n}\left(
t-t_{0}\right) \right) \exp \left( -\mathcal{E}_{n}/T\right) .  \notag
\end{eqnarray}

Using the periodicity and the normalization, Eq. (\ref{g12}), of the
functions $\tilde{\Phi}_{n}\left( t\right) $, one comes after integration
over $t_{0}$ to the following formula for the partition function
\begin{equation}
Z=\sum_{n}\exp \left( -\mathcal{E}_{n}/T\right) .  \label{g20}
\end{equation}

Equation (\ref{g20}) may generalize the standard formula for the partition
function, Eq. (\ref{g14}), to the case of the time-dependent wave functions.
It shows that, although one obtains time-dependent order parameters and wave
functions, the degeneracy with respect to the shift of time $t\rightarrow
t-t_{0}$ and the necessity of the averaging over $t_{0}$ allows one to
write, in particular, a reasonable formula for the partition function $Z.$

Of course, Eqs. (\ref{g19b}, \ref{g20}) are just a guess. It would be
interesting to prove the equivalence between such an approach and the
calculations in imaginary time $\tau $ carried out in the preceding sections
but this is beyond the scope of this work. At present, we merely demonstrate
that the appearance of the time-dependent wave functions does not contradict
thermodynamics.

Although single-time correlation functions do not depend on time, which
becomes clear after integration over $t_{0}$, two or more times correlation
functions can be dependent on differences of times.

\subsection{Why the `no-go' theorem cannot be applied to the present
scenario.}

The results obtained in the previous sections do not agree with conclusions
of the `no-go' theorem \cite{watanabe}, and it is worth discussing this
disagreement in details. Although it looks at first glance that the proof of
the theorem is general, it cannot be used in the situation considered in the
present work.

Actually, the publication \cite{watanabe} contains two different parts.
First, the authors prove that a two-time correlation function of two
arbitrary operators integrated over the volume of the system cannot have the
long-range time order at zero temperature at $T=0$. The consideration of
this part is based on an assumption of locality of the Hamiltonian but does
not imply a specific form of the latter. The second part containing the
proof for finite temperatures is different. It is based on rigorous results
of Ref. \cite{lieb} obtained for quantum spin models. Although this part is
rigorous, the results can be used for a certain class of spin models, while
models of interacting fermions have not been considered at all.

The results of the present work disagree with the results of both the parts
of the publication \cite{watanabe} and now we discuss these two parts
separately.

\subsubsection{Long-range time order at $T=0.$}

The proof of the theorem is performed at zero temperature $T=0$ using a
rather simple chain of inequalities applied for calculation of a two-time
correlation function of two operators (Eq. (6) of Ref. \cite{watanabe}). The
quantum mechanical average $\left\langle 0\right\vert ...\left\vert
0\right\rangle $ is performed using the ground state $\left\vert
0\right\rangle $. However, the final inequality (5) of that work has been
obtained assuming implicitly that the wave function of the ground state $%
\left\vert 0\right\rangle $ did not depend on time. Indeed, this looked
self-evident for a time-independent Hamiltonian, and the authors did not
even mention that this assumption had been made. Within the method of the
functional integration used in the present paper, the concept of the wave
functions of the ground state was not used for calculations. However, a
direct comparison becomes possible using the Hamiltonian approach developed
in this section.

The notion of the order parameter and spontaneous breaking of the
time-translation symmetry was not used in Ref. \cite{watanabe}, and
therefore the question about a time-dependence of the wave functions of the
ground state could not arise. Although all wave functions are time-dependent
even for static Hamiltonians due to the prefactor $\exp \left(
-iE_{n}t\right) $, Eqs. (\ref{f2a}, \ref{f3}), this time-dependent prefactor
cancels its complex conjugate in quantum mechanical averages.

However, if time-dependent periodic order parameters appear as a result of
the spontaneous breaking of the time-translation symmetry, the
eigenfunctions of the Hamiltonian acquire the Floquet-type form, Eq. (\ref%
{g19c}). The pre-factor $\exp \left( -i\mathcal{E}_{n}t\right) $ and its
complex conjugate cancel each other in the average over the ground state in
this case, too, but the periodic functions $\Phi _{0}\left( t\right) $, Eq. (%
\ref{g19c}), are still there. Finally, the average over the ground state
takes effectively a form like $\left\langle \Phi _{0}\left( t\right)
\right\vert ...\left\vert \Phi _{0}\left( t\right) \right\rangle $ with
time-dependent functions $\Phi _{0}\left( t\right) ,$ which invalidates the
proof given in the first part of Ref. \cite{watanabe}.

It is worth emphasizing that we consider a grand canonical ensemble with the
fixed chemical potential, while the standard form, Eq. (\ref{f2a}), of the
solution of the Schr\"{o}dinger equation is written for a Hamiltonian with
fixed number of particles. Therefore, the inequalities of Ref. \cite%
{watanabe} are correct, e.g. for a system of local spins but are not valid
for the model considered here.

In order to make a direct comparison of the present results with those
obtained in the first part of Ref. \cite{watanabe}, it is instructive to
consider the correlation functions of an operator $\hat{A}$ at two different
times $t_{1}$ and $t_{2}.$ We can write the correlation function $C\left(
t_{1},t_{2}\right) $ in the standard way
\begin{equation}
C\left( t_{1},t_{2}\right) =\frac{U_{\mathrm{0}}^{2}}{V^{2}}\left\langle
\Phi _{0}^{\ast }\right\vert \hat{A}\left( t_{1}\right) \hat{A}\left(
t_{2}\right) \left\vert \Phi _{0}\right\rangle ,  \label{g30}
\end{equation}%
where $\Phi _{0}$ is the wave function of the ground state of the
Hamiltonian $\hat{H}$, Eq. (\ref{g1}), and
\begin{equation}
\hat{A}\left( t\right) =e^{i\hat{H}t}\hat{A}e^{-i\hat{H}t}.  \label{g31}
\end{equation}%
Actually, Eqs. (\ref{g30}, \ref{g31}), is the starting point of the proof
given in the first part of Ref. \cite{watanabe}. At the same time, the
function $C\left( t_{1},t_{2}\right) $ coincides with the function $N\left(
t_{1}-t_{2}\right) $, Eq. (\ref{m1}), provided the operator $\hat{A}$ is
chosen in the following form
\begin{equation}
\hat{A}=\sum_{p}\left( c_{p}^{+}\Sigma _{2}c_{p}\right) ,  \label{g32}
\end{equation}%
where $c_{p}^{+}$ and $c_{p}$ are the fermion operators introduced in Eq. (%
\ref{g1}),
\begin{equation}
c_{p}\left( t\right) =e^{i\hat{H}t}c_{p}e^{-i\hat{H}t},  \label{g33}
\end{equation}%
and the Hamiltonian $\hat{H},$ Eq. (\ref{g1}), is used for the calculations.

The function $C\left( t_{1},t_{2}\right) ,$ Eq. (\ref{g30}), can be
rewritten in a form%
\begin{equation}
C\left( t_{1},t_{2}\right) =C_{0}\left( t_{1},t_{2}\right) +C_{1}\left(
t_{1},t_{2}\right) ,  \label{g34}
\end{equation}%
with
\begin{eqnarray}
C_{0}\left( t_{1},t_{2}\right) &=&\frac{U_{\mathrm{0}}^{2}}{V^{2}}\left(
A_{00}\right) ^{2},\quad  \label{g35} \\
C_{1}\left( t_{1},t_{2}\right) &=&\frac{U_{\mathrm{0}}^{2}}{V^{2}}%
\sum_{m\neq 0}A_{0m}A_{m0}e^{-\left( E_{m}-E_{0}\right) \left(
t_{1}-t_{2}\right) }.  \notag
\end{eqnarray}%
In Eqs. (\ref{g35}), $A_{0m}$ are matrix elements of the operator $\hat{A}$
between the ground state $0$ and a state $m.$ It is the correlation function
$C_{1}\left( t_{1},t_{2}\right) $ that has been estimated in Ref. \cite%
{watanabe}. Indeed, subtracting from the function $C_{1}\left(
t_{1},t_{2}\right) $ its value at $t_{1}=t_{2}$ and taking the absolute
value we obtain%
\begin{eqnarray}
&&\left\vert C_{1}\left( t_{1},t_{2}\right) -C_{1}\left( t_{2},t_{2}\right)
\right\vert  \label{g36} \\
&=&\frac{2U_{\mathrm{0}}^{2}}{V^{2}}\sum_{m\neq 0}A_{0m}A_{m0}\sin \frac{%
\left\vert E_{m}-E_{0}\right\vert \left\vert t_{1}-t_{2}\right\vert }{2}
\notag \\
&\leq &\frac{U_{\mathrm{0}}^{2}}{V^{2}}\left\vert t_{1}-t_{2}\right\vert
\sum_{m\neq 0}A_{0m}A_{m0}\left\vert E_{m}-E_{0}\right\vert .  \notag
\end{eqnarray}
In the model considered here, the sum over $m$ corresponds to a sum over
momenta $p$. Replacing the sum over $p$ by the integration with the help of
Eq. (\ref{g1a}) one can see that the inequality (\ref{g36}) agrees with the
inequality (5) of Ref. \cite{watanabe}, although it is obtained now in a
less rigorous way. However, this agreement does not mean that the time
crystals are impossible because the correlation function $C_{0}\left(
t_{1},t_{2}\right) $ in Eqs. (\ref{g35}) is not necessarily equal to zero.
The contribution $C_{0}\left( t_{1},t_{2}\right) $ was not considered in
Ref. \cite{watanabe} at all.

The Hamiltonian $\hat{H}$ is written in such a way that the mean field
theory becomes exact in the limit $V\rightarrow \infty $. This allows us to
make the replacement (\ref{g2}) and follow the subsequent steps, Eqs. (\ref%
{g5}-\ref{g12}). The quantum-mechanical average $C_{00}\left(
t_{1},t_{2}\right) $ in Eq. (\ref{g30}) equals the product of the averages
of the operators $\hat{A}\left( t_{1}\right) $ and $\hat{A}\left(
t_{2}\right) .$ We assume that the initial symmetry of the Hamiltonian $\hat{%
H}$ (\ref{g1}) is broken due to existence of the order parameters $B\left(
t\right) $ and $B_{1}\left( t\right) .$ Then, neglecting the function $%
C_{1}\left( t_{1},t_{2}\right) $ we write
\begin{eqnarray}
&&C\left( t_{1},t_{2}\right)  \label{g28} \\
&=&\frac{U_{\mathrm{0}}^{2}}{V^{2}}\sum_{s}\sum_{p_{1},p_{2}}\left\langle
c_{p_{1}}^{+}\left( t_{1}\right) \Sigma _{2}c_{p_{1}}\left( t_{1}\right)
\right\rangle _{s}\left\langle c_{p_{2}}^{+}\left( t_{2}\right) \Sigma
_{2}c_{p_{2}}\left( t_{2}\right) \right\rangle _{s},  \notag
\end{eqnarray}%
where the angular brackets stand for the quantum-mechanical averaging over
all degenerate states $s$ corresponding to the ground state. Using the
self-consistency equations (\ref{f4}, \ref{f5}) we obtain
\begin{equation}
C\left( t_{1},t_{2}\right) =\overline{B\left( t_{1}-t_{0}\right) B\left(
t_{2}-t_{0}\right) }  \label{g29}
\end{equation}%
and come finally to Eqs. (\ref{m22}-\ref{m25}). The bar in Eq. (\ref{g29})
stands for averaging over the time shift $t_{0}$. This corresponds to
summation over all degenerate states.

Now it is clear at which point the present derivation has deviated from the
arguments of Ref. \cite{watanabe}. The authors of the `no-go' theorem did
not consider the spontaneous breaking of the symmetry with the formation of
the time-dependent order parameter and therefore missed the possibility of
the periodic dependence of the two-time correlation function, Eq. (\ref{g29}%
).

\subsubsection{Quantum spin models.}

The `no-go' proof presented in the second part of Ref. \cite{watanabe} for
finite temperatures is based on rigorous results of Ref. \cite{lieb}. No
doubts, the results of Ref. \cite{lieb} are correct and rigorous but, as
follows already from the title of that publication, the authors considered
quantum spin models, which is sufficiently far away from what is considered
here. So, the rigor of the arguments of that work is not an argument against
the present results because the models considered there are different from
the model considered here. Indeed, the scheme developed here for the fermion
model simply cannot be applied to spin models. Many-body fermion models with
interaction can in some cases be `bosonized' but no general scheme making
this mapping in arbitrary dimensions exists. One-dimensional models are
rather an exception than a rule.

Although the arguments presented in Ref. \cite{lieb} are general, a very
important restriction is imposed on the models under consideration: the spin
operators are local. This means that their commutators decay fast when the
operators are taken at sufficiently distant points. This restriction is
formulated already in the abstract and in the beginning of Sec. 3 of Ref.
\cite{lieb}. The criterion of the locality has been formulated explicitly in
Sec. 2 of an earlier work by one of the authors of Ref. \cite{robinson}.

In fact, a pseudospin reformulation of the Hamiltonian $\hat{H},$ Eq. (\ref%
{g1}), can be carried out, and one has to answer the question why the proof
of Ref. \cite{lieb} cannot be applied to the present case. One can proceed
introducing pseudospin operators $\Sigma _{ip}$ similar to those introduced
by Anderson in Ref. \cite{anderson1} in the BCS model for superconductors.
In the present case, the pseudospin operators are introduced as follows%
\begin{eqnarray}
\hat{\Sigma}_{1p} &=&\left( c_{p}^{1+}c_{p}^{2}+c_{p}^{2+}c_{p}^{1}\right)
=c_{p}^{+}\Sigma _{1}c_{p}  \label{g41} \\
\text{ }\hat{\Sigma}_{2p} &=&-i\left(
c_{p}^{1+}c_{p}^{2}-c_{p}^{2+}c_{p}^{1}\right) =c_{p}^{+}\Sigma _{2}c_{p},%
\text{ }  \notag \\
\hat{\Sigma}_{3p} &=&c_{p}^{1+}c_{p}^{1}-c_{p}^{2+}c_{p}^{2}=c_{p}^{+}\Sigma
_{3}c_{p}.  \notag
\end{eqnarray}%
In a vector notation one writes a vector operator
\begin{equation}
\mathbf{\hat{\Sigma}}_{p}=c_{p}^{+}\mathbf{\Sigma }c_{p},  \label{g42}
\end{equation}%
where $\mathbf{\Sigma }$ is the vector of the Pauli matrices $\Sigma _{i},$ $%
i=1,2,3$.

It is important to emphasize that the pseudospin operators $\hat{\Sigma}$
are merely a convenient computational tool and have nothing to do with real
local spins.

It can easily be checked using the fermion anticommutation relations for the
operators $c_{p}$, $c_{p}^{+}$ that the operators $\hat{\Sigma}_{p}^{i}$
obey spin commutation relations
\begin{equation}
\left[ \hat{\Sigma}_{ip},\hat{\Sigma}_{jp^{\prime }}\right] =2ie_{ijk}\hat{%
\Sigma}_{kp^{\prime }}\delta _{pp^{\prime }},  \label{g43}
\end{equation}%
where $e_{ijk}$ is the antisymmetric tensor. The Hamiltonian $\hat{H}$, Eq. (%
\ref{g1}), can be rewritten in terms of the spin operators $\mathbf{\hat{%
\Sigma}}_{p}$ as%
\begin{eqnarray}
\hat{H} &=&\sum_{p}\left[ \varepsilon ^{+}\left( \mathbf{p}\right)
n_{p}+\varepsilon ^{-}\left( \mathbf{p}\right) \hat{\Sigma}_{3p}\right]
\label{g44} \\
&&-\frac{1}{4V}\left[ U_{\mathrm{0}}\left( \sum_{p}\hat{\Sigma}_{2p}\right)
^{2}-\tilde{U}_{\mathrm{0}}\left( \sum_{p}\hat{\Sigma}_{1p}\right) ^{2}%
\right] ,  \notag
\end{eqnarray}%
where
\begin{equation}
n_{p}=c_{p}^{1+}c_{p}^{1}+c_{p}^{2+}c_{p}^{2}=c_{p}^{+}c_{p}  \label{g45}
\end{equation}%
is the density operator. As the operator $n_{p}$ commutes with $\mathbf{\hat{%
\Sigma}}_{p}$, the term with $n_{p}$ in $\hat{H}$ is not important. The most
interesting correlation function that determines the long-range order in
space, Eq. (\ref{g28}), takes in the pseudospin representation the form
\begin{equation}
C\left( t_{1},t_{2}\right) =\frac{U_{\mathrm{0}}^{2}}{V^{2}}%
\sum_{p_{1},p_{2}}\left\langle \hat{\Sigma}_{2p_{1}}\left( t_{1}\right) \hat{%
\Sigma}_{2p_{2}}\left( t_{2}\right) \right\rangle .  \label{g45a}
\end{equation}%
Eq. (\ref{g44}), is quite convenient for calculations. For example, Eqs. (%
\ref{m12f}, \ref{m12h}, \ref{m12k}, \ref{m12l}) can be obtained from
equations of motion for the operators $\mathbf{\hat{\Sigma}}_{p}$ by the
quantum mechanical averaging of the operators $\mathbf{\hat{\Sigma}}_{p}.$
The derivations are very similar to those carried out in superconductivity
theory using the Anderson pseudospins \cite{altshuler,altshuler1,dzero}.

So, the fermion model specified by the Hamiltonian $\hat{H}$, Eq. (\ref{g1}%
), has been exactly rewritten in terms of the pseudospin operators $\hat{%
\Sigma}_{p}$. The choice of these operators is unambiguous because we are
interested in studying the long-range order in both space and time described
by the function $C\left( t_{1},t_{2}\right) ,$ Eqs. (\ref{g30}-\ref{g33}).
Equation (\ref{g45a}) allows one to use these operators directly for the
calculation of the correlation function of interest. Attempts to find other
spin operators do not make a sense because the function $C\left(
t_{1},t_{2}\right) ,$ Eq. (\ref{g30}), would be considerably more
complicated in terms of the other operators, while a correlation function of
several those operators will not be related to the long-range order. One
should really use the operators $\hat{\Sigma}_{p},$ Eqs. (\ref{g41}), and we
have a well defined model, Eq. (\ref{g44}), for that.

Can the rigorous estimates of Ref. \cite{lieb} repeated for the model, Eq. (%
\ref{g44}), invalidate the hope of obtaining the thermodynamic quantum
time-space crystal? The answer is definitely `no' because the operators $%
\hat{\Sigma}$ written in real space are not local, and the arguments of Ref.
\cite{lieb} cannot be repeated.

Indeed, writing the operators in space points as
\begin{equation}
\hat{\Sigma}_{i}\left( \mathbf{r}\right) =\frac{1}{V}\sum_{\mathbf{p}}\hat{%
\Sigma}_{i\mathbf{p}}e^{i\mathbf{pr}},  \label{g46}
\end{equation}%
we obtain for the Hamiltonian $\hat{H},$ Eq. (\ref{g44}), the following
formula (omitting the term with $n_{p}$)
\begin{equation}
\hat{H}=V\left[ \left( \varepsilon ^{-}\left( -i\mathbf{\nabla }\right) \hat{%
\Sigma}_{3}\left( \mathbf{r}\right) \right) _{\mathbf{r}=0}-\frac{1}{4}%
\left( U_{\mathrm{0}}\hat{\Sigma}_{2}^{2}\left( 0\right) -\tilde{U}_{\mathrm{%
0}}\hat{\Sigma}_{1}^{2}\left( 0\right) \right) \right] ,  \label{g46a}
\end{equation}%
while the correlation function $C\left( t_{1},t_{2}\right) $ takes the form
\begin{equation}
C\left( t_{1},t_{2}\right) =-U_{\mathrm{0}}^{2}\left\langle \hat{\Sigma}%
_{2}\left( t_{1},0\right) \hat{\Sigma}_{2}\left( t_{2},0\right)
\right\rangle ,  \label{g46b}
\end{equation}%
where
\begin{equation}
\hat{\Sigma}_{i}\left( t,\mathbf{r}\right) =e^{i\hat{H}t}\hat{\Sigma}%
_{i}\left( \mathbf{r}\right) e^{-i\hat{H}t}.  \label{g46c}
\end{equation}

The Hamiltonian $\hat{H},$ Eq. (\ref{g46a}), is very different from those
for local spin models. Moreover, we write the commutation relation for the
pseudospin operators $\Sigma _{i}\left( \mathbf{r}\right) $ and $\Sigma
_{j}\left( \mathbf{r}^{\prime }\right) $ at different space points $\mathbf{r%
}$ and $\mathbf{r}^{\prime }$ as
\begin{equation}
\left[ \hat{\Sigma}_{i}\left( \mathbf{r}\right) ,\hat{\Sigma}_{j}\left(
\mathbf{r}^{\prime }\right) \right] =\frac{1}{V^{2}}\sum_{p,p^{\prime
}}e^{i\left( \mathbf{pr+p}^{\prime }\mathbf{r}^{\prime }\right) }\left[ \hat{%
\Sigma}_{ip},\hat{\Sigma}_{jp^{\prime }}\right]  \label{g47}
\end{equation}%
Using Eq. (\ref{g43}) we obtain%
\begin{eqnarray}
\left[ \hat{\Sigma}_{i}\left( \mathbf{r}\right) ,\hat{\Sigma}_{j}\left(
\mathbf{r}^{\prime }\right) \right] &=&\frac{2ie_{ijk}}{V^{2}}\sum_{p}e^{i%
\mathbf{p}\left( \mathbf{r+r}^{\prime }\right) }\hat{\Sigma}_{kp}
\label{g48} \\
&=&\frac{2ie_{ijk}}{V}\hat{\Sigma}_{k}\left( \mathbf{r+r}^{\prime }\right) .
\notag
\end{eqnarray}%
One can see from Eq. (\ref{g48}) that the commutator of the pseudospins $%
\hat{\Sigma}_{i}\left( \mathbf{r}\right) $ at two different points $\mathbf{r%
}$ and $\mathbf{r}^{\prime }$ does not vanish in the limit $\left\vert
\mathbf{r-r}^{\prime }\right\vert \rightarrow \infty $. Presence of the
pre-factor $V^{-1}$ in Eq. (\ref{g48}) does not make the operators $\hat{%
\Sigma}_{i}\left( \mathbf{r}\right) $ local even in the limit $V\rightarrow
\infty $ because this pre-factor is compensated in equations of motion by
the factor $V$ in the Hamiltonian $\hat{H}$, Eq. (\ref{g46a}).

All this means that the operators $\hat{\Sigma}\left( t,\mathbf{r}\right) $
are not local, and applying the results of the work \cite{lieb} based on the
assumption of the locality of the spins to the model considered here is not
justified. As the BCS model of superconductivity can also be reformulated in
terms of the similar Anderson pseudospins, one cannot use the bounds of Ref.
\cite{lieb} for studying properties of superconductivity either. In
particular, amplitude (Higgs) modes \cite%
{vk,spivak,barankov,altshuler,altshuler1,dzero,moor} are not known in
systems of local spins. So, the local quantum spin models cannot in general
describe electron-electron or electron-hole pairing in fermion models with
interaction, and the bounds obtained in Ref. \cite{lieb} cannot be applied
to the model considered here.

\subsection{Concluding remarks to Section \protect\ref{sec:Mean_Field}.}

We conclude this Section with the statement that the `no-go' theorem of Ref.
\cite{watanabe} is not applicable to the phenomenon of the thermodynamic
quantum time-space crystal considered in the present work.

Strictly speaking, the results are obtained using the Hamiltonian $\hat{H}$
with a somewhat special interaction. Although the original electron-electron
interaction is short-ranged, the form of the interaction in Eq. (\ref{g1})
corresponds to an `infinite-range' interaction of the electron-hole pairs.
However, mean field theories for systems of interacting electrons are well
justified for small ratios of the order parameter to the Fermi energy, which
holds in most cases.

The fact that one can use the wave functions in the standard form of Eq. (%
\ref{f2a}) is applicable for models with a fixed number of particles. At the
same time, the grand canonical enesemble of interacting electrons is
discussed in the present work, and the wave functions have a more
complicated form that cannot be reduced to Eq. (\ref{f2a}). The spontaneous
breaking of the symmetry occurs in the grand canonical ensemble. So, the
non-trivial time-dependent form of the wave functions obtained here does not
contradict the standard form of the wave functions, Eq. (\ref{f2a}).

The consideration of the first part of Ref. \cite{watanabe} was based on the
assumption of the locality of the Hamiltonian. The authors of Ref. \cite%
{watanabe} write explicitly that their proof is not applicable to models
with an infinite-range interactions. Although the electron-electron
interaction in the Hamiltonian $\hat{H}$, Eq. (\ref{g1}), is short ranged,
it describes effectively an `infinite range' interaction of elecron-hole
pairs. Could it mean that the results obtained in the present paper are not
general and are specific to the considered model only? In this case, this
would mean that taking a more realistic interaction would destroy the
long-range time oscillations, and the results obtained here would not be as
interesting.

Fortunately, it does not seem to be so. Such a scenario would be possible
for a spin model with an infinite-range interaction. Indeed, according to
the estimates of Ref. \cite{watanabe} that are definitely correct for the
spin systems, making the radius of the interaction finite would destroy the
long-time oscillations but does not apply to the infinite range interaction.
However, the infinite range of the interaction would lead to, e.g.,
ferromagnetic or antiferomagnetic states rather to time crystals. I do not
see any chance to obtain the TQTC with a time-dependent order parameter in
the spin models, and they are not considered here. The situation with the
`infinite-range' interaction of the electron-hole pairs in Eq. (\ref{g1}) is
completely different because the electron-electron interaction is
short-ranged and the Hamiltonian $\hat{H}$ is local.

This property can easily be seen using an important estimate for local
Hamiltonians written on top of p. 3 in the proof of Ref. \cite{watanabe},
\begin{equation}
\left[ \hat{A}\left[ \hat{H},\hat{A}\right] \right] \propto V.  \label{g49}
\end{equation}

The proportionality to the volume $V$ instead of $V^{2}$ should indicate
according to Ref. \cite{watanabe} the absence of the time crystal behavior.
The proportionality to $V^{2}$ is possible for the infinite-range
interactions in the spin models.

On the other hand, it is not difficult to calculate the same commutator for
the Hamiltonian $\hat{H}$, Eqs. (\ref{g1}, \ref{g44}), and operator $\hat{A}$%
, Eq. (\ref{g32}). The computation is easy with the help of the pseudospins $%
\mathbf{\hat{\Sigma}}$, Eqs. (\ref{g41}-\ref{g43}), and one obtains%
\begin{eqnarray}
\left[ \hat{A}\left[ \hat{H},\hat{A}\right] \right] &=&-4\sum_{p}\varepsilon
^{-}\left( \mathbf{p}\right) \hat{\Sigma}_{3p}  \label{g50} \\
&&+\frac{8\tilde{U}_{\mathrm{0}}}{V}\sum_{p_{1}\neq p_{2}}\left( \hat{\Sigma}%
_{3p_{1}}\hat{\Sigma}_{3p_{2}}-\hat{\Sigma}_{1p_{1}}\hat{\Sigma}%
_{1p_{2}}\right) .  \notag
\end{eqnarray}%
Standard replacement of the sum over the momenta by the integral, Eq. (\ref%
{g1a}), leads to the conclusion that the estimate, Eq. (\ref{g49}), holds in
our case, too. This means that the Hamiltonian $\hat{H}$, Eq. (\ref{g1}),
with the `infinite-range' interactions of the elecron-hole pairs should be
classified as a short-range Hamiltonian in the proof of Ref. \cite{watanabe}%
. Existence of the TQTC is a consequence of the breaking of the
time-translation symmetry, and the `infinite-range' interaction of the
electron-hole pairs is helpful but not necessary for the latter phenomenon.
A more general form of the electron-electron interaction in the Hamiltonian $%
\hat{H}$ cannot change the estimate (\ref{g49}) because the interaction is
short-ranged anyway.

\section{\label{sec:Experiment}How to observe the time-space crystal
experimentally?}

As the average order parameter $B\left( t\right) $ equals zero, Eq. (\ref%
{m19}), one cannot expect oscillating quantities like, e.g., currents. The
oscillations should be seen in two-times correlation functions like $N\left(
t\right) ,$ Eqs. (\ref{m1}, \ref{m22}, \ref{m25}). Of course, possibility of
an experimental observation depends on systems that can be described by the
action, Eqs. (\ref{k24}- \ref{k30}, \ref{e0}-\ref{k1b}). Apparently, one can
find various models exhibiting the time crystal behavior. However, since the
action, Eqs. (\ref{k24}- \ref{k30}, \ref{e0}-\ref{k1b}), is suggested here
for description of cuprates and one can expect correlations of magnetic
moments oscillating in time and space (in contrast to static magnetic
moments in DDW theories), the polarized neutron spectroscopy can be a proper
tool. In this case, the Fourier-transform of the function $N\left( t\right) $
determines directly the cross-section of the inelastic scattering. It is
important that the magnetic moments are basically perpendicular to the
planes, which can help to distinguish them from the antiferromagnetic spin
excitations at $\left( \pi ,\pi \right) .$ Calculating the Fourier transform
$N\left( \omega \right) $ of the function $N\left( t\right) $, Eq. (\ref{m22}%
), and comparing it with the one for the time-independent DDW state $2\pi
\gamma ^{2}\delta \left( \omega \right) $ for the same model one can write
at low temperatures the ratio of the experimental responses at $\left( \pi
,\pi \right) $ for these two states as
\begin{equation}
\chi \left( \omega ,\mathbf{q}\right) =\chi _{0}\sum_{n=1}^{\infty
}f_{n}^{2}\delta \left( \omega -2n\gamma \right) \delta \left( \mathbf{q-Q}%
_{AF}\right) ,  \label{k23a}
\end{equation}%
where $f_{n}$ is determined by Eq. (\ref{m23}), and $\chi _{0}$ determines
the response $\chi _{DDW}$ of the DDW state, $\chi _{DDW}\left( \omega
\right) =\chi _{0}\delta \left( \omega \right) .$

Actually, anisotropic magnetic $\left( \pi ,\pi \right) $ excitations have
been observed \cite{hayden} in $YBa_{2}Cu_{3}O_{6.9}$. The response $\chi
_{c}\left( \omega ,\mathbf{q}\right) $ perpendicular to planes with wave
vector $\mathbf{q=}\left( 1.5,k,1.73\right) $ had a pronounced peak at $%
k=0.5 $ at $\omega =26$ $meV$ at temperature $94K,$ while this peak was
suppressed at $10K,$ which was below the superconducting temperature $T_{c}$
. At the same time, the parallel component $\chi _{a/b}$ was not so
sensitive to temperature. Using the results obtained in the present paper
one might argue that the peak in the susceptibility $\chi _{c}\left( \omega ,%
\mathbf{q}\right) $ was due the oscillations of magnetic moments. Within
this picture, the time-space crystal state had to be suppressed by the
superconductivity.

Being quite general, the model specified by Eqs. (\ref{e0}-\ref{k1b}) may
also be applied to other systems and one should design proper experiments
for each case.

\section{\label{sec:Conclusion}Discussion.}

The main result of the present study is that the quantum time crystals may
exist as a thermodynamically stable state in macroscopic systems even in the
limit of infinite volume, $V\rightarrow \infty $. The non-decaying
oscillations do not lead to an oscillating behavior of classical quantities
and the energy is conserved. The oscillations show up in correlation
functions of several times and can in principle be observed in e.g. quantum
scattering experiments. The order parameter of the thermodynamic quantum
time-space crystals is periodic in both real and imaginary times as well as
in space but its average over the phases of the oscillations vanish.

The procedure of the averaging over the phase of the oscillations is
equivalent to a new quantum description in terms of the operator order
parameter. In this picture, the oscillating behavior of correlation
functions originates from virtual transitions between states of an
oscillator. The distance between the energy levels equals the energy of
breaking the electron-hole pairs and does not decrease in the limit of
infinite volume, $V\rightarrow \infty .$

All calculations have been carried out in the limit of low temperatures for
a simplified version of the spin-fermion model with overlapping hot spots
relevant for the superconducting cuprates. In this simplified version, the
electron-electron interaction is written in a form of an infinite-range
interaction of electron-hole pairs. This is a standard approximation used
for studying new phase transitions in many-body electron systems first
suggested by Bardeen-Cooper-Schrieffer in their fundamental paper on theory
of superconductivity \cite{bcs}. This type of simplified models allows one
to solve the problem exactly and gives results that could be obtained from
more general models using a proper mean field approximation.

The electron-electron interaction of the original model is usually
short-ranged, and its explicit form is irrelevant for the phenomena studied
by this method. The possibility of using the mean field approximation in the
electron systems is different from the one used in, e.g., spin models. In
the latter case, the mean field approximation is usually justified by using
a long-range interaction between the spins, while in the electron models one
needs merely a small ratio of the order parameter to the Fermi energy.
Formation of charge density wave (CDW) and many other phase transitions have
been predicted and described in this way without assuming any special
long-range interaction.

Of course, one has to investigate fluctuations near the mean field solution.
Actually, it is possible to calculate the fluctuations in the same way as it
is usually done when studying new phases in electron systems. In practical
terms, one can perform Hubbard-Stratonovich decoupling of the interaction in
the original Lagrangian, Eqs. (\ref{k25}-\ref{k30}), and integrate out the
fermion fields $\chi ,\chi ^{\ast }.$ Then, one obtains a free energy
functional of the auxiliary fields $b$ and $b_{1}$ like the ones written in
Eq. (\ref{k10}). Generally, the fields $b$ and $b_{1}$ are functions of not
only time but also of coordinates. Following this scheme one finds the
minimum of the functional $\mathcal{F}\left[ b,b_{1}\right] $ and expands in
the deviations $\delta b$ and $\delta b_{1}$ from the functions $b^{\left(
0\right) }$ and $b_{1}^{\left( 0\right) }$ at the minimum. The functions $%
b^{\left( 0\right) }$ and $b_{1}^{\left( 0\right) }$ are just solutions of
the mean field equations.

As a result, one obtains a positive-definite quadratic form. Contribution of
higher order terms in $\delta b$ and $\delta b_{1}$ can be taken into
account by expansion in these variables and calculation of Gaussian
integrals. In low dimensions, some of these contributions can be divergent,
and one should use more sophisticated schemes of calculations. However, in
quasi-two and three-dimensional systems the corrections to the mean field
solutions are convergent, and unless they are very large the mean field
theory is a reasonable approximation. Preliminary consideration for the
present model shows that expansions near the minimum of the free energy
functional lead to convergent integrals and they are not very large. A more
detailed investigation of this problem is left for future but for now the
mean field theory does not look a bad approximation.

It is important to emphasize that a thermodynamically stable time-crystal
state corresponding to the minimum of the free energy has been obtained for
the first time. All previous works on time crystals have been performed for
non-equilibrium systems, which means that they were not at a minimum.
Therefore fluctuations could rather easily drive the systems away from the
initial state and destroy mean field solutions.

An important problem that awaits its resolution is finding the exact minimum
of the free energy functional $\mathcal{F}\left[ b,b_{1}\right] $, Eq. (\ref%
{k10}). Although the scheme of the calculations used for drawing Figs. \ref%
{fig:energy} is plausible, a more precise computation is certainly needed.
Of course, one can do this numerically but a more detailed analytical
investigation would be helpful. Alternatively, one could study Eqs. (\ref%
{m12f}, \ref{m12g}, \ref{m12h}-\ref{m12l}) for the pseudospins $\mathbf{S}_{%
\mathbf{p}}\left( t\right) .$ It is well-known that corresponding equations
for superconductors are integrable \cite{altshuler,altshuler1,dzero}, and
analyzing the possibility of the integrability of Eqs. (\ref{m12f}, \ref%
{m12g}, \ref{m12h}-\ref{m12l}) would be of a great interest.

One of the most exciting concepts of theoretical condensed matter and high
energy physics is the spontaneous breaking of symmetry. According to this
concept, original symmetries of the Hamiltonian can change below the phase
transition due to formation of the order parameter. For example, in the BCS
theory the effective Hamiltonian does not conserve the particle number,
appearance of CDW breaks the space-translation symmetry, etc. The
thermodynamic quantum time crystal (TQTC) proposed here is conceptually very
similar to CDW. The main difference is that now the time-translation
symmetry is broken instead of the space-translation one. Both the phenomena
are described within similar computational schemes (actually, in the present
model both the time and space symmetries are broken and one can speak of
time-space crystal).

The inapplicability of the `no-go' theorem of Ref. \cite{watanabe} to the
results obtained here follows from the fact that the authors did not imply
the possibility of the spontaneous breaking of the time-translation symmetry
in their proof. Their additional arguments based on the use of known
rigorous results for models of local spins cannot be applied to the electron
models considered here.

Although study of the time crystals is already a mature field of research
and a lot of interesting results have been obtained using various models and
techniques, TQTC is really a new phenomenon because, in all the previous
works, non-equilibrium phenomena were considered. Although the time crystal
as a thermodynamic state has been proposed by Wilczek in his pioneering work
\cite{wilczek}, a subsequent study has shown that his state was not in
thermodynamic equilibrium. As the `no-go' theorem of Ref. \cite{watanabe}
has been widely accepted, no attempts to obtain a thermodynamically stable
time crystal have been undertaken since then. The fact that the time crystal
obtained here is thermodynamically stable makes it different from all time
crystals obtained previously. One might argue that already a trivial 2-level
system like a spin in a magnetic field gives a similar 2-time correlation
function $N\left( t\right) $ \cite{watanabe}. However, it is demonstrated
here that this effect can exist in a macroscopic systems in the limit $%
V\rightarrow \infty $ in contrast to the statement of that work. The fact
that a phase transition effectively results in the formation of energy
levels with non-vanishing in the limit $V\rightarrow \infty $ level spacing
and quantum oscillations between the levels can in principle be observed in
physical quantities, is non-trivial and does not have any analogs.

The novelty and importance of the TQTC in macroscopic samples and its
difference with respect to the non-equilibrium time-crystals can be
understood in simple terms making a qualitative comparison with three
different types of electron systems in a magnetic field with non-decaying
currents: 1) an electron in an atom subjected to a magnetic field, 2) ideal
conductor in a magnetic field, and 3) superconductor in a magnetic field.

In case 1) a non-decaying electron current flows around the atom and
generates a constant magnetic moment.

In case 2) the applied magnetic field can also induce non-decaying
diamagnetic currents and a diamagnetic moment but these quantities are
sensitive to impurities phonons, etc. They depend also on the pre-history
because one obtains different results using different experiments. The
diamagnetic currents are induced when one applies the magnetic field at low
temperatures. On the other hand, one does not obtain any currents if one
starts with the system in the magnetic field at high temperatures, such that
the current has vanished due to the scattering of the electrons on phonons.
Then, one cools down the sample but does not obtain any current. The
dependence on the pre-history is typical for non-equilibrium phenomena. Of
course, one can maintain the currents applying an oscillating magnetic field
that would serve as driving force.

At last, in case 3) one has a genuine thermodynamic state because, as soon
as the applied magnetic field stops changing, the current evolves to a
constant value that does not depend on the pre-history (Meissner effect).
There are plenty of other interesting effects distinguishing the
superconductors from the ideal conductors. The superconductors are
characterized by the quantum coherence all over the sample and arise as a
result of the spontaneous breaking of the symmetry.

It is clear why the phenomena 1)-3) are different. The phenomenon 1) is
characterized by a constant value of the current but is microscopic. Both
the phenomena 2) and 3) are macroscopic but 2) is a non-equilibrium
phenomenon, while 3) is a thermodynamic quantum many-body state that appears
as the result of the spontaneous breaking of the symmetry. We see on this
example that although a non-decaying current can exist in all these systems,
their nature is different, and one should study additionally other physical
quantities to understand what phenomenon one deals with.

Now we can understand easily the difference between the three different
types of the time-crystal-like states discussed here. The microscopic
2-level system showing the time-oscillations is just trivial, and there is
no sense to call it `time crystals'. The non-equilibrium time-crystals can
be very non-trivial but they imply driving forces or they can be very
sensitive to various perturbations, etc. In contrast, TQTC is a genuine
thermodynamic state because the oscillations of the two-time correlation
functions $N\left( t\right) $ do not depend on time at all, and this
dependence is not sensitive to impurities or other static perturbations.
Fluctuations are also less important in TQTC than in systems out of
equilibrium, and the mean field approximation is better justified. In a
certain sense, TQTC is an analog of the superconductors but the analogy to
CDW is more direct.

The correlation functions $N\left( t\right) $ can, in principle, be
experimentally observed in scattering experiments, and study of the
properties of the TQTC looks interesting and important. The form of the
two-band Hamiltonian, Eq. (\ref{g1}), is quite general and one can
anticipate applications to other materials and devices. The non-decaying
oscillations is a very important property for qubits and one might think of
interesting applications.

Description of phase transitions between the time-space crystal state and
normal metal or the state with the time-independent order parameter can also
be of a great interest because they definitely differ from known phase
transitions.

The thermodynamic quantum time-space crystal may be a good candidate for the
still mysterious pseudogap state in superconducting cuprates. In many
respects, its properties like breaking time reversal symmetry, gap in the
electron spectrum, etc, resemble those of the DDW state. At the same time,
no static magnetic moments oscillating with the antiferromagnetic vector ($%
\pi ,\pi $) have been observed so far in agreement with the predictions for
the thermodynamic time-space crystal. Instead, the present results show that
correlations of the magnetic moments oscillate in time and can in principle
be studied in experiments with polarized inelastic neutron scattering.

\begin{acknowledgments}
I would like to thank S.I. Mukhin, B.Z. Spivak, P.A. Volkov, G.E. Volovik,
and P.B. Wiegmann for useful discussions.

Financial support of Deutsche Forschungsgemeinschaft (Projekt~EF~11/10-1)
and of the Ministry of Science and Higher Education of the Russian
Federation in the framework of Increase Competitiveness Program of NUST
\textquotedblleft MISiS\textquotedblright (Nr.~K2-2017-085") is greatly
appreciated.
\end{acknowledgments}

\appendix

\section*{Appendix.}

\section{Final formulas for the free energy of instantons and antiinstantons.%
}

The integrals (\ref{k67}, \ref{k69}-\ref{k72}) can be simplified changing
the variables of the integration as
\begin{eqnarray}
p_{x} &=&\pm p\sqrt{\frac{1+u}{2}},\;p_{y}=\pm p\sqrt{\frac{1-u}{2}},\;
\label{k75} \\
E &=&\varepsilon ^{-}\left( \mathbf{p}\right) =\frac{\alpha -\beta }{2}%
p^{2}u+P  \notag
\end{eqnarray}%
with $-1<u<1.$ Then, the integral $R$ for any non-singular function $f\left(
E\right) $,
\begin{equation}
R=\int f\left( \varepsilon ^{-}\left( \mathbf{p}\right) \right) \frac{d%
\mathbf{p}}{\left( 2\pi \right) ^{2}},  \label{k76}
\end{equation}%
can be written in the form
\begin{equation}
R=\frac{2}{\left( 2\pi \right) ^{2}}\frac{1}{\alpha +\beta }\int_{-1}^{1}%
\left[ \int_{P\left( T\right) }^{P\left( T\right) +\Lambda u}f\left(
E\right) dE\right] \frac{1}{u}\frac{1}{\sqrt{1-u^{2}}}du.  \label{k77}
\end{equation}%
Further, the integral over $u$ can be transformed integrating by parts and
we reduce $R$, Eq. (\ref{k76}), to the following integral over one variable%
\begin{equation}
R=\frac{2}{\left( 2\pi \right) ^{2}}\frac{\Lambda }{\alpha +\beta }%
\int_{-1}^{1}f\left( P\left( T\right) -u\Lambda \right) \ln \frac{\sqrt{%
1-u^{2}}+1}{\left\vert u\right\vert }du,  \label{k78}
\end{equation}%
which simplifies considerably the original integration over $\mathbf{p}$.

Finally, using parameters $\bar{P}=P/\gamma $, $\bar{\Lambda}=\Lambda
/\gamma $ we write the free energy $\Delta F$ as
\begin{equation}
\frac{\Delta F}{VT}=\frac{1}{\pi ^{2}}\frac{m\Lambda }{\alpha +\beta }\left(
s_{0}+s_{int}\right) .  \label{k79}
\end{equation}%
Herein,

\begin{eqnarray}
s_{0} &=&2\int_{-1}^{1}\left[ \ln \frac{\sqrt{\left( x-yu\right) ^{2}+1}+1}{%
\left\vert x-yu\right\vert }-\frac{1}{\sqrt{\left( x-yu\right) ^{2}+1}}%
\right]  \notag \\
&&\times \ln \frac{\sqrt{1-u^{2}}+1}{\left\vert u\right\vert }du,
\label{k80}
\end{eqnarray}%
and%
\begin{eqnarray}
s_{int} &=&-\frac{I_{0}}{2}\int_{0}^{1}\left[ K_{0}+\frac{1}{4}\left( K_{1}-%
\frac{M^{2}}{L}\right) \left( 1-v^{2}\right) ^{2}\right] ^{-1}  \notag \\
&&\times \left( 1-v^{2}\right) dv,  \label{k81}
\end{eqnarray}%
where%
\begin{equation*}
K_{0}=\left( 1+\frac{U_{0}}{\tilde{U}_{0}}\right) \int_{-1}^{1}\frac{1}{%
\sqrt{\left( x-yu\right) ^{2}+1}}\ln \frac{\sqrt{1-u^{2}}+1}{\left\vert
u\right\vert }du,
\end{equation*}%
\begin{equation*}
L=\int_{-1}^{1}\frac{1}{\left( \left( x-yu\right) ^{2}+1\right) ^{3/2}}\ln
\frac{\sqrt{1-u^{2}}+1}{\left\vert u\right\vert }du,
\end{equation*}%
\begin{eqnarray*}
M &=&\int_{-1}^{1}\frac{x-yu}{\left( \left( x-yu\right) ^{2}+\frac{\left(
1-k\right) ^{2}}{4}\right) \left( \left( x-yu\right) ^{2}+1\right) ^{3/2}} \\
&&\times \ln \frac{\sqrt{1-u^{2}}+1}{\left\vert u\right\vert }du,
\end{eqnarray*}%
\begin{eqnarray*}
K_{1} &=&\int \frac{1}{\left( \left( x-yu\right) ^{2}+1\right) ^{3/2}\left(
\left( x-yu\right) ^{2}+\frac{\left( 1-k\right) ^{2}}{4}\right) } \\
&&\times \ln \frac{\sqrt{1-u^{2}}+1}{\left\vert u\right\vert }du,
\end{eqnarray*}%
and%
\begin{eqnarray*}
I_{0} &=&\Big[\int \frac{sgn\left( x-yu\right) }{\sqrt{\left( x-yu\right)
^{2}+\frac{\left( 1-k\right) ^{2}}{4}}}\frac{1}{\sqrt{\left( x-yu\right)
^{2}+1}} \\
&&\times \ln \frac{\sqrt{1-u^{2}}+1}{\left\vert u\right\vert }du\Big]^{2}.
\end{eqnarray*}%
Equation (\ref{e19}) that determines the gap $\gamma $ takes the form%
\begin{equation}
\frac{\alpha +\beta }{U_{0}}=\frac{1}{2\pi ^{2}}\int_{-1}^{1}\frac{y}{\sqrt{%
\left( x-yu\right) ^{2}+1}}\ln \frac{\sqrt{1-u^{2}}+1}{\left\vert
u\right\vert }du.  \label{k82}
\end{equation}%
Equations (\ref{k80}, \ref{k81}, \ref{k82}) have been used for computation
of the surfaces in Figs. \ref{fig:energy}(a-e).

\end{document}